\documentclass[12pt,a4paper]{article}

\usepackage{jheppub}

\usepackage{etoolbox}
    \makeatletter
    \patchcmd{\maketitle}{\@fpheader}{}{}{}
    \makeatother

\usepackage{amssymb}
\usepackage{mathtools}

\usepackage[utf8]{inputenc}
\usepackage{microtype}
\usepackage{lmodern}
\usepackage{exscale}
\usepackage{upgreek}

\usepackage{enumitem}

\usepackage{mathrsfs}

\usepackage{graphicx}

\usepackage[dvipsnames]{xcolor}
\definecolor{yellowish}{rgb}{0.969, 0.788, 0.059}
\definecolor{bluish}{rgb}{0.329,0.757, 0.965}
\definecolor{reddish}{rgb}{1, 0.529, 0.529}
\definecolor{greenish}{rgb}{0.2, 0.8, 0.2}

\usepackage{array}

\usepackage{pgfplots}

\usepackage[low-sup]{subdepth}

\newcommand{\scr}{\scriptscriptstyle}
\newcommand{\bb}{\boldsymbol}
\newcommand{\widesim}[2][1.5]{\mathrel{\overset{#2}{\scalebox{#1}[1]{$\sim$}}}}

\usepackage{soul}

\title{Oscillating scalar dissipating in a medium}

\title{Oscillating scalar dissipating in a medium}

\author[1]{Wen-Yuan~Ai,}
\emailAdd{wenyuan.ai@uclouvain.be}
\author[1]{Marco~Drewes,}
\emailAdd{marco.drewes@uclouvain.be}
\author[2]{Dra\v{z}en~Glavan,}
\emailAdd{glavan@fzu.cz}
\author[\,3,1]{Jan~Hajer}
\emailAdd{jan.hajer@unibas.ch}

\affiliation[1]{Centre for Cosmology, Particle Physics and Phenomenology (CP3),
	\\
	Universit\'{e} catholique de Louvain, 
	\\
	Chemin du Cyclotron 2, 1348 Louvain-la-Neuve, Belgium}
\affiliation[2]{
	CEICO, Institute of Physics of the Czech Academy of Sciences (FZU), 
	\\
	Na Slovance 1999/2, 182 21 Prague 8, Czech Republic}
\affiliation[3]{Department of Physics, University of Basel, 
	\\
	Klingelbergstra{\ss}e 82, CH-4056 Basel, Switzerland}

\abstract{
We study how oscillations of a scalar field condensate are damped 
due to dissipative effects in a thermal medium.
Our starting point is a non-linear and non-local condensate equation of motion descending from
a 2PI-resummed effective action derived in the Schwinger-Keldysh formalism
appropriate for non-equilibrium quantum field theory. We solve this non-local equation 
by means of multiple-scale perturbation theory
appropriate for time-dependent systems, obtaining approximate analytic solutions
valid for very long times.
The non-linear effects lead to power-law damping of oscillations, that at late times transition
to exponentially damped ones characteristic for linear systems. 
These solutions describe the evolution very well, as we demonstrate numerically in a number
of examples.
We then approximate the non-local 
equation of motion by a 
Markovianised one, resolving the ambiguities appearing in the process, and solve it utilizing
the same methods to find the very same leading approximate solution.
This comparison justifies the use of Markovian equations at leading order. 
The standard time-dependent perturbation theory in comparison is not
capable of describing the non-linear condensate evolution beyond the early time regime of 
negligible damping. 
The macroscopic evolution of the condensate is interpreted in terms of microphysical
particle processes.
Our results have implications for the quantitative description of the decay of cosmological  
scalar fields in the 
early Universe, and may also be applied to other physical systems.

}

\preprint{{\tt CP3-21-48}}

\begin{document}

\maketitle

\section{Introduction}
\label{sec: Introduction}

Scalar fields may have played a key role in shaping the observable Universe. Cosmic inflation
\cite{Starobinsky:1980te,Guth:1980zm,Linde:1981mu, ArmendarizPicon:1999rj} provides the (probably) simplest explanation for the overall geometry of the cosmos as well as the small perturbations that seeded the formation of 
galaxies. In a large number of 
models~\cite{Martin:2013tda} the quasi-exponential expansion is driven by the 
potential energy of one or several scalar fields dubbed inflaton(s).
The transition from the inflationary phase to the radiation-dominated epoch occurs when the inflaton dissipates its energy into other degrees of freedom through particle production that fills the Universe with a dense plasma~\cite{Albrecht:1982mp,Dolgov:1989us,Traschen:1990sw,Shtanov:1994ce,Kofman:1994rk,Boyanovsky:1996sq,Kofman:1997yn}. Even though this process of cosmic reheating
is of uttermost importance for a complete understanding of the cosmic history as it sets
the initial conditions of the hot big bang, it has been studied far less than inflation itself, 
cf.~\cite{Bassett:2005xm,Allahverdi:2010xz,Amin:2014eta} for reviews. 
A quantitative understanding of the reheating phase is highly desirable because it affects the mapping between cosmological observables and inflationary model parameters~\cite{Martin:2010kz,Adshead:2010mc,Easther:2011yq,Martin:2014nya}, and fundamental physics parameters~\cite{Drewes:2015coa,Ueno:2016dim,Drewes:2017fmn,Drewes:2019rxn}.
Moreover, cosmological relics such as gravitational waves~\cite{Khlebnikov:1997di,Caprini:2018mtu} 
or Dark Matter~\cite{Kolb:2003ke,Gorbunov:2010bn,Harigaya:2014waa,Almeida:2018oid,Lebedev:2021zdh,Bezrukov:2020jmo}
can be produced during this period.

In many scenarios reheating occurs during coherent oscillations of the inflaton condensate around the 
minimum of its effective potential. 
For the most part this process takes place in a plasma of produced particles which 
practically constitute
a thermodynamic ensemble described by a von Neumann density operator $\varrho$.
In the present work we are interested in the background evolution,
in particular the evolution of the inflaton condensate
defined as an average over both quantum and statistical fluctuations,
\begin{equation}\label{AverageDef}
\varphi (t) = \bigl\langle \Phi(x) \bigr\rangle \equiv {\rm Tr} \bigl[ \varrho  \Phi(x) \bigr].
\end{equation}
Even though the underlying laws of physics are local, i.e. equations governing the evolution of
the statistical ensemble are local (as is the von Neumann equation for $\varrho$), when expressed 
in terms of a finite number of momenta 
of the statistical ensemble they generally exhibit 
non-Markovian (non-local) character. Such are the Kadanoff-Baym equations
supplemented by the effective condensate equation that arise in the two-particle-irreducible (2PI) effective
action formalism that we utilize here. In this work we shall specifically be analysing the non-Markovian equation,
\begin{equation}\label{FullEoM}
\ddot{\varphi}(t) + M^2 \varphi(t)
    + \frac{ \Lambda \varphi^3(t) }{6}
	+ \int_{t_0}^{t}\!
	    dt' \, \pi_{\scr R}(t\!-\!t') \varphi(t')
	+ \frac{ \varphi(t) }{6} \!
	\int_{t_0}^{t}\! dt' \,
	    v_{\scr R}(t\!-\!t') \varphi^2(t') = 0 \, ,
\end{equation}
describing the relaxation of~$\varphi$ as it oscillates around the minimum,
where for simplicity we neglect the expansion of the Universe.
This equation occurs for a broad class of interactions with both bosons and fermions,
and the information about specific interactions is encoded in the
time translation-invariant integral kernels~$\pi_{\scr R}(t\!-\!t')$ 
and~$v_{\scr R}(t\!-\!t')$, which 
are closely related to the retarded self-energies and 
proper four-vertices in a given particle physics model.
Equation~(\ref{FullEoM}) arises in the small field expansion, and the non-local 
terms can be seen as due to perturbative particle interactions and decays.
Therefore, the applicability of this equation is
limited to the {\it mildly non-linear regime}, where resonant particle production
is inefficient, but where non-linearities cannot be neglected.


Non-Markovian equations, such as~(\ref{FullEoM}) for the one-point function or similar equations
for two-point functions, 
 are analytically tractable in a limited number of simple cases,
such as in the linear regime when all non-linearities can be 
neglected~\cite{Anisimov:2008dz,Anisimov:2010dk,Garbrecht:2011xw,Garny:2011hg,Drewes:2012qw}
and the equation can be solved exactly.
However, this is not satisfactory for applications when condensate oscillations extend beyond the linear regime,
which they typically do, and thus a better understanding of the non-linear evolution
is highly desirable. In simplified models the problem can be addressed 
by resorting directly to lattice computations~\cite{Berges:2008wm}.
However, having a semi-analytic understanding of how the inflaton relaxes during reheating, and
how this is affected by different types of interactions and non-linear evolution is clearly advantageous.
This work is devoted to developing both qualitative and quantitative understanding of this problem.

\medskip

The first part of the paper is dedicated to
developing a systematic approximation scheme for solving non-Markovian 
equations of the type~(\ref{FullEoM})
in the mildly non-linear regime, based on applying the {\it multiple-scale perturbation theory}~\cite{Bender,Holmes}. 
This approximation method is designed to describe systems where physical processes happen
on different time scales. 
The system at hand exhibits such behaviour when the coupling constants are perturbatively small
and when the kernels of non-local terms in~Eq.~(\ref{FullEoM})  provide an effective window 
spanning no more than few oscillations: dissipative and dispersive effects
at play are locally small, but build up over long times and many oscillations
to appreciably alter the evolution. Such effects cannot
be captured by the naive time-dependent perturbation theory, which is plagued by secular artefacts
that grow without bound. Multiple-scale perturbation theory is designed to avoid such artefacts by effectively resumming
naive perturbation theory, thus yielding uniform approximations valid for very long times. 
Even the leading order approximation captures the evolution with great precision, as we confirmed 
by comparing to direct numerical solutions of Eq.~(\ref{FullEoM}) for a number of cases. 
The solutions exhibit interesting features: (i) a power-law damping behaviour arises from the 
cubic non-local term in Eq.~(\ref{FullEoM}), which competes with exponential damping arising from the 
linear non-local term, and (ii) there are in general time-dependent frequency corrections to the 
oscillation amplitude.

\medskip

The second part of the paper is devoted 
to considering other approximation schemes, and examining how they fare 
when compared to directly solving the non-local equation~(\ref{FullEoM}).
More complicated systems might be described by a complex system of coupled non-local equations,
which might be difficult to solve, even in the approximation scheme developed here. 
Instead of solving such equations directly,
a much more popular and intuitive approach to non-Markovian equations
is to derive their Markovian counterparts by making system-specific local approximations of the non-local 
terms.~\footnote{Here we consider an oscillating scalar condensate where the Markovianisation scheme
from~\cite{Greiner:1996dx} applies. For adiabatically evolving condensates the quasiadaibatic Markovianization scheme 
applies~\cite{Morikawa:1985mf,Morikawa:1986rp,Gleiser:1993ea} (cf.~\cite{Buldgen:2019dus} for a recent discussion).} 
These are Markovian
quantum kinetic equations (effective Boltzmann type equations) for quantities that describe 
important properties of the system under consideration, such as on-shell single particle distribution functions, 
cf.~e.g.~\cite{Calzetta:1986cq,Ivanov:1999tj,Arnold:2002zm,Juchem:2003bi,Berges:2005md,Lindner:2005kv,Herranen:2008hu,Millington:2012pf}.
In cosmology this approach has been heavily used in the treatment of 
baryogenesis (c.f. e.g.~\cite{Buchmuller:2000nd,Prokopec:2003pj,Beneke:2010wd,Frossard:2012pc}), 
and the case of 
a scalar field we examine
here~\cite{Morikawa:1986rp,Greiner:1996dx,Berera:2008ar,Mukaida:2013xxa}, where typically 
the following Markovian equation is considered,
\begin{equation}\label{LocalEquation}
\ddot{\varphi}
    + \Gamma(\varphi) \dot{\varphi}
    + \mathcal{V}'(\varphi) = 0 \, ,
\end{equation}
where~$\mathcal{V}(\varphi)$ is the effective potential, and~$\Gamma(\varphi)$ is the field-dependent {\it damping coefficient},
both containing thermal and quantum corrections. This equation is much easier to understand than the non-local one in~(\ref{FullEoM}),
as it is a generalization of the Duffing's equation
essentially describing a damped an-harmonic oscillator. 
While justified for adiabatically evolving condensate,
it is not exactly the form of the equation one finds when carefully applying 
the Markovainisation procedure appropriate for oscillating systems~\cite{Greiner:1996dx}, which also produces 
corrections to the kinetic term encoded by~$\mathcal{K}(\varphi)$,~\footnote{For more general non-renormalizable effective interactions
quantum and thermal corrections will produce Markovian equations with a more general structure,
\begin{equation*}
\biggl[ \frac{ \partial \mathcal{K}(\varphi,X) }{ \partial X} + 2 X \frac{ \partial^2 \mathcal{K}(\varphi,X) }{ \partial X^2} \biggr] \ddot{\varphi}
	+ X\frac{ \partial^2 \mathcal{K}(\varphi,X) }{ \partial \varphi \partial X}
	- \frac{1}{2} \frac{ \partial \mathcal{K}(\varphi,X) }{\partial \varphi}
	+ \Gamma(\varphi, X) \dot{\varphi}
	+ \mathcal{V}'(\varphi) = 0 \, ,
\end{equation*}
where~$X \!=\! \dot{\varphi}^2$.
Such equations can be seen to arise in Galileon theories~\cite{Nicolis:2008in} and 
Horndeski theories~\cite{Horndeski:1974wa}, hinting at a possibility 
that cosmologically interesting theories with non-canonical kinetic terms, such as 
e.g.~$k$-essence~\cite{ArmendarizPicon:1999rj}
or kinetic gravity braiding~\cite{Deffayet:2010qz}, 
could arise effectively as quantum-corrected canonical theories.
}
\begin{equation}
\mathcal{K}(\varphi) \ddot{\varphi} 
	+ \frac{1}{2} \mathcal{K}'(\varphi) \dot{\varphi}^2
	+ \Gamma(\varphi) \dot{\varphi}
	+ \mathcal{V}'(\varphi) = 0 \, .
\label{proper local eq}
\end{equation}
Neglecting quantum and thermal corrections to the kinetic term
effectively neglects some corrections to the inflaton oscillation frequency that might be relevant
when analysing the spectrum of produced particles. The equation of motion~(\ref{proper local eq})
correctly reproduces the leading order 
solutions when solved using multiple-scale perturbation theory. 
It is worth noting that the Markovian equation cannot correctly account for the subleading corrections
to the evolution, but it can nevertheless be used if one is not interested in that level of precision.
We examine in detail the Markovianisation of Eq.~(\ref{FullEoM})
and point out ambiguities in its construction, and its limitations 
by comparing the solutions obtained using multiple-scale perturbation theory
to the ones obtained by solving the non-local equation~(\ref{FullEoM}) directly.

Lastly, we examine to which extent the time-dependent perturbation theory can describe the evolution of damped inflaton oscillations, 
and demonstrate that it can only be applied to early times when damping is approximately linear in time.
It is in general not possible to extract how the oscillations dampen past this point, by using this
method only. An exception to this is the linear regime, where one can correctly guess that the
damping has to be exponential, and correctly resum the early time behaviour. This is not really an advantage
because in the linear regime we can solve for the evolution exactly. It is important to note
that in general we cannot resum the early time evolution in this way, as we do not know what
functional form the damping must take.

\medskip

In Sec.~\ref{sec: 2PI-resummed effective action} we present an explicit derivation of~\eqref{FullEoM} in a specific 
toy model starting from the  2PI effective action in the Schwinger-Keldysh formalism of nonequilibrium 
quantum field theory, and making a small field expansion. 
Explicit expressions for $\pi_{\scr R}$ and $v_{\scr R}$ in other models can be obtained by 
straightforward generalisations of the steps given there. 
In Sec.~\ref{sec: Solving condensate equations} we obtain analytic approximations for solutions
of~\eqref{FullEoM} by means of multiple-scale perturbation theory.
In Sec.~\ref{sec: Comparison to Markovian equations} we apply a 
Markovianisation procedure to \eqref{FullEoM} to obtain a local equation
of the form~(\ref{proper local eq}). We solve this Markovian equation using the very same 
approximation methods from Sec.~\ref{sec: Solving condensate equations}, allowing
us to establish limitations of Markovian equations. Lastly, in Sec.~\ref{sec: Limitations of time-dependent perturbation theory} 
we consider to which degree standard time-dependent perturbation theory 
is able to describe the relaxation of the 
oscillating scalar condensate,
confirming it is able to do so only at early times. The concluding section is devoted to discussing and comparing
different approaches considered in this work, with respect to their range of validity, 
their capability to compute subleading corrections, and the possibility to systematically include quantum statistical effects,
and also to interpreting the relaxation of the condensate in terms of microphysical 
on-shell processes.

It is worth remarking that the methods considered and results obtained here,
though motivated by a cosmological problem, are relevant for other fields, whenever
a system is modelled as an oscillating dissipating scalar field. Such descriptions arise
for example in studies of many-body theory, e.g.~\cite{Chatrchyan:2020syc,Prufer:2019kak,Weidinger_2017}.

\section{2PI-resummed effective action}
\label{sec: 2PI-resummed effective action}

Effective equations for the condensate evolution, such as Eq.~(\ref{FullEoM}), arise 
generically in a broad class of models.
Here we consider a simple one,
\begin{align}
S[\Phi, \chi]
	={}& \int\! d^{4}x \, \biggl[
	\frac{1}{2} (\partial_\mu \Phi) (\partial^\mu \Phi)
	+ \frac{1}{2} (\partial_\mu \chi) (\partial^\mu \chi)
	- \frac{m_\phi^2}{2} \Phi^2
	- \frac{m_\chi^2}{2} \chi^2
\nonumber \\
&
\hspace{3cm}
	- \frac{\lambda_\phi}{4!} \Phi^4
	- \frac{\lambda_\chi}{4!} \chi^4
	- \frac{g}{4} \Phi^2 \chi^2
	\biggr] \, ,
\label{action}
\end{align}
with two scalar fields~$\Phi$ and~$\chi$, to demonstrate how to derive Eq.~(\ref{FullEoM})
in the Schwinger-Keldysh (in-in, closed-time-path) 
formalism~\cite{Schwinger:1960qe,Keldysh:1964ud,Berges:2004yj,NonEqLectures},
by constructing the 2PI-resummed effective action in the small field expansion.
Our starting point is the 2PI effective action~\cite{Cornwall:1974vz,Berges:2004yj},
dependent on the condensate~$\varphi(x)\!=\!\bigl\langle \Phi(x) \bigr\rangle$
(the expectation value of~$\chi$ is set to vanish from the beginning),
and the two-point
functions~$\Delta_\varphi$ and~$\Delta_\chi$,
\begin{align}
\Gamma_{\rm 2PI} \bigl[ \varphi , \Delta_\phi, \Delta_\chi \bigr]
	={}&
	S\bigl[\varphi, \chi\!=\!0 \bigr]
	+ \frac{i}{2} {\rm Tr} \ln \bigl( \Delta_\phi^{-1} \bigr)
	+ \frac{i}{2} {\rm Tr} \bigl( G^{-1}_{\phi}  \Delta_\phi \bigr)
\nonumber \\
&	\hspace{1cm}
	+ \frac{i}{2} {\rm Tr} \ln \bigl( \Delta_\chi^{-1} \bigr)
	+ \frac{i}{2} {\rm Tr} \bigl( G^{-1}_{\chi}  \Delta_\chi \bigr)
	+ \Gamma_2\bigl[\varphi, \Delta_\phi^{}, \Delta_\chi \bigr] \, ,
\label{general 2PI action}
\end{align}
where the kinetic operators are,
\begin{subequations}
\begin{align}
&
-iG_\phi^{-1} = \square + m_\phi^2
    + \frac{\lambda_\phi}{2} \varphi^2
	= -iG^{-1}_{\phi,0}
	    + \frac{\lambda_\phi}{2} \varphi^2 \, ,
\label{Gphi}
\\
&
-iG_\chi^{-1} = \square + m_\chi^2
    + \frac{g}{2} \varphi^2
	= -iG^{-1}_{\chi,0}
	    + \frac{g}{2} \varphi^2 \, ,
\label{Gchi}
\end{align}
\label{Gs}%
\end{subequations}
and where~$\Gamma_2$ contains all the 2PI Feynman diagrams with two and more loops. For simplicity
we consider its two-loop truncation only,
\begin{align}
&
\Gamma_2 =
-
\
\frac{i}{8}
\
\begin{tikzpicture}[baseline={0cm-0.5*height("$=$")}]
\draw[thick, double] (-0.51,0) circle (0.5) ;
\draw[thick, double] (0.51,0) circle (0.5) ;
\filldraw (0,0) circle (2.5pt) node {} ;
\end{tikzpicture}
\
-
\
\frac{i}{4}
\
\begin{tikzpicture}[baseline={0cm-0.5*height("$=$")}]
\draw[thick, double] (-0.51,0) circle (0.5) ;
\draw[thick, double,dashed] (0.51,0) circle (0.5) ;
\filldraw[fill=black] (0,0) circle (2.5pt) node {} ;
\end{tikzpicture}
\
-
\
\frac{i}{8}
\
\begin{tikzpicture}[baseline={0cm-0.5*height("$=$")}]
\draw[thick, double,dashed] (-0.51,0) circle (0.5) ;
\draw[thick, double,dashed] (0.51,0) circle (0.5) ;
\filldraw[fill=black] (0,0) circle (2.5pt) node {} ;
\end{tikzpicture}
%
\nonumber \\[1.5ex]
&
\hspace{5em}
-
\
\frac{i}{12}
\
\begin{tikzpicture}[baseline={0cm-0.5*height("$=$")}]
\draw[thick, double] (0,0) circle (0.5) ;
\draw[thick, double] (-0.5,0) -- (0.5,0) ;
\filldraw (-0.5,0) circle (2.5pt) node {} ;
\draw[thick] (0.5,0) -- (0.8,0) ;
\draw[thick] (0.95,0) circle (0.15) ;
\draw[thick] (0.95-0.707*0.15,0.707*0.15) -- (0.95+0.707*0.15,-0.707*0.15) ;
\draw[thick] (0.95-0.707*0.15,-0.707*0.15) -- (0.95+0.707*0.15,+0.707*0.15) ;
\filldraw (0.5,0) circle (2.5pt) node {} ;
\draw[thick] (-0.8,0) -- (-0.5,0) ;
\draw[thick] (-0.95,0) circle (0.15) ;
\draw[thick] (-0.95-0.707*0.15,0.707*0.15) -- (-0.95+0.707*0.15,-0.707*0.15) ;
\draw[thick] (-0.95-0.707*0.15,-0.707*0.15) -- (-0.95+0.707*0.15,+0.707*0.15) ;
\end{tikzpicture}
\
-
\
\frac{i}{4}
\
\begin{tikzpicture}[baseline={0cm-0.5*height("$=$")}]
\draw[thick, double,dashed] (0,0) circle (0.5) ;
\draw[thick, double] (-0.5,0) -- (0.5,0) ;
\filldraw (-0.5,0) circle (2.5pt) node {} ;
\draw[thick] (0.5,0) -- (0.8,0) ;
\draw[thick] (0.95,0) circle (0.15) ;
\draw[thick] (0.95-0.707*0.15,0.707*0.15) -- (0.95+0.707*0.15,-0.707*0.15) ;
\draw[thick] (0.95-0.707*0.15,-0.707*0.15) -- (0.95+0.707*0.15,+0.707*0.15) ;
\filldraw (0.5,0) circle (2.5pt) node {} ;
\draw[thick] (-0.8,0) -- (-0.5,0) ;
\draw[thick] (-0.95,0) circle (0.15) ;
\draw[thick] (-0.95-0.707*0.15,0.707*0.15) -- (-0.95+0.707*0.15,-0.707*0.15) ;
\draw[thick] (-0.95-0.707*0.15,-0.707*0.15) -- (-0.95+0.707*0.15,+0.707*0.15) ;
\end{tikzpicture}
\label{Gamma2def}
\ \
.
\end{align}
All the~$\pm$ polarity indices carried by lines and vertices
of the Schwinger-Keldysh diagrammatic expansion
are implied and summed over when not explicitly denoted.
Double lines stand for the connected two-point functions,
\begin{equation}
\Delta_\phi^{\! ab}(x;x') =
\begin{tikzpicture}[baseline={0cm-0.5*height("$=$")}]
\draw[thick,double] (-0.6,0) -- (0.6,0) ;
\filldraw (-0.6,0) circle (2.5pt) node[below] {$\scriptstyle(x,a)$} ;
\filldraw (0.6,0) circle (2.5pt) node[below] {$\scriptstyle(x'\!,b)$} ;
\end{tikzpicture}
\ ,
\qquad
\Delta_\chi^{\!ab} (x;x') =
\begin{tikzpicture}[baseline={0cm-0.5*height("$=$")}]
\draw[thick,double,dashed] (-0.6,0) -- (0.6,0) ;
\filldraw (-0.6,0) circle (2.5pt) node[below] {$\scriptstyle(x,a)$} ;
\filldraw (0.6,0) circle (2.5pt) node[below] {$\scriptstyle(x'\!,b)$} ;
\end{tikzpicture}
\ ,
\qquad 
(a,b\!=\!\pm) \, ,
\end{equation}
crossed circles stand for the~$\varphi$ insertion, and vertices stand for,
\begin{subequations}
\begin{align}
\begin{tikzpicture}[baseline={0cm-0.5*height("$=$")}]
\draw[thick,double] (-0.35,-0.35) -- (0.35,0.35) ;
\draw[thick,double] (0.35,-0.35) -- (-0.35,0.35) ;
\filldraw (0,0) circle (2.5pt) node[below] {$\scriptstyle a$} ;
\end{tikzpicture}
\
\!&=
a (-i)\lambda_\phi \! \int\!\! d^{4}x \  , &
\begin{tikzpicture}[baseline={0cm-0.5*height("$=$")}]
\draw[thick,double] (-0.35,-0.35) -- (0,0) ;
\draw[thick] (0,0) -- (0.35,0.35) ;
\draw[thick,double] (0.35,-0.35) -- (-0.35,0.35) ;
\filldraw (0,0) circle (2.5pt) node[below] {$\scriptstyle a$} ;
\end{tikzpicture}
\
\!&=
a (-i)\lambda_\phi \! \int\!\! d^{4}x \  , &
\begin{tikzpicture}[baseline={0cm-0.5*height("$=$")}]
\draw[thick] (-0.35,-0.35) -- (0,0) ;
\draw[thick] (0,0) -- (0.35,0.35) ;
\draw[thick] (0.35,-0.35) -- (-0.35,0.35) ;
\filldraw (0,0) circle (2.5pt) node[below] {$\scriptstyle a$} ;
\end{tikzpicture}
\
\!&=
a (-i)\lambda_\phi \! \int\!\! d^{4}x \  ,
\\
\begin{tikzpicture}[baseline={0cm-0.5*height("$=$")}]
\draw[thick,double, dashed] (-0.35,-0.35) -- (0.35,0.35) ;
\draw[thick,double,dashed] (-0.35,0.35) -- (0.35,-0.35) ;
\filldraw (0,0) circle (2.5pt) node[below] {$\scriptstyle a$} ;
\end{tikzpicture}
\
\! &=
a (-i)\lambda_\chi \! \int\!\! d^{4}x \  , &
\begin{tikzpicture}[baseline={0cm-0.5*height("$=$")}]
\draw[thick,dashed] (-0.35,-0.35) -- (0.35,0.35) ;
\draw[thick,dashed] (-0.35,0.35) -- (0.35,-0.35) ;
\filldraw (0,0) circle (2.5pt) node[below] {$\scriptstyle a$} ;
\end{tikzpicture}
\
\! &=
a (-i)\lambda_\chi \! \int\!\! d^{4}x \  , 
& &
\qquad \quad (a \!=\! \pm)
\\
\begin{tikzpicture}[baseline={0cm-0.5*height("$=$")}]
\draw[thick,double] (-0.35,-0.35) -- (0.35,0.35) ;
\draw[thick,double,dashed] (-0.35,0.35) -- (0.35,-0.35) ;
\filldraw (0,0) circle (2.5pt) node[below] {$\scriptstyle a$} ;
\end{tikzpicture}
\
&=
a (-i)g \! \int\!\! d^{4}x \ , &
\begin{tikzpicture}[baseline={0cm-0.5*height("$=$")}]
\draw[thick,double] (-0.35,-0.35) -- (0,0) ;
\draw[thick] (0,0) -- (0.35,0.35) ;
\draw[thick,double, dashed] (0.35,-0.35) -- (-0.35,0.35) ;
\filldraw (0,0) circle (2.5pt) node[below] {$\scriptstyle a$} ;
\end{tikzpicture}
\
\!&=
a (-i)g \! \int\!\! d^{4}x \  , &
\begin{tikzpicture}[baseline={0cm-0.5*height("$=$")}]
\draw[thick] (-0.35,-0.35) -- (0,0) ;
\draw[thick] (0,0) -- (0.35,0.35) ;
\draw[thick,dashed] (0.35,-0.35) -- (-0.35,0.35) ;
\filldraw (0,0) circle (2.5pt) node[below] {$\scriptstyle a$} ;
\end{tikzpicture}
\
\!&=
a (-i)g \! \int\!\! d^{4}x \  ,
\end{align}
\end{subequations}
As evident from definitions above, a given vertex is determined by how many solid
and dashed lines connect to it, regardless of whether they are double lines,
or single lines to be defined in Sec.~\ref{subsec: Solving for two-point functions} below.
We are interested in writing down the effective theory for just the condensate.
This is accomplished by first solving on-shell for the 2-point functions as functionals of the condensate,
\begin{equation}
0 =
\frac{ \delta \Gamma_\text{2PI} }{ \delta \Delta_\phi } \biggr|_{\Delta_\phi = \Delta_\phi[\varphi]} \ ,
\qquad\quad
0 =
\frac{ \delta \Gamma_\text{2PI} }{ \delta \Delta_\chi } \biggr|_{\Delta_\chi = \Delta_\chi[\varphi]} \, ,
\label{symbolic 2pt eom}
\end{equation}
and then plugging these solutions back into the 2PI action, which produces the 2PI-resummed effective action for 
the condensate,
\begin{equation}
\Gamma_{\rm 2PI}^{\rm res.}[\varphi]
	\equiv \Gamma_{\rm 2PI} \bigl[ \varphi, \Delta_\phi[\varphi] , \Delta_\chi[\varphi] \bigr] \, .
\label{general 2PI Gamma resummed}
\end{equation}
This is the form of the effective action that we derive in the following subsection,
and is our starting point for studying the dynamics of the oscillating condensate,
whose equation of motion then follows from the variation of~(\ref{general 2PI Gamma resummed}),
\begin{equation}
\frac{ \delta \Gamma_{\rm 2PI}^{\rm res.}[\varphi] }{ \delta \varphi }= 0 \, .
\end{equation}
We are interested in describing the evolution of the condensate in the regime where
non-linearities are small, but not negligible. A scheme well adapted to such a regime is the
small field expansion, where the 2PI-resummed effective action is organized in
powers of the condensate. In particular, we will be constructing~$\Gamma_{\rm 2PI}^{\rm res.}[\varphi]$ 
up to quartic order in~$\varphi$.

\subsection{Solving for two-point functions}
\label{subsec: Solving for two-point functions}

Constructing the 2PI-resummed effective action 
requires us to solve for the two-point functions up to quadratic order in the condensate.
We do this we diagrammatically, starting from the equations of motion~(\ref{symbolic 2pt eom})
for the two-point functions,
\begin{align}
 G_{\phi}^{-1}(x) \!
\begin{tikzpicture}[baseline={0cm-0.5*height("$=$")}]
\draw[thick,double] (-0.6,0) -- (0.6,0) ;
\filldraw (-0.6,0) circle (2.5pt) node[below] {$\scriptstyle(x,a)$} ;
\filldraw (0.6,0) circle (2.5pt) node[below] {$\scriptstyle (x'\!,b)$};
\end{tikzpicture}
\!
={}&
\
a \delta_{ab} \delta^4(x\!-\!x')
\
+
\
\frac{a(-i)\lambda_\phi}{2}
\
\begin{tikzpicture}[baseline={0cm-0.5*height("$=$")}]
\draw[thick, double] (0,0.53) circle (0.5) ;
\draw[thick,double] (0,0) -- (1.2,0) ;
\filldraw (0,0) circle (2.5pt) node[below] {$\scriptstyle(x,a)$} ;
\filldraw (1.2,0) circle (2.5pt) node[below] {$\scriptstyle(x'\!,b)$} ;
\end{tikzpicture}
+
\
\frac{a(-i)g}{2}
\
\begin{tikzpicture}[baseline={0cm-0.5*height("$=$")}]
\draw[thick, double,dashed] (0,0.53) circle (0.5) ;
\draw[thick,double] (0,0) -- (1.2,0) ;
\filldraw (1.2,0) circle (2.5pt) node[below] {$\scriptstyle(x'\!,b)$} ;
\filldraw (0,0) circle (2.5pt) node[below] {$\scriptstyle(x,a)$} ;
\end{tikzpicture}
%
\nonumber \\[1ex]
&\hspace{-3cm}
+
\
\frac{a(-i)\lambda_\phi}{2}
\
\begin{tikzpicture}[baseline={0cm-0.5*height("$=$")}]
\draw[thick,double] (0.,0) arc (0:180:0.6);
\draw[thick,double] (0,0) -- (-1.2,0) ;
\draw[thick,double] (0,0) -- (1.2,0) ;
\filldraw (1.2,0) circle (2.5pt) node[below] {$\scriptstyle(x'\!,b)$} ;
\filldraw (-1.2,0) circle (2.5pt) node[left] {$\scriptstyle(x,a)$} ;
\filldraw (0,0) circle (2.5pt) node {} ;
\draw[thick] (0,0) -- (0,-0.3) ;
\draw[thick] (0,-0.45) circle (0.15) ;
\draw[thick] (0-0.707*0.15,-0.45+0.707*0.15) -- (0+0.707*0.15,-0.45-0.707*0.15) ;
\draw[thick] (0-0.707*0.15,-0.45-0.707*0.15) -- (0+0.707*0.15,-0.45+0.707*0.15) ;
\draw[thick] (-1.2,0) -- (-1.2,-0.3) ;
\draw[thick] (-1.2,-0.45) circle (0.15) ;
\draw[thick] (-1.2-0.707*0.15,-0.45+0.707*0.15) -- (-1.2+0.707*0.15,-0.45-0.707*0.15) ;
\draw[thick] (-1.2-0.707*0.15,-0.45-0.707*0.15) -- (-1.2+0.707*0.15,-0.45+0.707*0.15) ;
\end{tikzpicture}
\
+
\
\frac{a(-i)g}{2}
\
\begin{tikzpicture}[baseline={0cm-0.5*height("$=$")}]
\draw[thick,double,dashed] (0.,0) arc (0:180:0.6);
\draw[thick,double,dashed] (0,0) -- (-1.2,0) ;
\draw[thick,double] (0,0) -- (1.2,0) ;
\filldraw (1.2,0) circle (2.5pt) node[below] {$\scriptstyle(x'\!,b)$} ;
\filldraw (0,0) circle (2.5pt) node {} ;
\filldraw (-1.2,0) circle (2.5pt) node[left] {$\scriptstyle(x,a)$} ;
\draw[thick] (0,0) -- (0,-0.3) ;
\draw[thick] (0,-0.45) circle (0.15) ;
\draw[thick] (0-0.707*0.15,-0.45+0.707*0.15) -- (0+0.707*0.15,-0.45-0.707*0.15) ;
\draw[thick] (0-0.707*0.15,-0.45-0.707*0.15) -- (0+0.707*0.15,-0.45+0.707*0.15) ;
\draw[thick] (-1.2,0) -- (-1.2,-0.3) ;
\draw[thick] (-1.2,-0.45) circle (0.15) ;
\draw[thick] (-1.2-0.707*0.15,-0.45+0.707*0.15) -- (-1.2+0.707*0.15,-0.45-0.707*0.15) ;
\draw[thick] (-1.2-0.707*0.15,-0.45-0.707*0.15) -- (-1.2+0.707*0.15,-0.45+0.707*0.15) ;
\end{tikzpicture}
\ , 
\label{prop eq 1}
\\[2ex]
 G_{\chi}^{-1}(x) \!
\begin{tikzpicture}[baseline={0cm-0.5*height("$=$")}]
\draw[thick,double,dashed] (-0.6,0) -- (0.6,0) ;
\filldraw (-0.6,0) circle (2.5pt) node[below] {$\scriptstyle(x,a)$} ;
\filldraw (0.6,0) circle (2.5pt) node[below] {$\scriptstyle(x' \! ,b)$} ;
\end{tikzpicture}
\!
={}&
\
a \delta_{ab} \delta^4(x\!-\!x')
\
+
\frac{ a (-i) \lambda_\chi }{2}
\
\begin{tikzpicture}[baseline={0cm-0.5*height("$=$")}]
\draw[thick, double,dashed] (0,0.53) circle (0.5) ;
\draw[thick,double,dashed] (0,0) -- (1.2,0) ;
\filldraw (1.2,0) circle (2.5pt) node[below] {$\scriptstyle(x' \! ,b)$} ;
\filldraw (0,0) circle (2.5pt) node[below] {$\scriptstyle(x,a)$} ;
\end{tikzpicture}
\
+
\frac{a(-i)g}{2}
\
\begin{tikzpicture}[baseline={0cm-0.5*height("$=$")}]
\draw[thick, double] (0,0.53) circle (0.5) ;
\draw[thick,double,dashed] (0,0) -- (1.2,0) ;
\filldraw (1.2,0) circle (2.5pt) node[below] {$\scriptstyle(x' \! ,b)$} ;
\filldraw (0,0) circle (2.5pt) node[below] {$\scriptstyle(x,a)$} ;
\end{tikzpicture}
%
\nonumber \\[.5ex]
& \hspace{3em}
+
\
a(-i)g
\
\begin{tikzpicture}[baseline={0cm-0.5*height("$=$")}]
\draw[thick,double] (0.,0) arc (0:180:0.6);
\draw[thick,double,dashed] (0,0) -- (-1.2,0) ;
\draw[thick,double,dashed] (0,0) -- (1.2,0) ;
\filldraw (1.2,0) circle (2.5pt) node[below] {$\scriptstyle(x'\!,b)$} ;
\filldraw (0,0) circle (2.5pt) node {} ;
\filldraw (-1.2,0) circle (2.5pt) node[left] {$\scriptstyle(x,a)$} ;
\draw[thick] (0,0) -- (0,-0.3) ;
\draw[thick] (0,-0.45) circle (0.15) ;
\draw[thick] (0-0.707*0.15,-0.45+0.707*0.15) -- (0+0.707*0.15,-0.45-0.707*0.15) ;
\draw[thick] (0-0.707*0.15,-0.45-0.707*0.15) -- (0+0.707*0.15,-0.45+0.707*0.15) ;
\draw[thick] (-1.2,0) -- (-1.2,-0.3) ;
\draw[thick] (-1.2,-0.45) circle (0.15) ;
\draw[thick] (-1.2-0.707*0.15,-0.45+0.707*0.15) -- (-1.2+0.707*0.15,-0.45-0.707*0.15) ;
\draw[thick] (-1.2-0.707*0.15,-0.45-0.707*0.15) -- (-1.2+0.707*0.15,-0.45+0.707*0.15) ;
\end{tikzpicture}
\  .
\label{prop eq 2}
\end{align}
Note that vertices labeled by coordinates
and Schwinger-Keldysh polarities are not integrated over, while the unlabelled ones are
both integrated and summed over.
Since we are interested in the small field expansion of the 2PI-resummed effective action,
it suffices to solve for the two-point functions above up to
quadratic order. The leading contributions to the two-point functions in the small field expansion,
which we denote by single solid and dashed lines, are independent of the condensate,
\begin{equation}
\begin{tikzpicture}[baseline={0cm-0.5*height("$=$")}]
\draw[thick, double] (-0.6,0) -- (0.6,0) ;
\filldraw (-0.6,0) circle (2.5pt) node[below] {$\scriptstyle(x,a)$} ;
\filldraw (0.6,0) circle (2.5pt) node[below] {$\scriptstyle(x' \! ,b)$} ;
\end{tikzpicture}
\
=
\
\begin{tikzpicture}[baseline={0cm-0.5*height("$=$")}]
\draw[thick] (-0.6,0) -- (0.6,0) ;
\filldraw (-0.6,0) circle (2.5pt) node[below] {$\scriptstyle(x,a)$} ;
\filldraw (0.6,0) circle (2.5pt) node[below] {$\scriptstyle(x' \! ,b)$} ;
\end{tikzpicture}
\!
+
\
\mathcal{O}(\varphi^2) \, ,
\quad
\begin{tikzpicture}[baseline={0cm-0.5*height("$=$")}]
\draw[thick, double, dashed] (-0.6,0) -- (0.6,0) ;
\filldraw (-0.6,0) circle (2.5pt) node[below] {$\scriptstyle(x,a)$} ;
\filldraw (0.6,0) circle (2.5pt) node[below] {$\scriptstyle(x' \! ,b)$} ;
\end{tikzpicture}
\
=
\
\begin{tikzpicture}[baseline={0cm-0.5*height("$=$")}]
\draw[thick, dashed] (-0.6,0) -- (0.6,0) ;
\filldraw (-0.6,0) circle (2.5pt) node[below] {$\scriptstyle(x,a)$} ;
\filldraw (0.6,0) circle (2.5pt) node[below] {$\scriptstyle(x' \! ,b)$} ;
\end{tikzpicture}
\!
+
\
\mathcal{O}(\varphi^2) \, ,
\end{equation}
They are determined from Eqs.~(\ref{prop eq 1}) and~(\ref{prop eq 2}) where
the condensate-dependent parts are dropped,
\begin{align}
 G_{\phi,0}^{-1}(x)
\begin{tikzpicture}[baseline={0cm-0.5*height("$=$")}]
\draw[thick] (-0.6,0) -- (0.6,0) ;
\filldraw (-0.6,0) circle (2.5pt) node[below] {$\scriptstyle(x,a)$} ;
\filldraw (0.6,0) circle (2.5pt) node[below] {$\scriptstyle(x' \!,b)$} ;
\end{tikzpicture}
=&{}
\
a \delta_{ab} \delta^{4}(x \!-\! x')
\ +
\nonumber \\
&+ \
\frac{a(-i)\lambda_\phi}{2}
\
\begin{tikzpicture}[baseline={0cm-0.5*height("$=$")}]
\draw[thick] (0,0.5) circle (0.5) ;
\draw[thick] (0,0) -- (1.2,0) ;
\filldraw (1.2,0) circle (2.5pt) node[below] {$\scriptstyle(x' \! ,b)$} ;
\filldraw (0,0) circle (2.5pt) node[below] {$\scriptstyle(x,a)$} ;
\end{tikzpicture}
\
+
\
\frac{a(-i)g}{2}
\
\begin{tikzpicture}[baseline={0cm-0.5*height("$=$")}]
\draw[thick, dashed] (0,0.5) circle (0.5) ;
\draw[thick] (0,0) -- (1.2,0) ;
\filldraw (1.2,0) circle (2.5pt) node[below] {$\scriptstyle(x'\!,b)$} ;
\filldraw (0,0) circle (2.5pt) node[below] {$\scriptstyle(x,a)$} ;
\end{tikzpicture}
\ ,
\label{leading propagator eq 1}
\\[2ex]
 G_{\chi,0}^{-1}(x)
\begin{tikzpicture}[baseline={0cm-0.5*height("$=$")}]
\draw[thick,dashed] (-0.6,0) -- (0.6,0) ;
\filldraw (-0.6,0) circle (2.5pt) node[below] {$\scriptstyle(x,a)$} ;
\filldraw (0.6,0) circle (2.5pt) node[below] {$\scriptstyle(x'\!,b)$} ;
\end{tikzpicture}
\
=&{}
\
a\delta_{ab}\delta^{4}(x\!-\!x')
\
+
\nonumber \\
&+
\
\frac{a (-i)\lambda_\chi}{2}
\
\begin{tikzpicture}[baseline={0cm-0.5*height("$=$")}]
\draw[thick,dashed] (0,0.5) circle (0.5) ;
\draw[thick,dashed] (0,0) -- (1.2,0) ;
\filldraw (1.2,0) circle (2.5pt) node[below] {$\scriptstyle(x' \!,b)$} ;
\filldraw (0,0) circle (2.5pt) node[below] {$\scriptstyle(x,a)$} ;
\end{tikzpicture}
\
+
\
\frac{a(-i)g}{2}
\
\begin{tikzpicture}[baseline={0cm-0.5*height("$=$")}]
\draw[thick] (0,0.5) circle (0.5) ;
\draw[thick,dashed] (0,0) -- (1.2,0) ;
\filldraw (1.2,0) circle (2.5pt) node[below] {$\scriptstyle(x'\!,b)$} ;
\filldraw (0,0) circle (2.5pt) node[below] {$\scriptstyle(x,a)$} ;
\end{tikzpicture}
\  ,
\label{leading propagator eq 2}
\end{align}
with the kinetic operators on the left hand sides defined
in~(\ref{Gphi})--(\ref{Gchi}).
The solutions to these equations,
\begin{equation}
\begin{tikzpicture}[baseline={0cm-0.5*height("$=$")}]
\draw[thick] (-0.6,0) -- (0.6,0) ;
\filldraw (-0.6,0) circle (2.5pt) node[below] {$\scriptstyle(x,a)$} ;
\filldraw (0.6,0) circle (2.5pt) node[below] {$\scriptstyle(x' \!,b)$} ;
\end{tikzpicture}
=
D_\phi^{ab}(x\!-\!x') \, ,
\qquad \qquad
\begin{tikzpicture}[baseline={0cm-0.5*height("$=$")}]
\draw[thick,dashed] (-0.6,0) -- (0.6,0) ;
\filldraw (-0.6,0) circle (2.5pt) node[below] {$\scriptstyle(x,a)$} ;
\filldraw (0.6,0) circle (2.5pt) node[below] {$\scriptstyle(x' \!,b)$} ;
\end{tikzpicture}
=
D_\chi^{ab}(x\!-\!x') \, ,
\label{thermal propagators}
\end{equation}
are free thermal equilibrium propagators,
\begin{align}
D_{\phi,\chi}^{\scr ++}(x\!-\!x')
={}&
    \int\! \frac{d^4k}{(2\pi)^4} \,
    e^{-ik\cdot (x-x')}
    \biggl[
    \frac{i}{k^2 \!-\! M_{\phi,\chi}^2 \!+\! i\varepsilon}
    +
    2\pi f_{\scr B}(|k_0|)
    \delta \bigl( k^2 \!-\! M_{\phi,\chi}^2 \bigr) \biggr] \, ,\label{D++ thermal}
\\
D_{\phi,\chi}^{\scr +-}(x\!-\!x')
={}&
    \int\! \frac{d^4k}{(2\pi)^4 } \,
    e^{-ik\cdot (x-x')}
    \biggl[
    2\pi f_{\scr B}(k_0) {\rm sgn}(k_0) \,
        \delta
        \bigl( k^2 \!-\! M_{\phi,\chi}^2 \bigr) \biggr] \, ,\label{D+- thermal}
\end{align}
with the remaining components being complex
conjugates,~$D^{\scr --}_{\phi,\chi}(x\!-\!x') \!=\! \bigl[ D^{\scr ++}_{\phi,\chi}(x\!-\!x') \bigr]^*$, $D^{\scr -+}_{\phi,\chi}(x\!-\!x') \!=\! \bigl[ D^{\scr +-}_{\phi,\chi}(x\!-\!x') \bigr]^*$,
where~$f_{\scr B}(\omega) \!=\! 1/\bigl(e^{\omega/T} \!-\! 1 \bigr)$
is the Bose-Einstein distribution,
and where the thermal
masses in the large temperature limit are,
\begin{subequations}
\begin{align}
&
M_\phi^2
\ = \
m_\phi^2
\
-
\
\frac{i \lambda_\phi}{2}
\
\begin{tikzpicture}[baseline={0cm-0.5*height("$=$")}]
\draw[thick] (0,0) circle (0.5) ;
\filldraw (0.5,0) circle (2.5pt) node[right] {$\scriptstyle (x,+)$} ;
\end{tikzpicture}
\
-
\
\frac{i g}{2}
\
\begin{tikzpicture}[baseline={0cm-0.5*height("$=$")}]
\draw[thick, dashed] (0,0) circle (0.5) ;
\filldraw (0.5,0) circle (2.5pt) node[right] {$\scriptstyle (x,+)$} ;
\end{tikzpicture}
\
=
\
m_\phi^2 +\frac{ (\lambda_\phi \!+\! g) T^2}{24}
\, ,
\label{thermal Mphi}
\\
&
M_\chi^2
\ = \
m_\chi^2
\
-
\
\frac{i \lambda_\chi}{2}
\
\begin{tikzpicture}[baseline={0cm-0.5*height("$=$")}]
\draw[thick,dashed] (0,0) circle (0.5) ;
\filldraw (0.5,0) circle (2.5pt) node[right] {$\scriptstyle (x,+)$} ;
\end{tikzpicture}
\
-
\
\frac{i g}{2}
\
\begin{tikzpicture}[baseline={0cm-0.5*height("$=$")}]
\draw[thick] (0,0) circle (0.5) ;
\filldraw (0.5,0) circle (2.5pt) node[right] {$\scriptstyle (x,+)$} ;
\end{tikzpicture}
\
=
\
m_\chi^2 +\frac{ (\lambda_\chi \!+\! g) T^2}{24}
\, .
\label{thermal Mchi}
\end{align}
\label{thermal masses}%
\end{subequations}
Note that the coincidence limits of equilibrium thermal propagators are independent
of space-time coordinates, hence~$M_\phi$
and~$M_\chi$ are constants.
Including the widths of the propagator
would require going to two loops in Eqs.~(\ref{leading propagator eq 1}) and~(\ref{leading propagator eq 2}),
generated by the  three-loop effective action, which should be a straightforward extension.
Solving for the condensate-dependent corrections now gives,
\begin{align}
\begin{tikzpicture}[baseline={0cm-0.5*height("$=$")}]
\draw[thick,double] (-0.6,0) -- (0.6,0) ;
\filldraw (-0.6,0) circle (2.5pt) node[below] {$\scriptstyle(x,a)$} ;
\filldraw (0.6,0) circle (2.5pt) node {}node[below] {$\scriptstyle(x'\!,b)$} ;
\end{tikzpicture}
\
&=
\
\begin{tikzpicture}[baseline={0cm-0.5*height("$=$")}]
\draw[thick] (-0.6,0) -- (0.6,0) ;
\filldraw (-0.6,0) circle (2.5pt) node[below] {$\scriptstyle(x,a)$} ;
\filldraw (0.6,0) circle (2.5pt) node {}node[below] {$\scriptstyle(x'\!,b)$} ;
\end{tikzpicture}
\
+ \
\frac{1}{2}
\begin{tikzpicture}[baseline={0cm-0.5*height("$=$")}]
\draw[thick] (-1.,0) -- (1.,0) ;
\filldraw (-1.,0) circle (2.5pt) node[below] {$\scriptstyle(x,a)$} ;
\filldraw (1.,0) circle (2.5pt) node[below] {$\scriptstyle(x'\!,b)$} ;
\filldraw (0,0) circle (2.5pt) node {} ;
\draw[thick] (0,0) -- (0,-0.3) ;
\draw[thick] (0,-0.45) circle (0.15) ;
\draw[thick] (0-0.707*0.15,-0.45+0.707*0.15) -- (0+0.707*0.15,-0.45-0.707*0.15) ;
\draw[thick] (0-0.707*0.15,-0.45-0.707*0.15) -- (0+0.707*0.15,-0.45+0.707*0.15) ;
\draw[thick] (0,0) -- (0,0.3) ;
\draw[thick] (0,0.45) circle (0.15) ;
\draw[thick] (0-0.707*0.15,0.45+0.707*0.15) -- (0+0.707*0.15,0.45-0.707*0.15) ;
\draw[thick] (0-0.707*0.15,0.45-0.707*0.15) -- (0+0.707*0.15,0.45+0.707*0.15) ;
\end{tikzpicture}
\
+
\
\frac{1}{4}
\begin{tikzpicture}[baseline={0cm-0.5*height("$=$")}]
\draw[thick] (-1.,0) -- (2.,0) ;
\filldraw (-1.,0) circle (2.5pt) node[below] {$\scriptstyle(x,a)$} ;
\filldraw (2.,0) circle (2.5pt) node[below] {$\scriptstyle(x'\!,b)$} ;
\filldraw (0,0) circle (2.5pt) node {} ;
\draw[thick] (0,0) -- (0,-0.3) ;
\draw[thick] (0,-0.45) circle (0.15) ;
\draw[thick] (0-0.707*0.15,-0.45+0.707*0.15) -- (0+0.707*0.15,-0.45-0.707*0.15) ;
\draw[thick] (0-0.707*0.15,-0.45-0.707*0.15) -- (0+0.707*0.15,-0.45+0.707*0.15) ;
\draw[thick] (0,0) -- (0,0.3) ;
\draw[thick] (0,0.45) circle (0.15) ;
\draw[thick] (0-0.707*0.15,0.45+0.707*0.15) -- (0+0.707*0.15,0.45-0.707*0.15) ;
\draw[thick] (0-0.707*0.15,0.45-0.707*0.15) -- (0+0.707*0.15,0.45+0.707*0.15) ;
\filldraw (1,0) circle (2.5pt) node {} ;
\draw[thick] (1,0) -- (1,-0.3) ;
\draw[thick] (1,-0.45) circle (0.15) ;
\draw[thick] (1-0.707*0.15,-0.45+0.707*0.15) -- (1+0.707*0.15,-0.45-0.707*0.15) ;
\draw[thick] (1-0.707*0.15,-0.45-0.707*0.15) -- (1+0.707*0.15,-0.45+0.707*0.15) ;
\draw[thick] (1,0) -- (1,0.3) ;
\draw[thick] (1,0.45) circle (0.15) ;
\draw[thick] (1-0.707*0.15,0.45+0.707*0.15) -- (1+0.707*0.15,0.45-0.707*0.15) ;
\draw[thick] (1-0.707*0.15,0.45-0.707*0.15) -- (1+0.707*0.15,0.45+0.707*0.15) ;
\end{tikzpicture}
%
\nonumber \\[.5ex]
&\hspace{-3.5em}
+
\
\frac{1}{4}
\begin{tikzpicture}[baseline={0cm-0.5*height("$=$")}]
\draw[thick] (-1.,0) -- (1.,0) ;
\filldraw (-1.,0) circle (2.5pt) node[below] {$\scriptstyle(x,a)$} ;
\filldraw (1.,0) circle (2.5pt) node[below] {$\scriptstyle(x'\!,b)$} ;
\filldraw (0,0) circle (2.5pt) node {} ;
\draw[thick] (0,0.51) circle (0.5) ;
\filldraw (0,1.01) circle (2.5pt) node {} ;
\draw[thick] (0,1.01) -- (+0.3*0.707,1.01+0.3*0.707) ;
\draw[thick] (+0.45*0.707,1.01+0.45*0.707) circle (0.15) ;
\draw[thick] (+0.3*0.707,1.01+0.3*0.707) -- (+0.6*0.707,1.01+0.6*0.707) ;
\draw[thick] (+0.6*0.707,1.01+0.3*0.707) -- (+0.3*0.707,1.01+0.6*0.707) ;
\draw[thick] (0,1.01) -- (-0.3*0.707,1.01+0.3*0.707) ;
\draw[thick] (-0.45*0.707,1.01+0.45*0.707) circle (0.15) ;
\draw[thick] (-0.3*0.707,1.01+0.3*0.707) -- (-0.6*0.707,1.01+0.6*0.707) ;
\draw[thick] (-0.6*0.707,1.01+0.3*0.707) -- (-0.3*0.707,1.01+0.6*0.707) ;
\end{tikzpicture}
\
+
\
\frac{1}{4}
\begin{tikzpicture}[baseline={0cm-0.5*height("$=$")}]
\draw[thick] (-1.,0) -- (1.,0) ;
\filldraw (-1.,0) circle (2.5pt) node[below] {$\scriptstyle(x,a)$} ;
\filldraw (1.,0) circle (2.5pt) node[below] {$\scriptstyle(x'\!,b)$} ;
\filldraw (0,0) circle (2.5pt) node {} ;
\draw[thick,dashed] (0,0.51) circle (0.5) ;
\filldraw (0,1.01) circle (2.5pt) node {} ;
\draw[thick] (0,1.01) -- (+0.3*0.707,1.01+0.3*0.707) ;
\draw[thick] (+0.45*0.707,1.01+0.45*0.707) circle (0.15) ;
\draw[thick] (+0.3*0.707,1.01+0.3*0.707) -- (+0.6*0.707,1.01+0.6*0.707) ;
\draw[thick] (+0.6*0.707,1.01+0.3*0.707) -- (+0.3*0.707,1.01+0.6*0.707) ;
\draw[thick] (0,1.01) -- (-0.3*0.707,1.01+0.3*0.707) ;
\draw[thick] (-0.45*0.707,1.01+0.45*0.707) circle (0.15) ;
\draw[thick] (-0.3*0.707,1.01+0.3*0.707) -- (-0.6*0.707,1.01+0.6*0.707) ;
\draw[thick] (-0.6*0.707,1.01+0.3*0.707) -- (-0.3*0.707,1.01+0.6*0.707) ;
\end{tikzpicture}
\
+
\
\frac{1}{2}
\begin{tikzpicture}[baseline={0cm-0.5*height("$=$")}]
\draw[thick] (0.,0) arc (0:180:0.6);
\draw[thick] (0,0) -- (-2.2,0) ;
\draw[thick] (0,0) -- (1.,0) ;
\filldraw (-2.2,0) circle (2.5pt) node[below] {$\scriptstyle(x,a)$} ;
\filldraw (1.,0) circle (2.5pt) node[below] {$\scriptstyle(x'\!,b)$} ;
\filldraw (-1.2,0) circle (2.5pt) node {} ;
\filldraw (0,0) circle (2.5pt) node {} ;
\draw[thick] (0,0) -- (0,-0.3) ;
\draw[thick] (0,-0.45) circle (0.15) ;
\draw[thick] (0-0.707*0.15,-0.45+0.707*0.15) -- (0+0.707*0.15,-0.45-0.707*0.15) ;
\draw[thick] (0-0.707*0.15,-0.45-0.707*0.15) -- (0+0.707*0.15,-0.45+0.707*0.15) ;
\draw[thick] (-1.2,0) -- (-1.2,-0.3) ;
\draw[thick] (-1.2,-0.45) circle (0.15) ;
\draw[thick] (-1.2-0.707*0.15,-0.45+0.707*0.15) -- (-1.2+0.707*0.15,-0.45-0.707*0.15) ;
\draw[thick] (-1.2-0.707*0.15,-0.45-0.707*0.15) -- (-1.2+0.707*0.15,-0.45+0.707*0.15) ;
\end{tikzpicture}
%
\nonumber \\[2ex]
&\hspace{-3.5em}
+
\
\frac{1}{2}
\begin{tikzpicture}[baseline={0cm-0.5*height("$=$")}]
\draw[thick] (-1.2,0) -- (-2.2,0) ;
\draw[thick,dashed] (0.,0) arc (0:180:0.6);
\draw[thick,dashed] (0,0) -- (-1.2,0) ;
\draw[thick] (0,0) -- (1.,0) ;
\filldraw (1.,0) circle (2.5pt) node[below] {$\scriptstyle(x'\!,b)$} ;
\filldraw (0,0) circle (2.5pt) node {} ;
\filldraw (-1.2,0) circle (2.5pt) node {} ;
\filldraw (-2.2,0) circle (2.5pt) node[below] {$\scriptstyle(x,a)$} ;
\draw[thick] (0,0) -- (0,-0.3) ;
\draw[thick] (0,-0.45) circle (0.15) ;
\draw[thick] (0-0.707*0.15,-0.45+0.707*0.15) -- (0+0.707*0.15,-0.45-0.707*0.15) ;
\draw[thick] (0-0.707*0.15,-0.45-0.707*0.15) -- (0+0.707*0.15,-0.45+0.707*0.15) ;
\draw[thick] (-1.2,0) -- (-1.2,-0.3) ;
\draw[thick] (-1.2,-0.45) circle (0.15) ;
\draw[thick] (-1.2-0.707*0.15,-0.45+0.707*0.15) -- (-1.2+0.707*0.15,-0.45-0.707*0.15) ;
\draw[thick] (-1.2-0.707*0.15,-0.45-0.707*0.15) -- (-1.2+0.707*0.15,-0.45+0.707*0.15) ;
\end{tikzpicture}
\ ,
\label{solid propagator solution}
\\[2ex]
\begin{tikzpicture}[baseline={0cm-0.5*height("$=$")}]
\draw[thick,double,dashed] (-0.6,0) -- (0.6,0) ;
\filldraw (-0.6,0) circle (2.5pt) node[below] {$\scriptstyle(x,a)$} ;
\filldraw (0.6,0) circle (2.5pt) node {}node[below] {$\scriptstyle(x'\!,b)$} ;
\end{tikzpicture}
\
&=
\
\begin{tikzpicture}[baseline={0cm-0.5*height("$=$")}]
\draw[thick,dashed] (-0.6,0) -- (0.6,0) ;
\filldraw (-0.6,0) circle (2.5pt) node[below] {$\scriptstyle(x,a)$} ;
\filldraw (0.6,0) circle (2.5pt) node {}node[below] {$\scriptstyle(x'\!,b)$} ;
\end{tikzpicture}
\
+
\
\frac{1}{2}
\begin{tikzpicture}[baseline={0cm-0.5*height("$=$")}]
\draw[thick,dashed] (-1.,0) -- (1.,0) ;
\filldraw (-1.,0) circle (2.5pt) node[below] {$\scriptstyle(x,a)$} ;
\filldraw (1.,0) circle (2.5pt) node {}node[below] {$\scriptstyle(x'\!,b)$} ;
\filldraw (0,0) circle (2.5pt) node {} ;
\draw[thick] (0,0) -- (0,-0.3) ;
\draw[thick] (0,-0.45) circle (0.15) ;
\draw[thick] (0-0.707*0.15,-0.45+0.707*0.15) -- (0+0.707*0.15,-0.45-0.707*0.15) ;
\draw[thick] (0-0.707*0.15,-0.45-0.707*0.15) -- (0+0.707*0.15,-0.45+0.707*0.15) ;
\draw[thick] (0,0) -- (0,0.3) ;
\draw[thick] (0,0.45) circle (0.15) ;
\draw[thick] (0-0.707*0.15,0.45+0.707*0.15) -- (0+0.707*0.15,0.45-0.707*0.15) ;
\draw[thick] (0-0.707*0.15,0.45-0.707*0.15) -- (0+0.707*0.15,0.45+0.707*0.15) ;
\end{tikzpicture}
\
+
\
\frac{1}{4}
\begin{tikzpicture}[baseline={0cm-0.5*height("$=$")}]
\draw[thick, dashed] (-1.,0) -- (2.,0) ;
\filldraw (-1.,0) circle (2.5pt) node[below] {$\scriptstyle(x,a)$} ;
\filldraw (2.,0) circle (2.5pt) node[below] {$\scriptstyle(x'\!,b)$} ;
\filldraw (0,0) circle (2.5pt) node {} ;
\draw[thick] (0,0) -- (0,-0.3) ;
\draw[thick] (0,-0.45) circle (0.15) ;
\draw[thick] (0-0.707*0.15,-0.45+0.707*0.15) -- (0+0.707*0.15,-0.45-0.707*0.15) ;
\draw[thick] (0-0.707*0.15,-0.45-0.707*0.15) -- (0+0.707*0.15,-0.45+0.707*0.15) ;
\draw[thick] (0,0) -- (0,0.3) ;
\draw[thick] (0,0.45) circle (0.15) ;
\draw[thick] (0-0.707*0.15,0.45+0.707*0.15) -- (0+0.707*0.15,0.45-0.707*0.15) ;
\draw[thick] (0-0.707*0.15,0.45-0.707*0.15) -- (0+0.707*0.15,0.45+0.707*0.15) ;
\filldraw (1,0) circle (2.5pt) node {} ;
\draw[thick] (1,0) -- (1,-0.3) ;
\draw[thick] (1,-0.45) circle (0.15) ;
\draw[thick] (1-0.707*0.15,-0.45+0.707*0.15) -- (1+0.707*0.15,-0.45-0.707*0.15) ;
\draw[thick] (1-0.707*0.15,-0.45-0.707*0.15) -- (1+0.707*0.15,-0.45+0.707*0.15) ;
\draw[thick] (1,0) -- (1,0.3) ;
\draw[thick] (1,0.45) circle (0.15) ;
\draw[thick] (1-0.707*0.15,0.45+0.707*0.15) -- (1+0.707*0.15,0.45-0.707*0.15) ;
\draw[thick] (1-0.707*0.15,0.45-0.707*0.15) -- (1+0.707*0.15,0.45+0.707*0.15) ;
\end{tikzpicture}
%
\nonumber \\[.5ex]
&\hspace{-4.5em}
\
+
\
\frac{1}{4}
\begin{tikzpicture}[baseline={0cm-0.5*height("$=$")}]
\draw[thick,dashed] (-1.,0) -- (1.,0) ;
\filldraw (-1.,0) circle (2.5pt) node[below] {$\scriptstyle(x,a)$} ;
\filldraw (1.,0) circle (2.5pt) node {}node[below] {$\scriptstyle(x'\!,b)$} ;
\filldraw (0,0) circle (2.5pt) node {} ;
\draw[thick,dashed] (0,0.51) circle (0.5) ;
\filldraw (0,1.01) circle (2.5pt) node {} ;
\draw[thick] (0,1.01) -- (+0.3*0.707,1.01+0.3*0.707) ;
\draw[thick] (+0.45*0.707,1.01+0.45*0.707) circle (0.15) ;
\draw[thick] (+0.3*0.707,1.01+0.3*0.707) -- (+0.6*0.707,1.01+0.6*0.707) ;
\draw[thick] (+0.6*0.707,1.01+0.3*0.707) -- (+0.3*0.707,1.01+0.6*0.707) ;
\draw[thick] (0,1.01) -- (-0.3*0.707,1.01+0.3*0.707) ;
\draw[thick] (-0.45*0.707,1.01+0.45*0.707) circle (0.15) ;
\draw[thick] (-0.3*0.707,1.01+0.3*0.707) -- (-0.6*0.707,1.01+0.6*0.707) ;
\draw[thick] (-0.6*0.707,1.01+0.3*0.707) -- (-0.3*0.707,1.01+0.6*0.707) ;
\end{tikzpicture}
\
+
\
\frac{1}{4}
\begin{tikzpicture}[baseline={0cm-0.5*height("$=$")}]
\draw[thick,dashed] (-1.,0) -- (1.,0) ;
\filldraw (-1.,0) circle (2.5pt) node[below] {$\scriptstyle(x,a)$} ;
\filldraw (1.,0) circle (2.5pt) node {}node[below] {$\scriptstyle(x'\!,b)$} ;
\filldraw (0,0) circle (2.5pt) node {} ;
\draw[thick] (0,0.51) circle (0.5) ;
\filldraw (0,1.01) circle (2.5pt) node {} ;
\draw[thick] (0,1.01) -- (+0.3*0.707,1.01+0.3*0.707) ;
\draw[thick] (+0.45*0.707,1.01+0.45*0.707) circle (0.15) ;
\draw[thick] (+0.3*0.707,1.01+0.3*0.707) -- (+0.6*0.707,1.01+0.6*0.707) ;
\draw[thick] (+0.6*0.707,1.01+0.3*0.707) -- (+0.3*0.707,1.01+0.6*0.707) ;
\draw[thick] (0,1.01) -- (-0.3*0.707,1.01+0.3*0.707) ;
\draw[thick] (-0.45*0.707,1.01+0.45*0.707) circle (0.15) ;
\draw[thick] (-0.3*0.707,1.01+0.3*0.707) -- (-0.6*0.707,1.01+0.6*0.707) ;
\draw[thick] (-0.6*0.707,1.01+0.3*0.707) -- (-0.3*0.707,1.01+0.6*0.707) ;
\end{tikzpicture}
\
+
\
\begin{tikzpicture}[baseline={0cm-0.5*height("$=$")}]
\draw[thick,dashed] (-2.2,0) -- (1.,0) ;
\draw[thick] (0.,0) arc (0:180:0.6);
\filldraw (1.,0) circle (2.5pt) node[below] {$\scriptstyle(x'\!,b)$} ;
\filldraw (0,0) circle (2.5pt) node {} ;
\filldraw (-1.2,0) circle (2.5pt) node {} ;
\filldraw (-2.2,0) circle (2.5pt) node[below] {$\scriptstyle(x,a)$} ;
\draw[thick] (0,0) -- (0,-0.3) ;
\draw[thick] (0,-0.45) circle (0.15) ;
\draw[thick] (0-0.707*0.15,-0.45+0.707*0.15) -- (0+0.707*0.15,-0.45-0.707*0.15) ;
\draw[thick] (0-0.707*0.15,-0.45-0.707*0.15) -- (0+0.707*0.15,-0.45+0.707*0.15) ;
\draw[thick] (-1.2,0) -- (-1.2,-0.3) ;
\draw[thick] (-1.2,-0.45) circle (0.15) ;
\draw[thick] (-1.2-0.707*0.15,-0.45+0.707*0.15) -- (-1.2+0.707*0.15,-0.45-0.707*0.15) ;
\draw[thick] (-1.2-0.707*0.15,-0.45-0.707*0.15) -- (-1.2+0.707*0.15,-0.45+0.707*0.15) ;
\end{tikzpicture}
\ .
\label{dashed propagator solution}
\end{align}
Note that we have solved for the propagators up to (i) quartic order in the condensate at zero loops,
and (ii) quadratic order at one loop. This is sufficient for the order to which we compute the
2PI-resummed effective action, which we obtain by plugging the
solutions~(\ref{solid propagator solution}) and~(\ref{dashed propagator solution})
back into the 2PI effective action~(\ref{general 2PI action}),
\begin{subequations}
\begin{align}
\Gamma_{\rm 2PI}^{\rm res.} \bigl[ \varphi_{\scr +} , \varphi_{\scr -} \bigr]
	&= S[ \varphi_{\scr +} ] - S[ \varphi_{\scr -} ]
\\[.5ex]
&	\hspace{0.5cm}
-
\
\frac{i}{4}
\
\begin{tikzpicture}[baseline={0cm-0.5*height("$=$")}]
\draw[thick] (0,0) circle (0.5) ;
\filldraw (0.5,0) circle (2.5pt) node {} ;
\draw[thick] (0.5,0) -- (0.5+0.3*0.707,0+0.3*0.707) ;
\draw[thick] (0.5+0.45*0.707,+0.45*0.707) circle (0.15) ;
\draw[thick] (0.5+0.3*0.707,+0.3*0.707) -- (0.5+0.6*0.707,+0.6*0.707) ;
\draw[thick] (0.5+0.6*0.707,+0.3*0.707) -- (0.5+0.3*0.707,+0.6*0.707) ;
\draw[thick] (0.5,0) -- (0.5+0.3*0.707,-0.3*0.707) ;
\draw[thick] (0.5+0.45*0.707,-0.45*0.707) circle (0.15) ;
\draw[thick] (0.5+0.3*0.707,-0.3*0.707) -- (0.5+0.6*0.707,-0.6*0.707) ;
\draw[thick] (0.5+0.6*0.707,-0.3*0.707) -- (0.5+0.3*0.707,-0.6*0.707) ;
\end{tikzpicture}
\
-
\
\frac{i}{4}
\
\begin{tikzpicture}[baseline={0cm-0.5*height("$=$")}]
\draw[thick, dashed] (0,0) circle (0.5) ;
\filldraw (0.5,0) circle (2.5pt) node {} ;
\draw[thick] (0.5,0) -- (0.5+0.3*0.707,0+0.3*0.707) ;
\draw[thick] (0.5+0.45*0.707,+0.45*0.707) circle (0.15) ;
\draw[thick] (0.5+0.3*0.707,+0.3*0.707) -- (0.5+0.6*0.707,+0.6*0.707) ;
\draw[thick] (0.5+0.6*0.707,+0.3*0.707) -- (0.5+0.3*0.707,+0.6*0.707) ;
\draw[thick] (0.5,0) -- (0.5+0.3*0.707,-0.3*0.707) ;
\draw[thick] (0.5+0.45*0.707,-0.45*0.707) circle (0.15) ;
\draw[thick] (0.5+0.3*0.707,-0.3*0.707) -- (0.5+0.6*0.707,-0.6*0.707) ;
\draw[thick] (0.5+0.6*0.707,-0.3*0.707) -- (0.5+0.3*0.707,-0.6*0.707) ;
\end{tikzpicture}
\label{1loop local}
\\[.5ex]
&	\hspace{1.5cm}
-
\
\frac{i}{12}
\
\begin{tikzpicture}[baseline={0cm-0.5*height("$=$")}]
\draw[thick] (0,0) circle (0.5) ;
\draw[thick] (-0.5,0) -- (0.5,0) ;
\filldraw (-0.5,0) circle (2.5pt) node {} ;
\draw[thick] (0.5,0) -- (0.8,0) ;
\draw[thick] (0.95,0) circle (0.15) ;
\draw[thick] (0.95-0.707*0.15,0.707*0.15) -- (0.95+0.707*0.15,-0.707*0.15) ;
\draw[thick] (0.95-0.707*0.15,-0.707*0.15) -- (0.95+0.707*0.15,+0.707*0.15) ;
\filldraw (0.5,0) circle (2.5pt) node {} ;
\draw[thick] (-0.8,0) -- (-0.5,0) ;
\draw[thick] (-0.95,0) circle (0.15) ;
\draw[thick] (-0.95-0.707*0.15,0.707*0.15) -- (-0.95+0.707*0.15,-0.707*0.15) ;
\draw[thick] (-0.95-0.707*0.15,-0.707*0.15) -- (-0.95+0.707*0.15,+0.707*0.15) ;
\end{tikzpicture}
\
-
\
\frac{i}{4}
\
\begin{tikzpicture}[baseline={0cm-0.5*height("$=$")}]
\draw[thick,dashed] (0,0) circle (0.5) ;
\draw[thick] (-0.5,0) -- (0.5,0) ;
\draw[thick] (0.5,0) -- (0.8,0) ;
\draw[thick] (0.95,0) circle (0.15) ;
\draw[thick] (0.95-0.707*0.15,0.707*0.15) -- (0.95+0.707*0.15,-0.707*0.15) ;
\draw[thick] (0.95-0.707*0.15,-0.707*0.15) -- (0.95+0.707*0.15,+0.707*0.15) ;
\filldraw[fill=black] (0.5,0) circle (2.5pt) node {} ;
\draw[thick] (-0.8,0) -- (-0.5,0) ;
\draw[thick] (-0.95,0) circle (0.15) ;
\draw[thick] (-0.95-0.707*0.15,0.707*0.15) -- (-0.95+0.707*0.15,-0.707*0.15) ;
\draw[thick] (-0.95-0.707*0.15,-0.707*0.15) -- (-0.95+0.707*0.15,+0.707*0.15) ;
\filldraw[fill=black] (-0.5,0) circle (2.5pt) node {} ;
\end{tikzpicture}
\label{quadratic non-local}
\\[.5ex]
&	\hspace{2.5cm}
-
\
\frac{i}{16}
\
\begin{tikzpicture}[baseline={0cm-0.5*height("$=$")}]
\draw[thick] (0,0) circle (0.5) ;
\filldraw (0.5,0) circle (2.5pt) node {} ;
\filldraw (-0.5,0) circle (2.5pt) node {} ;
\draw[thick] (0.5,0) -- (0.5+0.3*0.707,0+0.3*0.707) ;
\draw[thick] (0.5+0.45*0.707,+0.45*0.707) circle (0.15) ;
\draw[thick] (0.5+0.3*0.707,+0.3*0.707) -- (0.5+0.6*0.707,+0.6*0.707) ;
\draw[thick] (0.5+0.6*0.707,+0.3*0.707) -- (0.5+0.3*0.707,+0.6*0.707) ;
\draw[thick] (0.5,0) -- (0.5+0.3*0.707,-0.3*0.707) ;
\draw[thick] (0.5+0.45*0.707,-0.45*0.707) circle (0.15) ;
\draw[thick] (0.5+0.3*0.707,-0.3*0.707) -- (0.5+0.6*0.707,-0.6*0.707) ;
\draw[thick] (0.5+0.6*0.707,-0.3*0.707) -- (0.5+0.3*0.707,-0.6*0.707) ;
\draw[thick] (-0.5,0) -- (-0.5-0.3*0.707,0-0.3*0.707) ;
\draw[thick] (-0.5-0.45*0.707,-0.45*0.707) circle (0.15) ;
\draw[thick] (-0.5-0.3*0.707,-0.3*0.707) -- (-0.5-0.6*0.707,-0.6*0.707) ;
\draw[thick] (-0.5-0.6*0.707,-0.3*0.707) -- (-0.5-0.3*0.707,-0.6*0.707) ;
\draw[thick] (-0.5,0) -- (-0.5-0.3*0.707,+0.3*0.707) ;
\draw[thick] (-0.5-0.45*0.707,+0.45*0.707) circle (0.15) ;
\draw[thick] (-0.5-0.3*0.707,+0.3*0.707) -- (-0.5-0.6*0.707,+0.6*0.707) ;
\draw[thick] (-0.5-0.6*0.707,+0.3*0.707) -- (-0.5-0.3*0.707,+0.6*0.707) ;
\end{tikzpicture}
\
-
\
\frac{i}{16}
\
\begin{tikzpicture}[baseline={0cm-0.5*height("$=$")}]
\draw[thick, dashed] (0,0) circle (0.5) ;
\filldraw (0.5,0) circle (2.5pt) node {} ;
\filldraw (-0.5,0) circle (2.5pt) node {} ;
\draw[thick] (0.5,0) -- (0.5+0.3*0.707,0+0.3*0.707) ;
\draw[thick] (0.5+0.45*0.707,+0.45*0.707) circle (0.15) ;
\draw[thick] (0.5+0.3*0.707,+0.3*0.707) -- (0.5+0.6*0.707,+0.6*0.707) ;
\draw[thick] (0.5+0.6*0.707,+0.3*0.707) -- (0.5+0.3*0.707,+0.6*0.707) ;
\draw[thick] (0.5,0) -- (0.5+0.3*0.707,-0.3*0.707) ;
\draw[thick] (0.5+0.45*0.707,-0.45*0.707) circle (0.15) ;
\draw[thick] (0.5+0.3*0.707,-0.3*0.707) -- (0.5+0.6*0.707,-0.6*0.707) ;
\draw[thick] (0.5+0.6*0.707,-0.3*0.707) -- (0.5+0.3*0.707,-0.6*0.707) ;
\draw[thick] (-0.5,0) -- (-0.5-0.3*0.707,0-0.3*0.707) ;
\draw[thick] (-0.5-0.45*0.707,-0.45*0.707) circle (0.15) ;
\draw[thick] (-0.5-0.3*0.707,-0.3*0.707) -- (-0.5-0.6*0.707,-0.6*0.707) ;
\draw[thick] (-0.5-0.6*0.707,-0.3*0.707) -- (-0.5-0.3*0.707,-0.6*0.707) ;
\draw[thick] (-0.5,0) -- (-0.5-0.3*0.707,+0.3*0.707) ;
\draw[thick] (-0.5-0.45*0.707,+0.45*0.707) circle (0.15) ;
\draw[thick] (-0.5-0.3*0.707,+0.3*0.707) -- (-0.5-0.6*0.707,+0.6*0.707) ;
\draw[thick] (-0.5-0.6*0.707,+0.3*0.707) -- (-0.5-0.3*0.707,+0.6*0.707) ;
\end{tikzpicture}
\ \ ,
\label{quartic non-local}
\end{align}
\label{truncated eff action}%
\end{subequations}
where we neglect writing constant terms which do not contribute to the dynamics.
The reasoning behind the above truncation of the 2PI-resummed effective action is as follows.
Different classes of diagrams induce qualitatively rather different behaviour for the condensate. 
The first big difference is the one between local and non-local diagrams. The local ones
cannot induce dissipative effects, while non-local ones can. Local diagrams can only be quadratic in the fields,
and we truncate them to the leading one-loop order
in~(\ref{1loop local}).
The non-local diagrams are ones where condensate fields appear at non-equal space-time points.
The kinds of effects that non-local diagrams can induce depend on the power of condensate fields
appearing in them. Here we keep lowest order quadratic non-local diagrams, which are the
two-loop ones in~(\ref{quadratic non-local}), and the lowest order quartic diagrams,
which are the one-loop ones in~(\ref{quartic non-local}). Thus, in each qualitatively different class
we keep only the diagrams of lowest loop order.
The truncated 2PI-resumed effective action~(\ref{truncated eff action})
is for our purposes conveniently written as,
\begin{align}
\MoveEqLeft[2]
\Gamma_\text{2PI}^\text{res.} \bigl[ \varphi_{\scr +}, \varphi_{\scr -} \bigr]
	=
	\int\! d^{4}x \, \biggl[
	\frac{1}{2} \partial^\mu \varphi_{\scr +}(x)  \, \partial_\mu \varphi_{\scr +}(x)
	- \frac{M_\phi^2}{2} \varphi_{\scr +}^2(x)
	- \frac{\lambda_\phi}{4!} \varphi_{\scr +}^4(x)
\label{2PI resummed general}
\\
&	\hspace{4cm}
	- \frac{1}{2} \partial^\mu \varphi_{\scr -}(x) \, \partial_\mu \varphi_{\scr -}(x)
	+ \frac{M_\phi^2}{2} \varphi_{\scr -}^2(x)
	+ \frac{\lambda_\phi}{4!} \varphi_{\scr -}^4(x)
	\biggr]
\nonumber \\
&
	- \int\! d^{4}x \int\! d^{4}x' \sum_{a,b=\pm}
	\biggl[
	\frac{1}{2} \varphi_{a}(x) \, \Pi_{ab}(x\!-\!x') \, \varphi_{b}(x')
	+
	\frac{1}{4!}
	\varphi_{a}^2(x) \, V_{ab}(x\!-\!x') \, \varphi_{b}^2(x')
	\biggr] \, ,
\nonumber
\end{align}
where~$M_\phi$ was given
in~(\ref{thermal Mphi}), and where,
\begin{equation}\label{PiDef}
\Pi_{ab}(x\!-\!x')
	=
\
-
\,
\frac{i(ab)\lambda_\phi^2}{6}
\,
\begin{tikzpicture}[baseline={0cm-0.5*height("$=$")}]
\draw[thick] (0,0) circle (0.5) ;
\draw[thick] (-0.5,0) -- (0.5,0) ;
\filldraw (-0.5,0) circle (2.5pt) node[left] {$\scriptstyle(x,a)$} ;
\filldraw (0.5,0) circle (2.5pt) node[right] {$\scriptstyle(x'\!,b)$} ;
\end{tikzpicture}
\
-
\
\frac{i(ab)g^2}{2}
\,
\begin{tikzpicture}[baseline={0cm-0.5*height("$=$")}]
\draw[thick,dashed] (0,0) circle (0.5) ;
\draw[thick] (-0.5,0) -- (0.5,0) ;
\filldraw[fill=black] (0.5,0) circle (2.5pt) node[right] {$\scriptstyle(x'\!,b)$} ;
\filldraw[fill=black] (-0.5,0) circle (2.5pt) node[left] {$\scriptstyle(x,a)$} ;
\end{tikzpicture}
\ ,
\end{equation}
is the two-loop non-local self-energy and,
\begin{equation}\label{VDef}
V_{ab}(x\!-\!x')
=
\
-
\
\frac{3i (ab) \lambda_\phi^2}{2}
\,
\begin{tikzpicture}[baseline={0cm-0.5*height("$=$")}]
\draw[thick] (0,0) circle (0.5) ;
\filldraw[fill=black] (0.5,0) circle (2.5pt) node[right] {$\scriptstyle(x'\!,b)$} ;
\filldraw[fill=black] (-0.5,0) circle (2.5pt) node[left] {$\scriptstyle(x,a)$} ;
\end{tikzpicture}
\
-
\
\frac{3i (ab) g^2}{2}
\,
\begin{tikzpicture}[baseline={0cm-0.5*height("$=$")}]
\draw[thick, dashed] (0,0) circle (0.5) ;
\filldraw[fill=black] (0.5,0) circle (2.5pt) node[right] {$\scriptstyle(x'\!,b)$} ;
\filldraw[fill=black] (-0.5,0) circle (2.5pt) node[left] {$\scriptstyle(x,a)$} ;
\end{tikzpicture}
\
,
\end{equation}
is the one-loop non-local proper 4-vertex function, with the lines in the diagrams
being thermal
propagators~(\ref{thermal propagators}).
Note that, seemingly, at this order of truncation
there is no dependence on the self-coupling~$\lambda_\chi$ of the~$\chi$-field. This is,
however, not the case as the thermal
masse~\eqref{thermal Mchi}
of the corresponding resummed propagator depends on this coupling constant.

\subsection{Homogeneous isotropic condensate}
\label{subsec: Homogeneous isotropic condensate}

When the 2PI-resummed effective
action~(\ref{2PI resummed general})
is specialized to a homogeneous
and isotropic scalar condensate that we are interested in here,
it reads,
\begin{align}
\MoveEqLeft[2]
\Gamma_\text{2PI}^\text{res.} \bigl[ \varphi_{\scr +} , \varphi_{\scr -} \bigr]
\nonumber \\
={}& V \int \! dt \, \biggl[
		\frac{1}{2} \dot{\varphi}_{\scr +}^2(t)
		- \frac{M_\phi^2}{2} \varphi_{\scr +}^2(t)
		- \frac{\lambda_\phi}{4!} \varphi_{\scr +}^4(t)
		- \frac{1}{2} \dot{\varphi}_{\scr -}^2(t)
		+ \frac{M_\phi^2}{2} \varphi_{\scr -}^2(t)
		+ \frac{\lambda_\phi}{4!} \varphi_{\scr -}^4(t)  \biggr]
\nonumber \\
&
	- V \int\! dt \int\! dt'
	\sum_{a,b=\pm} \biggl[
	\frac{1}{2} \varphi_a(t) \pi_{ab}(t\!-\!t') \varphi_b(t')
	+
	\frac{1}{4!} \varphi_a^2(t) v_{ab}(t\!-\!t') \varphi_b^2(t')
	\biggr] \, ,
\end{align}
where~$V\!=\!\int\! d^{3}x \!=\! (2\pi)^3 \delta^3(\bb{0})$ is an irrelevant spatial volume factor,
and where,
\begin{equation}
\pi_{ab}(t\!-\!t')
	= \int\! d^{3\!}x \, \Pi_{ab}(x\!-\!x') \, ,
\qquad
v_{ab}(t\!-\!t')
	= \int\! d^{3\!}x \, V_{ab}(x\!-\!x') \, ,
\end{equation}
are, respectively, the self-energy and the proper four-vertex in the mixed (two-time) representation
evaluated at the vanishing spatial momentum. In the following we
study the dynamics
of the condensate encoded by this effective action,
described by the equation of motion,
\begin{equation}
0 = \frac{\delta \Gamma_\text{2PI}^\text{res.}}{\delta \varphi_{\scr +}(t)}
\bigg|_{\varphi_{\scr +} = \, \varphi_{\scr -} = \, \varphi} \, .
\label{symbolic condensate eom}
\end{equation}
The condensate equation of motion takes the explicit form~\eqref{FullEoM},
\begin{equation}
\ddot{\varphi}(t) + M_\phi^2 \varphi(t)
    + \frac{ \lambda_\phi \varphi^3(t) }{6}
	+ \int_{t_0}^{t}\!
	    dt' \, \pi_{\scr R}(t\!-\!t') \varphi(t')
	+ \frac{ \varphi(t) }{6}
	\int_{t_0}^{t}\! dt' \,
	    v_{\scr R}(t\!-\!t') \varphi^2(t') = 0 \, ,
\label{condensate eom}
\end{equation}
where~$t_0\!=\!0$ is the initial time at which initial conditions~$\varphi(0)$ and~$\dot{\varphi}(0)$
are specified, and where the retarded self-energy and the retarded proper four-vertex function are,
respectively,
\begin{align}
&
\pi_{\scr R}(t\!-\!t')
    = \pi_{\scr ++}(t\!-\!t') + \pi_{\scr +-}(t\!-\!t')
	= \theta(t\!-\!t')
	    \Bigl[ - \pi_{\scr -+}(t\!-\!t')
	        + \pi_{\scr +-}(t\!-\!t')  \Bigr] \, ,
\label{piR def}
\\
&
v_{\scr R}(t\!-\!t')
    = v_{\scr ++}(t\!-\!t') + v_{\scr +-}(t\!-\!t')
	= \theta(t\!-\!t')
	    \Bigl[ - v_{\scr -+}(t\!-\!t')
	        + v_{\scr +-}(t\!-\!t')  \Bigr] \, .
\label{vR def}
\end{align}
Important quantities that appear in the following sections are Fourier transforms of
the retarded self-energy and the proper four-vertex, 
\begin{equation}\label{PiTildeDef}
\widetilde{\pi}_{\scr R}(\omega) = \int_{-\infty}^{\infty} \!
	dt' \, e^{i\omega(t-t')} \, \pi_{\scr R}(t\!-\!t') \, ,
\qquad 
\widetilde{v}_{\scr R}(\omega) = \int_{-\infty}^{\infty} \!
	dt' \, e^{i\omega(t-t')} \, v_{\scr R}(t\!-\!t') \, .
\end{equation}
Note that we just as well could have written~$t$ as the upper limit of integration
due to the step functions in~(\ref{piR def}) and~(\ref{vR def}).
The diagrams making up the quantities above appear in different contexts in related studies,
e.g.~\cite{Chou:1984es,Morikawa:1986rp,Yokoyama:2004pf,Mukaida:2013xxa,Cheung:2015iqa,Buldgen:2019dus}.

For the two scalar model at hand~(\ref{action}), and at the level of truncation of the
2PI-resummed effective action~(\ref{truncated eff action}),  the imaginary part of the 
quantity~$\widetilde{v}_{\scr R}(\omega)$ can be computed analytically~\cite{Boyanovsky:2004dj},
\begin{align}
   {\rm Im} \bigl[ \widetilde{v}_{\scr R}(\omega) \bigr] ={}&
    -
    \theta\bigl(\omega^2 \!-\! 4 M_\phi^2 \bigr)
	\frac{ 3 \lambda_\phi^2 }{ 32\pi }
		\sqrt{ 1 \!-\! \frac{4 M_\phi^2}{\omega^2} }
		\,
		\Bigl[ 1 + 2f_{\scr B} \Bigl( \frac{\omega}{2} \Bigr) \Bigr] \, ,
\nonumber \\
&   \hspace{1.5cm}
	- \theta\bigl(\omega^2 \!-\! 4
	    M_\chi^2 \bigr)
		\frac{ 3 g^2 }{ 32\pi }
		\sqrt{ 1 \!-\! \frac{4 M_\chi^2}{\omega^2} }
		\,
		\Bigl[ 1 + 2f_{\scr B} \Bigl( \frac{\omega}{2} \Bigr) \Bigr] \, .
\label{Im vR omega T}
\end{align}
The real part of the proper four-vertex has been studied in e.g.~\cite{Drewes:2013bfa}. 
Even though this contribution is often neglected as it does not contribute to dissipation,
we do take it into account here, and
find that it does produce an interesting effect of time-dependent corrections to the
oscillation frequency of the condensate.
For the comparison with time-dependent perturbation theory
in Sec.~\ref{sec: Limitations of time-dependent perturbation theory} we shall require the~$T\!\to\!0$
limit of expression~(\ref{Im vR omega T}) above,
\begin{equation}
{\rm Im} \bigl[ \widetilde{v}_{\scr R}(\omega) \bigr] 
	= - \theta\bigl(\omega^2 \!-\! 4 m_\phi^2 \bigr)
	\frac{ 3 \lambda_\phi^2 }{ 32\pi }
		\sqrt{ 1 \!-\! \frac{4 m_\phi^2}{\omega^2} }
	- \theta\bigl(\omega^2 \!-\! 4
	m_\chi^2 \bigr)
		\frac{ 3 g^2 }{ 32\pi }
		\sqrt{ 1 \!-\! \frac{4 m_\chi^2}{\omega^2} }
		\, .
\label{Im vR omega}
\end{equation}

The leading contribution to $\widetilde{\pi}_{\scr R}(\omega)$ comes from the ``setting sun'' two-loop diagrams
in Eq.~(\ref{PiDef}).
We are not aware of any closed-form solutions
for these diagrams. For the self-interaction piece
there are estimates for its imaginary part
in the following regimes~\cite{Parwani:1991gq,Drewes:2013iaa},
\begin{equation} \label{ThermalScatterings}
{\rm Im} \bigl[ \widetilde{\pi}_{\scr R}(\omega) \bigr] \simeq
\begin{cases}
\displaystyle
    - \frac{\lambda_\phi M_\phi^2}{32\pi}
    = - \frac{\lambda_\phi^2 T^2}{768\pi}
        & \quad
      (\omega\simeq M_\phi \ll T) \, ,
\\[2ex]
\displaystyle
    - \frac{\lambda_\phi^2 \omega^2 }{6(2\pi)^4}\frac{T^2}{M_\phi^2}
    \biggl[ 1 + \ln \biggl(
        \frac{81}{8}\frac{M_\phi}{\omega}
        \biggr)
    \biggr]
        & \quad
        ( \omega\ll M_\phi \ll T) \,.
\end{cases}
\end{equation}
For the $g\Phi^2\chi^2/4$ interaction one can expect a similar behaviour in the high-temperature regime.

The results for diagrams we quote in this section pertain to the simple two-field model from~\eqref{action}. 
If, for instance, $\chi$ happens to couple to other degrees of freedom, the collision terms in~\eqref{condensate eom} 
in the high-temperature regime can look much more complicated, cf.~e.g. Refs.~\cite{Mukaida:2012qn,Mukaida:2013xxa,Drewes:2013iaa}.

\section{Solving non-local condensate equations}
\label{sec: Solving condensate equations}

In this section we consider a
non-local (integro-differential)
equation,
\begin{equation}
\ddot{\varphi}(t) + M^2 \varphi(t)
    + \frac{ \Lambda \varphi^3(t) }{6}
	+ \int_{t_0}^{t}\!
	    dt' \, \pi_{\scr R}(t\!-\!t') \varphi(t')
	+ \frac{ \varphi(t) }{6} \!
	\int_{t_0}^{t}\! dt' \,
	    v_{\scr R}(t\!-\!t') \varphi^2(t') = 0 \, ,
\label{sec3eq}
\end{equation}
corresponding to the quantum corrected equation
of motion for the condensate~(\ref{condensate eom})
derived from the 2PI effective action,
where~$M\!=\!M_\phi$ and~$\lambda\!=\!\Lambda_\phi$,
and~$t_0\!=\!0$ is the initial time.
We systematically construct approximate solutions of this equation
in four cases of increasing complexity. The approximation scheme relies on 
a couple of assumptions. The equation of motion~(\ref{sec3eq})
was derived in the preceding section in the small field expansion,
assuming that nonlinear terms can be considered small.  In addition we make two
more assumptions: (i)
that the coupling constants appearing in kernels are perturbatively small
so that the linear non-local term in~(\ref{sec3eq}) can also be considered small,
and (ii)
that the kernels of the non-local terms are such that they provide an effective window 
that spans no longer than several oscillation periods, which in thermal theories is 
typically true.

Given that the last three terms in Eq.~(\ref{sec3eq}) are small by assumption, 
it might be tempting to approach the problem using standard perturbation theory.
However, this approach fails completely for time-dependent problems, even when applied
to familiar local equations~(cf.~\cite{Berges:2004yj} for an illustrative example).
The reason behind this is a feature of secular growth of subleading corrections in
perturbation theory applied to nonequilibrium problems. This secular growth causes
the corrections to grow without bound, eventually becoming comparable to the
leading approximation, thus invalidating the perturbative expansion. Such features are ubiquitous,
and appear even for systems with a bounded evolution, and were explored in~\cite{Boyanovsky:1994me}
for the case of an oscillating scalar condensate we consider here.
These issues are artefacts of the perturbative expansion, where the solution to
the previous order equation provides a resonant driving force for the equation for the following order,
thus leading to secular growth. 
For these reasons a more sophisticated approximation scheme needs to be applied 
to Eq.~(\ref{sec3eq}).

Our method of choice is the multiple-scale perturbation theory 
(multiple-scale analysis)~\cite{Bender,Holmes}
(for a particularly lucid and insightful
conceptual introduction to the method cf. Ref.~\cite{Ramnath:1969}),
which we adapt to non-local equations. This method is perfectly
suited for treating time-dependent problems consisting of processes happening on different time scales,
which is typically the case when some parts of the dynamical equation can be treated locally as 
perturbations, such as in the problem at hand. Even though locally small,
these can build up over longer times to appreciably alter the evolution. It is this physical observation,
together with realising the technical limitations of standard perturbation theory, that motivate
the biggest conceptual leap of multiple-scale perturbation theory: the solution is assumed to depend
on two times,~$\varphi(t) \!=\! F(t,\tau\!=\!\varepsilon t)$, and the two times~$t$ and~$\tau$ are 
treated as formally independent, the former accounting for processes taking place on shorter time scales, and the latter for ones evolving on longer time scales. 
Mathematically, this turns our eqaution of motion~(\ref{sec3eq})
from an ordinary differential equation in single time into a partial differential equation in two 
times, and thus turns a well-posed initial value problem into an underdetermined
one, since no additional initial conditions are being specified. It is this introduced ambiguity that is
crucial when applying perturbation theory to the evolution in two times, as 
it is fixed by requiring
that spurious secular growth be absent from subleading corrections, and for perturbative expansion
to be valid uniformly for late times. This step effectively resums an infinite subclass of naive
perturbative corrections.

In the interest of clarity,
we treat all cases in an algorithmic form. All steps are explained in the simplest case
example of the linear equation. For all other cases the algorithm is applied without further
explanation. 
The approximate solutions for the condensate evolution in the cases we consider
all take the same form,
\begin{equation}
\varphi(t) \approx A(t) \cos\biggl[ \int_{t_0}^{t} \!\! dt' \, \Omega(t') + {\rm const.} \biggr] \, ,
\end{equation}
where~$\Omega(t)$ is the {\it local frequency} of condensate oscillations, and
the amplitude~$A(t)$
is a monotonically decreasing function of time describing damping of the condensate oscillations,
conveniently expressed in terms of a {\it local damping rate},
\begin{equation}
\Upsilon(t) \equiv - \frac{1}{A(t)} \frac{d A(t)}{dt} \, .
\label{upsilon definition}
\end{equation}
The evolution of~$\Omega(t)$ and~$\Upsilon(t)$ depends crucially on which interactions we consider,
and on the coupling governing those interactions.
The physical phenomena exhibited in individual cases are commented at the 
end of each subsection.

\medskip

\begin{figure}[h!]
\vspace{0.2cm}
\includegraphics[width=7.2cm]{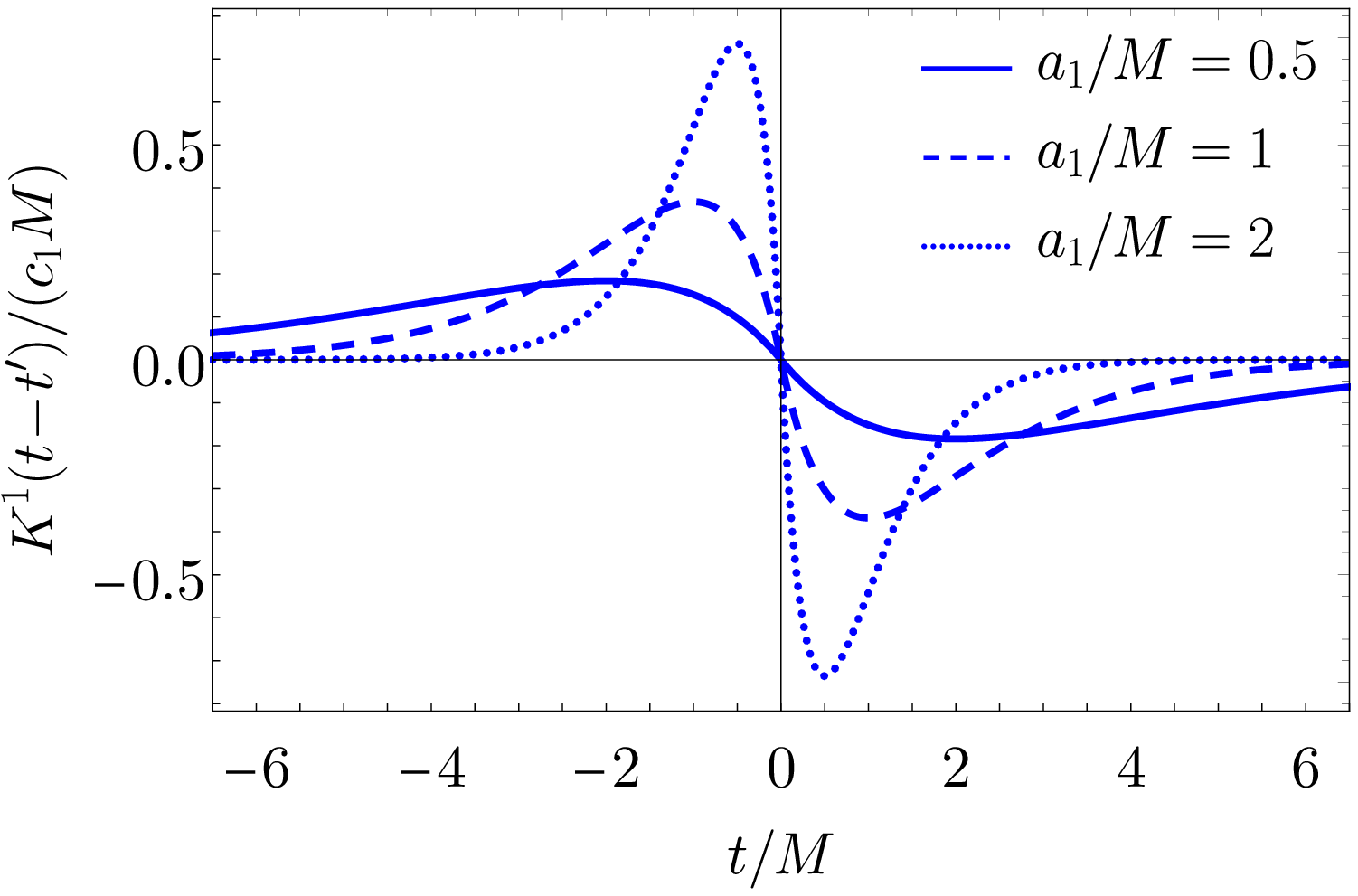}
\hfill
\includegraphics[width=7.2cm]{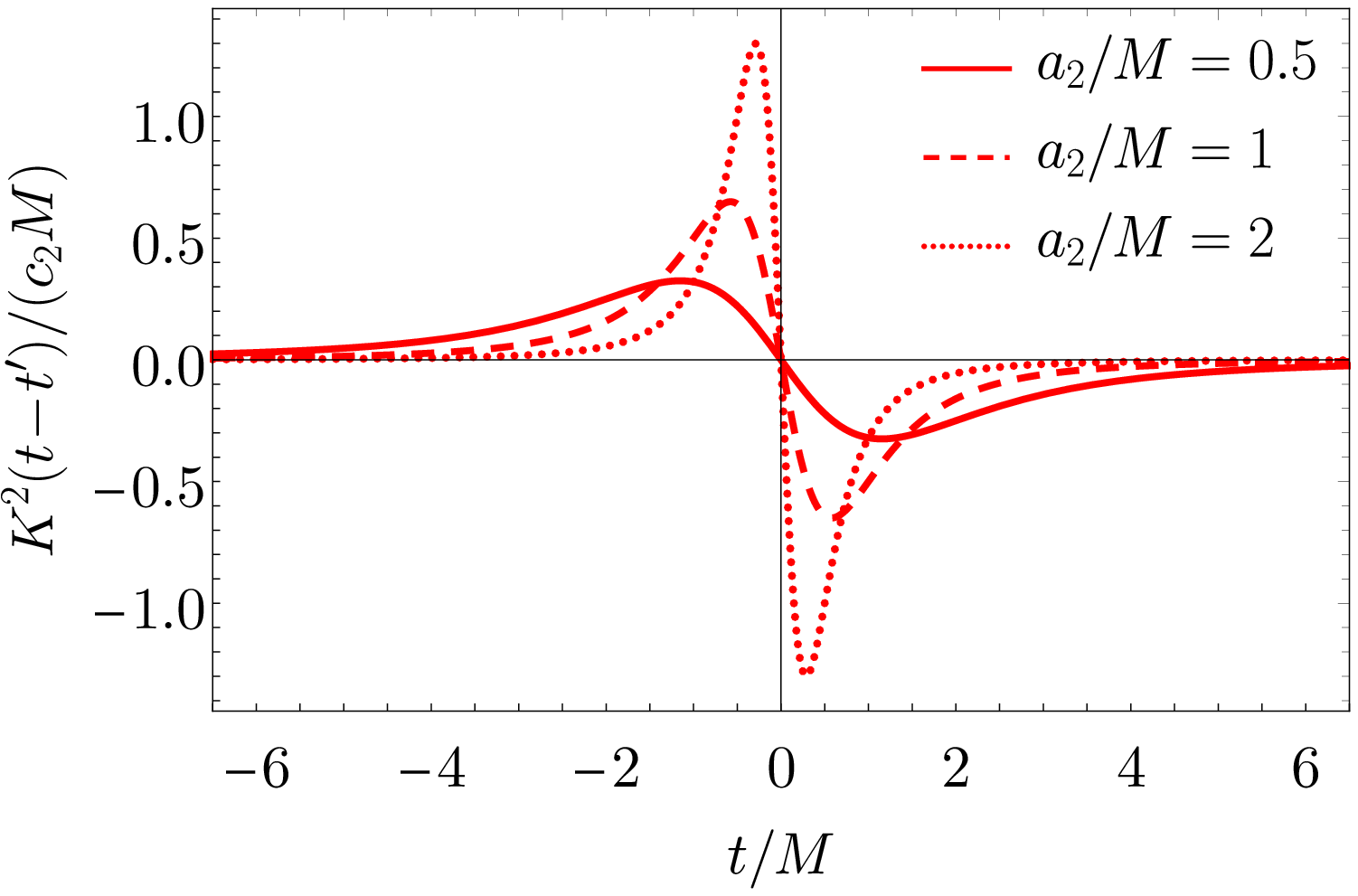}
\vspace{-0.3cm}
\caption[Kernels]{
{\it Left:} illustration of the exponential mock kernel from Eq.~(\ref{eq:kernelExp})
for several values of the kernel width parameter;
{\it Right:} illustration of the the Lorentzian mock kernel from Eq.~(\ref{eq:kernelLor})
for several values of the kernel width parameter.
} \label{fig:kernels}
\end{figure}

Approximate solutions constructed using multiple-scale perturbation theory are checked
versus direct numerical solutions of the non-local equation~(\ref{sec3eq}).
Since the approximation scheme does not depend on the specificities of
the two integral kernels (self-energy and proper four-vertex) in non-local terms, apart from 
them providing a finite window, for numerical checks we make use of simplified mock kernels instead,
\begin{equation}
\frac{\pi_{\scr R}(t\!-\!t')}{M^2}\, ,\
v_{\scr R}(t\!-\!t')
\longrightarrow
K_{\scr R}(t\!-\!t')
=
\theta(t\!-\!t') K(t\!-\!t') \, ,
\end{equation}
which satisfy the basic property
of being anti-symmetric,~$K(t\!-\!t')\!=\! - K(t'\!-\!t)$.
The two specific kernels we use are the exponential and the Lorentzian,
\begin{align}
K^{1}(t\!-\!t')
    ={}& - c_1 \times a_1^2
        (t\!-\!t') e^{-a_1|t-t'|} \, ,
\label{eq:kernelExp}
\\
K^{2}(t\!-\!t')
    ={}& - c_2 \times
        \frac{ 2 a_2^2  ({ t\!-\!t'})}
            { \bigl[ 1 + a_2^2 ({ t\!-\!t'})^2 \bigr]^2 } \, ,
\label{eq:kernelLor}
\end{align}
illustrated in~Fig.~\ref{fig:kernels}.
They provide a window of
a typical inverse width~$a_{1,2} \!>\! 0$, and~$c_{1,2}\!>\!0$ play the
role of the coupling constants in the sense that kernels are
normalized to it,
\begin{equation}
- \int_{-\infty}^{\infty} \! dt' \,
    K_{\scr R}^{1,2}(t\!-\!t')
    =
    - \int_{-\infty}^{t} \! dt' \,
    K^{1,2}(t\!-\!t')
    = c_{1,2} \, .
\end{equation}
What eventually appears in the analytic approximate solutions of Eq.~(\ref{sec3eq}) are the Fourier transforms
of integral kernels; for the mock kernels they read,
\begin{align}
\widetilde{K}_{\scr R}^1(\omega)
    ={}& c_1 \frac{ a_1 ^2 \bigl( \omega^2 \!-\! a_1^2 \bigr) }{ \bigl( \omega^2 \!+\! a_1^2 \bigr)^{\!2} }
		- i c_1 \frac{ 2 a_1^3 \omega}{ \bigl( \omega^2 \!+\! a_1^2 \bigr)^{\!2} } \, ,
\\
\widetilde{K}_{\scr R}^2(\omega)
    ={}&
    c_2 \biggl[
		\frac{\omega}{a_2}
		    {\rm ch} \Bigl( \frac{\omega}{a_2} \Bigr) \,
		    {\rm Shi} \Bigl( \frac{\omega}{a_2} \Bigr)
		- \frac{\omega}{a_2}
		    {\rm sh} \Bigl( \frac{\omega}{a_2} \Bigr) \,
		    {\rm Chi} \Bigl( \frac{\omega}{a_2} \Bigr)
		- 1 \biggr]
	- i c_2
	\frac{ \pi \omega e^{ - \frac{\omega}{a_2} } }{ 2 a} \, .
\end{align}
where the hyperbolic sine and cosine integrals are,
\begin{equation}
{\rm Shi}(x) = \int_{0}^{x} \! dz \, \frac{{\rm sh}(z) }{ z } \, ,
\qquad \quad
{\rm Chi}(x) = \gamma_\text{\tiny E} + \ln(x) +  \int_{0}^{x} \! dz \, \frac{ {\rm ch}(z) \!-\! 1 }{ z } \, ,
\end{equation}
respectively, and~$\gamma_\text{\tiny E}$
is the Euler-Mascheroni constant.

\bigskip

\subsection{Case 1: Linear equation}
\label{subsec: Case 1: Linear equation}

The first case we consider is the simplest, where in~(\ref{condensate eom}) we only keep 
terms linear in the condensate,
\begin{equation}
\ddot{\varphi}(t) + M^2 \varphi(t)
	+ \int_{t_0}^{t}\! dt' \, \pi_{\scr R}(t\!-\!t') \, \varphi(t') = 0 \, .
\label{linear eq}
\end{equation}
The approximate solution is constructed in a series of steps described below.
\begin{enumerate}[label=(\roman*)]
\item
\textbf{Identify small terms.}
The terms that can be treated as small are multiplied by
a formal parameter~$\varepsilon$ for bookkeeping purposes.
The solution is then constructed as an asymptotic series
in~$\varepsilon$, which is set back to one at the
very end. It is natural to treat the non-local term as
a small correction,
\begin{equation}
\ddot{\varphi}(t) + M^2 \varphi(t)
	+ \varepsilon \int_{t_0}^{t} \! dt' \, \pi_{\scr R}(t\!-\!t') \varphi(t') = 0 \, .
\label{linear epsilon}
\end{equation}

\item
\textbf{Assume two physical time scales.}
When there are hierarchies between typical physical scales of a given time-dependent problem,
it is reasonable to expect processes taking place on different time scales.
For small contributions in the equation of motion it will typically take a long time until their
influence has built up appreciably. This observation is formally implemented by assuming the
system evolves in two independent times --- {\it fast time}~$t$ and slow time~$\tau\!=\!\varepsilon t$ 
--- so that the solution takes the following form,
\begin{equation}
\varphi(t) = F(t,\tau;\varepsilon) \, .
\end{equation}
This assumption allows us to write the equation of motion~(\ref{linear eq}) as,
\begin{align}
\MoveEqLeft[6]
\frac{\partial^2 F(t,\tau;\epsilon) }{\partial t^2}
	+ 2 \varepsilon \frac{\partial^2 F(t,\tau; \epsilon) }{\partial t \partial \tau}
	+ \varepsilon^2 \frac{\partial^2 F(t,\tau; \epsilon )}{\partial\tau^2}
\nonumber \\
&
	+ M^2 F(t,\tau; \epsilon )
	+ \varepsilon \int_{t_0}^{t} \! dt' \,
		\pi_{\scr R}(t\!-\!t') F(t',\tau'; \varepsilon) = 0 \, ,
\label{EqTwoTime}
\end{align}
where we have used the chain rule to write out the derivatives.
Note that inside of the integral we still have to consider that~$\tau'\!=\!\varepsilon t'$.

\item
\textbf{Look for perturbative solution.}
Typically non-local equation such as~(\ref{linear epsilon}) are not analytically tractable,
and we have to resort to perturbation theory in the
formally small parameter~$\varepsilon$. The crucial step is to {\it not}
treat the~$\varepsilon$-dependence of the slow time~$\tau$
as a perturbative parameter, but only the~$\varepsilon$-dependence outside of it.
Thus, we look for a power series solution of the form,
\begin{equation}
F(t,\tau;\varepsilon)
	= F_0(t,\tau) + \varepsilon F_1(t,\tau) + \varepsilon^2 F_1(t,\tau) + \dots \, ,
\label{F expansion}
\end{equation}
where the coefficient functions are,
\begin{equation}
F_n(t,\tau) = \frac{1}{n!} \frac{\partial^n F(t,\tau;\varepsilon)}{\partial \varepsilon^n} \bigg|_{\varepsilon = 0} \, .
\end{equation}
In a sense, we have assumed a partially resummed ansatz,
where the~$\tau$ dependence of the solution eventually
accounts for the resummation of naive perturbation theory,
extending the validity of perturbative solution to late times.

\item\label{it:assumption}
\textbf{Assume limited extent of non-locality.} If the non-locality effectively extends
over a window of time  which the solution varies only slightly
in~$\tau$, it is appropriate to Taylor-expand the~$\tau$-dependence of the integrand,
\begin{align}
F(t',\tau';\varepsilon)
	&=
	F\bigl(t', \tau \!+\! (\tau'\!-\!\tau) ;\varepsilon \bigr)
\\
	&=
	F(t',\tau;\varepsilon)
	+ (\tau'\!-\!\tau) \frac{ \partial F(t',\tau;\varepsilon)}{\partial \tau}
	+ \frac{1}{2} (\tau'\!-\!\tau)^2 \frac{ \partial^2 F(t',\tau;\varepsilon)}{\partial \tau^2}
	+ \dots
\nonumber \\
	&=
	F(t',\tau;\varepsilon)
	+ \varepsilon (t'\!-\!t) \frac{ \partial F(t',\tau;\varepsilon)}{\partial \tau}
	+ \frac{\varepsilon^2}{2} (t'\!-\!t)^2 \frac{ \partial^2 F(t',\tau;\varepsilon)}{\partial \tau^2}
	+ \dots
\nonumber 
\end{align}
This effectively localizes the~$\tau$-dependence of the equation of motion, by recasting 
the non-local term 
as a power series expansion in~$\varepsilon$ which is introduced as an expansion parameter
in~(\ref{F expansion}),
\begin{align}
\MoveEqLeft[3]
\varepsilon \int_{t_0}^{t} \! dt' \,
		\pi_{\scr R}(t\!-\!t') F(t',\tau';\varepsilon)
\label{nonlocal expansion}
\\
={}&
	\varepsilon \int_{t_0}^{t} \! dt' \,
		\pi_{\scr R}(t\!-\!t') F(t',\tau;\varepsilon)
	+
	\varepsilon^2 \int_{t_0}^{t} \! dt' \, (t\!-\!t')
		\pi_{\scr R}(t\!-\!t') \frac{\partial F(t',\tau;\varepsilon) }{\partial \tau}
	+ \dots
\nonumber
\end{align}
This step is novel compared to the usual multiple-scale perturbation theory applied to local equations.
In physical terms, it corresponds to assuming the amplitude and the frequency 
of the condensate oscillations change only slightly during the window provided by the kernel
of the non-local term.

\item
\textbf{Organize equation of motion in powers of~$\boldsymbol{\varepsilon}$}.
Introducing the expansion~(\ref{nonlocal expansion}) for the non-local term, and
the perturbative expansion of the solution~(\ref{F expansion}) into the equation of
motion~(\ref{EqTwoTime}) allows to collect the terms of the same order in~$\varepsilon$,
which defines the perturbative tower of equations that have to be satisfied independently,
and are in general easier to solve. The resulting equations are solved order by order, where
the solution of the previous one provides a source for the following one.
We need only the leading and first subleading order,
\begin{align}
\MoveEqLeft[7]
0 =
\biggl[ \frac{\partial^2 F_0(t,\tau)}{\partial t^2} + M^2 F_0(t,\tau) \biggr]
+ \varepsilon \biggl[ \frac{\partial^2 F_1(t,\tau)}{\partial t^2} + M^2 F_1(t,\tau)
\\
&
	+ 2 \frac{ \partial^2 F_0(t,\tau) }{\partial t \partial \tau}
	+\! \int_{t_0}^{t} \! dt' \,
		\pi_{\scr R} (t\!-\!t') F_0(t',\tau) \biggr]
	+ \mathcal{O}(\varepsilon^2)
	 \, .
\nonumber
\end{align}

\item
\textbf{Solve leading order equation.} This is a harmonic oscillator equation,
\begin{equation}
\frac{\partial^2 F_0(t,\tau)}{ \partial t^2 }
	+ M^2 F_0(t,\tau) = 0 \, ,
\end{equation}
with a well known solution,
\begin{equation}
F_0(t,\tau) = {\rm Re} \Bigl[ R(\tau) e^{-iM t} \Bigr] \, .
\label{leading solution 1}
\end{equation}
Note that the leading equation is blind to the~$\tau$-dependence, resulting in~$R(\tau)$ being
left arbitrary as a constant of integration. This is a crucial feature of multiple-scale analysis
that plays an important role at the next step.

\item
{\bf Simplify sources in subleading equation.} The leading solution~(\ref{leading solution 1}) now
provides a source for the subleading equation,
\begin{align}
\MoveEqLeft[4]
\frac{\partial^2 F_1(t,\tau)}{ \partial t^2} + M^2 F_1(t,\tau)
	= - 2 \frac{\partial^2 F_0(t,\tau)}{ \partial t \partial \tau}
	- \int_{t_0}^{t} \! dt' \,
	\pi_{\scr R} (t\!-\!t') F_0(t',\tau)
\nonumber \\
={}&
	2 \, {\rm Re} \biggl\{e^{-iM t}  \biggl[ i M \frac{dR(\tau)}{d\tau}
	- \frac{1}{2} R(\tau) \int_{t_0}^{t} \! dt' \,
		\pi_{\scr R}(t\!-\!t') e^{iM (t-t')} \biggr] \biggr\}  \, .
\end{align}
Bearing in mind the assumption of a finite extent of non-locality, here we make the approximation 
of neglecting any early time transient effects due to initial conditions. This amounts to 
eliminating explicit~$t_0$ dependence of the source appearing in the lower limit of integration
by taking~$t_0\!\to\!-\infty$,
\begin{equation}
\int_{t_0}^{t} \! dt' \, \pi_{\scr R}(t\!-\!t') \, e^{i\omega (t-t')}
	\approx
	\int_{-\infty}^{t} \! dt' \, \pi_{\scr R}(t\!-\!t') \, e^{i\omega (t-t')}
	=
	\widetilde{\pi}_{\scr R}(\omega) \, ,
\end{equation}
which allows us to recognize the integrals as Fourier transforms 
defined in~(\ref{PiTildeDef}).
It is convenient to define separately real and imaginary parts of the Fourier transform,
\begin{equation}
\mu = \frac{ {\rm Re} \bigl[ \widetilde{\pi}_{\scr R}(M) \bigr]}{2M}  \, ,
\qquad\qquad
\gamma = - \frac{ {\rm Im} \bigl[ \widetilde{\pi}_{\scr R}(M) \bigr]}{M}  \, .
\end{equation}
as they will be responsible for different physical effects.

\item
\textbf{Remove spurious resonances from subleading correction.}
The equation of motion for the subleading correction now reads,
\begin{align}
\frac{\partial^2 F_1(t,\tau)}{ \partial t^2} + M^2 F_1(t,\tau)
	={}&
	2 \, {\rm Re} \biggl\{ i M e^{-iM t} \biggl[  \frac{dR(\tau)}{d\tau} 
		+ \frac{\gamma}{2} R(\tau) + i \mu R(\tau)  \biggr] \biggr\} \, .
\label{linear resonant forces}
\end{align}
Note that the~$\tau$ dependence appears only parametrically in this equation, which can
be considered as an ordinary differential equation in time~$t$.
It describes a harmonic oscillator~$F_1$ with a natural frequency~$M$, being driven by 
an oscillating force of the same frequency~$M$. 
It is well known that such driven oscillator exhibits resonant behaviour where its amplitude
grows without bound. However, this resonance is not a physical effect, but an artefact of
organizing the computation in a perturbative expansion. Physically, the amplitude of oscillations
can only decrease in this system. This is where the assumption of two times comes into play.
In order to avoid the spurious resonances that invalidate the perturbative
expansion at late times and give unphysical answers for the evolution, we require that~$R(\tau)$,
which is thus far unspecified, to be such that resonant driving forces are absent from 
Eq.~(\ref{linear resonant forces}), implying,
\begin{equation}
\frac{dR(\tau)}{d\tau} + \frac{\gamma}{2} R(\tau) + i \mu R(\tau) = 0 \, .
\label{ex1 R eq}
\end{equation}
This simultaneously fixes the ambiguity from the leading solution~(\ref{leading solution 1})
and removes unphysical secular behaviour from the subleading order.

When solving for the complex function~$R(\tau)$,
it is convenient to represent it in the polar basis
in terms of two real functions,~$R(\tau) \!=\! A(\tau) e^{-i f(\tau)}$,
which then breaks the complex equation~(\ref{ex1 R eq}) into two real ones,
\begin{equation}
\frac{dA(\tau)}{d\tau} = \frac{\gamma}{2} A(\tau) \, ,
\qquad\qquad
\frac{d f(\tau)}{d\tau} = \mu \, .
\end{equation}
which are readily solved to give,
\begin{equation}
R(\tau) = A_0 \exp\biggl[
	- i f_0
	- i \mu \tau - \frac{\gamma}{2} \tau \biggr] \, ,
\end{equation}
where~$A_0$ and~$f_0$ are real constants of integration. Removing the spurious resonances from the 
subleading correction thus fully fixes the leading order approximation as it determines~$R(\tau)$.
Note that here we did not have to solve for the subleading correction to accomplish that.

%
\begin{figure}[h!]
\footnotesize
\centering
(a) Exponential kernel \hfill
\setlength{\tabcolsep}{7.5pt}
\hfill
\begin{tabular}{c c c c c c}
	&	$a_1/M$		&	$c_1$	&	$\mu/M$	&	$\gamma/M$
\\
\cline{1-5}
$\color{yellowish} \blacksquare$		
    &	$0.5$	&	$0.05$	&	$0.003$		&	$0.008$
\\
$\color{bluish} \blacksquare$
    &	$0.5$	&	$0.1$	&	$0.006$		&	$0.016$
\\
$\color{reddish} \blacksquare$
    &	$2$		&	$0.05$	&	$-0.012$	&	$0.032$
\\
$\color{greenish} \blacksquare$
    &	$2$		&	$0.1$	&	$-0.024$	&	$0.064$
\end{tabular}
\includegraphics[width=\linewidth]{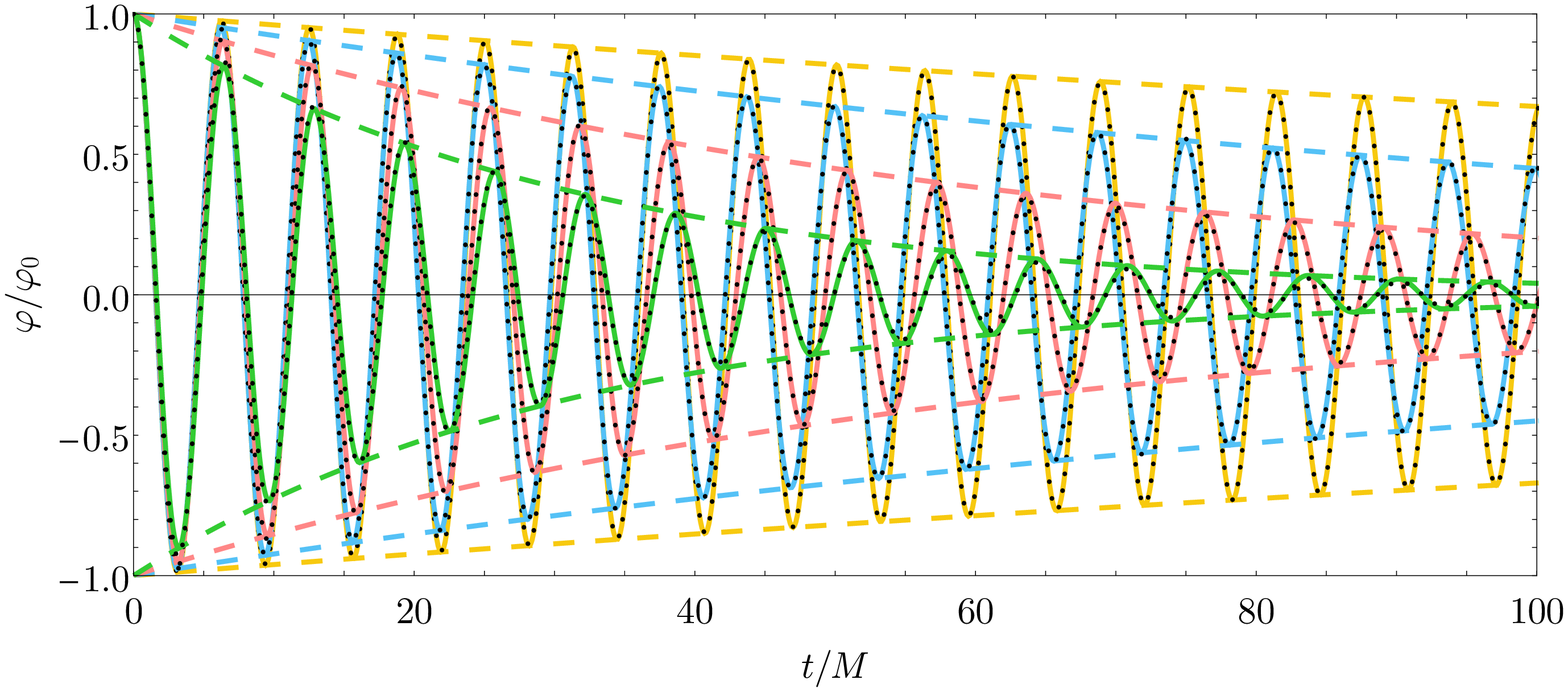}
\\
\vspace{0.35cm}
(b) Lorentzian kernel
\hfill
\begin{tabular}{c c c c c }
    &	$a_2/M$		&	$c_2$	&	$\mu/M$	&	$\gamma/M$
\\
\hline
$\color{yellowish} \blacksquare$
    &	$0.25$	&	$0.05$	&	$0.0033$	&	$0.0058$
\\
$\color{bluish} \blacksquare$
    &	$0.25$	&	$0.1$	&	$0.0066$	&	$0.0115$
\\
$\color{reddish} \blacksquare$
    &	$2$		&	$0.05$	&	$-0.0175$	&	$0.0238$
\\
$\color{greenish} \blacksquare$
    &	$2$		&	$0.1$	&	$-0.0350$	&	$0.0476$
\end{tabular}
\\
\includegraphics[width=\linewidth]{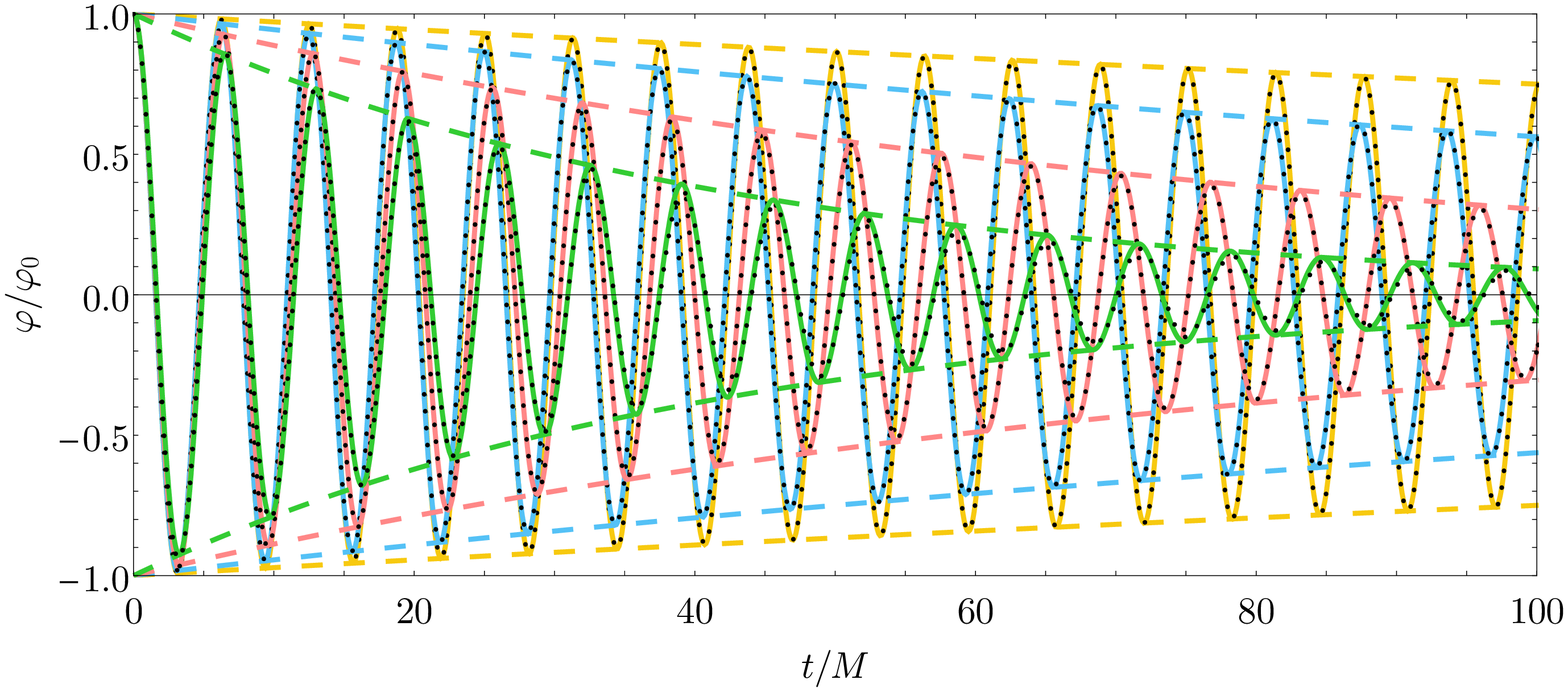}
\caption{
Solutions of the linear equation of motion~(\ref{linear eq}), where
the self-energy is taken to be (a) the exponential mock kernel from~(\ref{eq:kernelExp}),
and (b) the Lorentzian mock kernel from~(\ref{eq:kernelLor}), for different choices of parameters.
Analytic approximations~(\ref{ex1 solution}) are drawn in coloured solid curves with respective amplitudes 
in coloured dashed curves; corresponding numerical solutions of Eq.~(\ref{linear eq}) are indicated by 
black dotted curves.
}
\label{linear_plot}
\end{figure}
%

\item
\textbf{Obtain leading solution and determine initial conditions.}
Removing spurious resonances from the subleading
equation in the previous step fixes completely the leading approximation,
\begin{equation}
\varphi(t) \approx F_0(t,t)
	= A_0 e^{- \frac{1}{2} \gamma t}
	\cos\Bigl[ f_0 + ( M \!+\! \mu ) t \, \Bigr] \, ,
\label{ex1 solution}
\end{equation}
where~$A_0$ and~$f_0$ are constants of integration related to
initial conditions~$\varphi(t_0)\!=\!\varphi_0$
and~$\dot{\varphi}(t_0)\!=\!\dot{\varphi}_0$,
\begin{align}
A_0 = \sqrt{ \varphi_0^2
	+ \biggl[ \frac{
		\dot{\varphi}_0 \!+\! \frac{\gamma}{2} \varphi_0 }{ M \!+\! \mu }
		 \biggr]^{\!2} } \, ,
\qquad \quad
f_0 = - \arctan\biggl[ \frac{ \frac{\gamma}{2}
		\!+\! \frac{\dot{\varphi}_0}{\varphi_0 } }{ M \!+\! \mu } \biggr] \, .
\end{align}
This connection is obtained by evaluating the approximate
solution~(\ref{ex1 solution}) and its time derivative
at the initial time~$t_0\!=\!0$. 

\end{enumerate}

\medskip

\noindent Comparison between the approximate
solution~(\ref{ex1 solution}) and a direct numerical solution of Eq.~(\ref{linear eq})
is given in Fig.~\ref{linear_plot} demonstrating excellent agreement.
This is not surprising for this example, since the linear equation can be solved exactly yielding
the functional form captured by the approximation in~(\ref{ex1 solution}).
The effect that the linear non-local correction in the equation of motion~(\ref{linear eq}) 
has on the oscillating dynamics of the condensate is a constant shift of the oscillation
frequency, 
\begin{equation}
\Omega(t) = M + \mu \, ,
\end{equation}
and the damping of the oscillation amplitude which takes exponential form with the
rate~$\gamma/2$, giving a constant local damping rate,
\begin{equation}
\Upsilon(t) 
	= \frac{\gamma}{2} \, .
\end{equation}
The approximate solution~(\ref{ex1 solution}) has the precise form
of the solution one obtains for a linear damped harmonic oscillator described by a local equation.
This is not accidental, and is related to the assumption on the kernel providing a finite time window.
There exists a local equation approximation to the problem which captures the leading solution
we found here, which we discuss in detail in Sec.~\ref{sec: Comparison to Markovian equations}.

\subsection{Case 2: Quartic potential}
\label{subsec: Case 2: Quartic potential}

Taking into account the non-linear local potential
in addition to the linear non-local correction, the equation
of motion for the condensate reads,
\begin{equation}
\ddot{\varphi}(t) + M^2 \varphi(t)
	+ \frac{\Lambda}{6} \varphi^3(t)
	+ \int_{t_0}^{t}\! dt' \, \pi_{\scr R}(t\!-\!t') \varphi(t') = 0 \, .
\label{quartic eq}
\end{equation}
We find the approximate solution to this equation applying the algorithm developed
in Sec.~\ref{subsec: Case 1: Linear equation}.
\begin{enumerate}[label=(\roman*)]

\item
Consider all non-linear and non-local terms as small
corrections,
\begin{equation}
\ddot{\varphi}(t) + M^2 \varphi(t)
	+ \varepsilon \frac{\Lambda}{6} \varphi^3(t)
	+ \varepsilon \int_{t_0}^{t}\! dt' \, \pi_{\scr R}(t\!-\!t') \varphi(t') = 0 \, .
\end{equation}

\item
Assume the solution depends on the two time
scales,~$t$ and~$\tau\!=\!\varepsilon t$,
\begin{equation}
\varphi(t) = F(t,\tau;\varepsilon) \, ,
\end{equation}
for which the equation of motion is,
\begin{align}
\MoveEqLeft[6]
\frac{\partial^2 F(t,\tau;\varepsilon)}{\partial t^2}
	+ 2 \varepsilon \frac{\partial^2 F(t,\tau;\varepsilon) }{\partial t \partial \tau}
	+ \varepsilon^2 \frac{\partial^2 F(t,\tau;\varepsilon)}{\partial\tau^2}
	+ M^2 F(t,\tau;\varepsilon)
\nonumber \\
&
    + \varepsilon \frac{\Lambda}{6}
        \bigl[ F(t,\tau;\varepsilon) \bigr]^3
	+ \varepsilon \int_{t_0}^{t} \! dt' \,
		\pi_{\scr R}(t\!-\!t') F(t',\tau';\varepsilon) = 0 \, .
\end{align}

\item
Assume a perturbative series solution of the form,
\begin{equation}
F(t,\tau;\varepsilon)
	= F_0(t,\tau)
	+ \varepsilon F_1(t,\tau)
	+ \varepsilon^2 F_2(t,\tau) + \dots
\end{equation}

\item
The~$\tau$ dependence of the integrand in the non-local
term is expanded as,
\begin{equation}
\varepsilon \int_{t_0}^{t} \! dt' \, \pi_{\scr R}(t\!-\!t') F(t',\tau';\varepsilon)
	= \varepsilon \int_{t_0}^{t} \! dt' \, \pi_{\scr R}(t\!-\!t') F(t',\tau;\varepsilon)
	+ \mathcal{O}(\varepsilon^2) \, .
\end{equation}

\item
The equation of motion is organized in powers
of~$\varepsilon$ as,
\begin{align}
\MoveEqLeft[3]
\biggl[ \frac{\partial^2 F_0(t,\tau) }{ \partial t^2 } + M^2 F_0(t,\tau) \biggr]
	+ \varepsilon \biggl[
	\frac{\partial^2 F_1(t,\tau)}{\partial t^2} + M^2 F_1(t,\tau)
	+ 2 \frac{\partial^2 F_0(t,\tau) }{ \partial t \partial\tau }
\qquad
\nonumber \\
&
	+ \frac{\Lambda}{6}
	    \bigl[ F_0(t,\tau) \bigr]^3
	+ \int_{t_0}^{t} \! dt'\, \pi_{\scr R}(t\!-\!t') F_0(t',\tau)
	\biggr]
	+ \mathcal{O}(\varepsilon^2)
	= 0 \, .
\end{align}

\item
The solution of the leading equation is,
\begin{equation}
F_0(t,\tau) = {\rm Re} \Bigl[ R(\tau) e^{-iM t} \Bigr] \, .
\end{equation}

\item
The subleading equation is,
\begin{align}
\MoveEqLeft[4]
\frac{\partial^2 F_1(t,\tau)}{ \partial t^2} + M^2 F_1(t,\tau)
	= 2 \, {\rm Re} \biggl\{ e^{-iM t}  \biggl[ i M\frac{dR(\tau)}{d\tau}
		- \frac{\Lambda}{16}
		   \bigl[ R(\tau) \bigr]^2 R^*(\tau)
\nonumber \\
&
		- \frac{ R(\tau)}{2} \! \int_{t_0}^{t} \! dt' \,
		\pi_{\scr R}(t\!-\!t') e^{iM (t-t')} \biggr] \biggr\}
	- \frac{\Lambda}{24} \, {\rm Re} \biggl\{ e^{-3iM t} \bigl[ R(\tau) \bigr]^3 \biggr\}
\, .
\end{align}
We neglect early time transient effects by taking~$t_0\!\to\!-\infty$,
recognizing the Fourier transform from~(\ref{PiTildeDef})
\begin{equation}
\int_{t_0}^{t} \! dt' \, \pi_{\scr R}(t\!-\!t') e^{i\omega (t-t')} 
\approx
\int_{-\infty}^{t} \! dt' \, \pi_{\scr R}(t\!-\!t') e^{i\omega (t-t')} 
	=
	\widetilde{\pi}_{\scr R}(\omega) \, ,
\end{equation}
so that the subleading equation reads,
\begin{align}
\MoveEqLeft[4]
\frac{\partial^2 F_1(t,\tau)}{ \partial t^2} + M^2 F_1(t,\tau)
	= 2 \, {\rm Re} \biggl\{ i M e^{-iM t}  \biggl[ \frac{dR(\tau)}{d\tau}
		+ \frac{ i \Lambda}{16 M} \bigl[ R(\tau) \bigr]^2 R^*(\tau)
\nonumber \\
&
		+ \Bigl( \frac{ \gamma }{2} \!+\! i \mu \Bigr) R(\tau)
		\biggr]
		\biggr\}
	- \frac{\Lambda}{24} \, {\rm Re} \biggl\{ e^{-3iM t} \bigl[ R(\tau) \bigr]^3 \biggr\} \, ,
\label{ex2 subleading eq}
\end{align}
where we have defined,
\begin{equation}
\mu = \frac{ {\rm Re} \bigl[ \widetilde{\pi}_{\scr R}(M) \bigr]}{2M}  \, ,
\qquad\qquad
\gamma = - \frac{ {\rm Im} \bigl[ \widetilde{\pi}_{\scr R}(M) \bigr]}{M}  \, .
\end{equation}

\item
Removing spurious resonances from the subleading equation requires removing sources
in~(\ref{ex2 subleading eq}) that oscillate with the frequency~$M$, which is accomplished by,
\begin{equation}
\frac{dR(\tau)}{d\tau}
		+ \frac{ i \Lambda}{16 M} \bigl[ R(\tau) \bigr]^2 R^*(\tau)
		+ \Bigl( \frac{ \gamma }{2} \!+\! i \mu \Bigr) R(\tau) = 0 \, .
\end{equation}
It is best to adopt a polar representation for the solution,~$R(\tau) \!=\! A(\tau) e^{-if(\tau)}$,
upon which the complex equation above breaks into two real ones,
\begin{equation}
\frac{dA(\tau)}{d\tau}
	= - \frac{\gamma}{2} A(\tau) \, ,
\qquad \qquad
\frac{df(\tau)}{d\tau}
	=
	\mu
	+ \frac{\Lambda}{16 M} A^2(\tau) \, .
\end{equation}
Integrating first the left equation, and then the right one gives,
\begin{equation}
A(\tau) = A_0 e^{-\frac{1}{2} \gamma \tau} \, ,
\qquad \quad
f(\tau) = f_0 + \mu \tau
	- \frac{\Lambda A_0^2}{16 M \gamma}
	    \bigl( e^{-\gamma \tau} \!-\! 1 \bigr) \, ,
\end{equation}
where~$A_0$ and~$f_0$ are constants of integration.

\item
The leading approximation in this case is therefore,
\begin{equation}
\varphi(t) 
	\approx F_0(t,t)
	= A_0 e^{- \frac{1}{2} \gamma t}
	\cos\biggl[ f_0 + ( M \!+\! \mu ) t
		 + \frac{\Lambda A_0^2}{16 M \gamma} \bigl( 1 \!-\! e^{-\gamma t} \bigr) \biggr] \, ,
\label{ex2 solution}
\end{equation}
where~$A_0$ and~$f_0$ are related to initial conditions via,
\begin{align}
0 ={}&
\bigl( \varphi_0^2 \!-\! A_0^2 \bigr)
    \biggl( M \!+\! \mu \!+\! \frac{ \Lambda A_0^2 }{ 16M } \biggr)^{\!2}
    +
    \biggl( \dot{\varphi}_0 \!+\! \frac{\gamma}{2} \varphi_0 \biggr)^{\!2}
     \, ,
\\
f_0 ={}& - \arctan\biggl[ \frac{ \frac{\gamma}{2}
		\!+\!
		\frac{\dot{\varphi}_0}{\varphi_0 } }
		    { M \!+\! \mu \!+\! \frac{\Lambda A_0^2}{16M} }
		    \biggr] \, .
\end{align}

\end{enumerate}

\medskip

\begin{figure}[h!]
\footnotesize
\centering
(a) Exponential kernel \hfill
\setlength{\tabcolsep}{7.5pt}
\hfill
\begin{tabular}{c c c c c c}
	&	$a_1/M$		&	$c_1$	&	$\mu/M$	&	$\gamma/M$
\\
\cline{1-5}
$\color{yellowish} \blacksquare$		&	$0.5$	&	$0.05$	&	$0.003$		&	$0.008$
\\
$\color{bluish} \blacksquare$			&	$0.5$	&	$0.1$	&	$0.006$		&	$0.016$
\\
$\color{reddish} \blacksquare$			&	$2$		&	$0.05$	&	$-0.012$		&	$0.032$
\\
$\color{greenish} \blacksquare$		&	$2$		&	$0.1$	&	$-0.024$		&	$0.064$
\end{tabular}
\includegraphics[width=\linewidth]{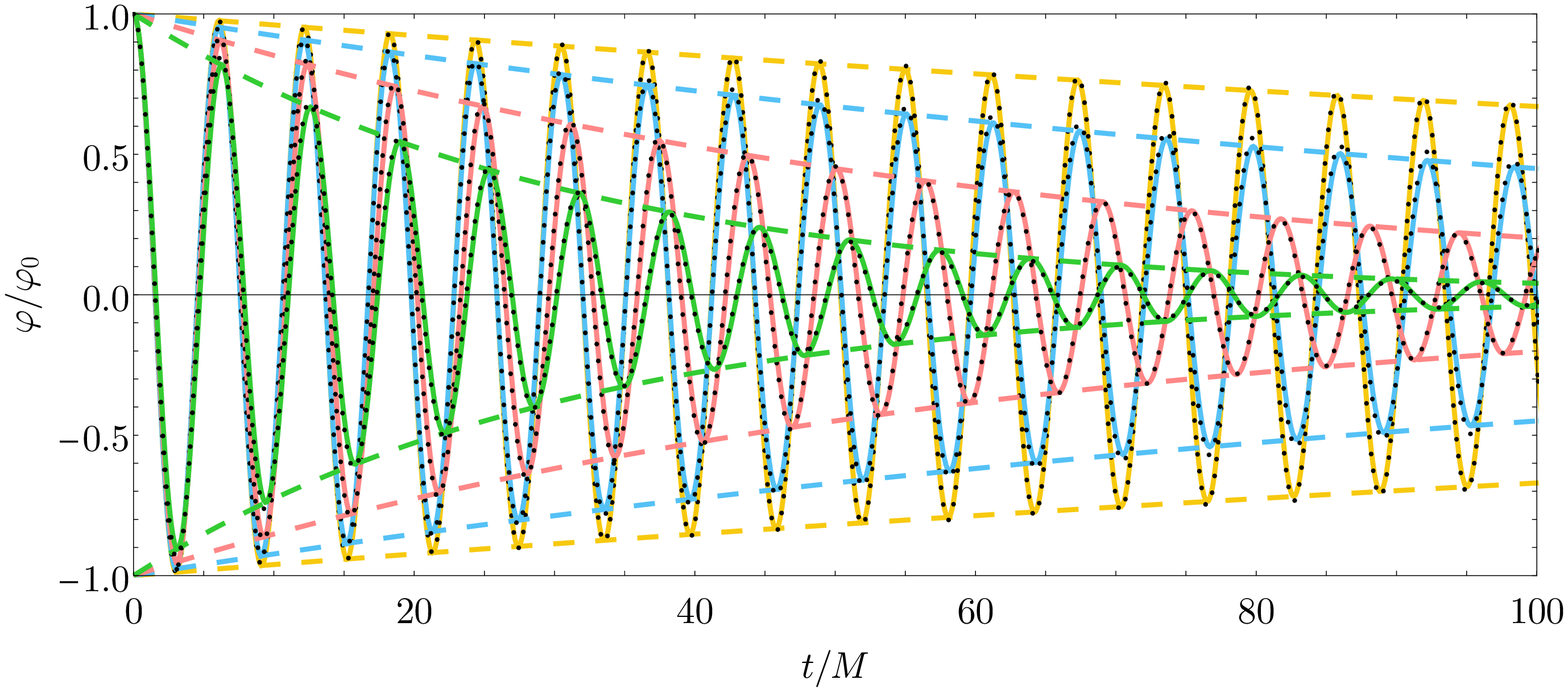}
\\
\vspace{0.25cm}
(b) Lorentzian kernel
\hfill
\begin{tabular}{c c c c c }
&	$a_2/M$		&	$c_2$	&	$\mu/M$	&	$\gamma/M$
\\
\hline
$\color{yellowish} \blacksquare$		&	$0.25$	&	$0.05$	&	$0.0033$	&	$0.0058$
\\
$\color{bluish} \blacksquare$			&	$0.25$	&	$0.1$ 	&	$0.0066$	&	$0.0115$
\\
$\color{reddish} \blacksquare$			&	$2$		&	$0.05$	&	$-0.0175$	&	$0.0238$
\\
$\color{greenish} \blacksquare$		&	$2$		&	$0.1$	&	$-0.0350$	&	$0.0476$
\end{tabular}
\\
\includegraphics[width=\linewidth]{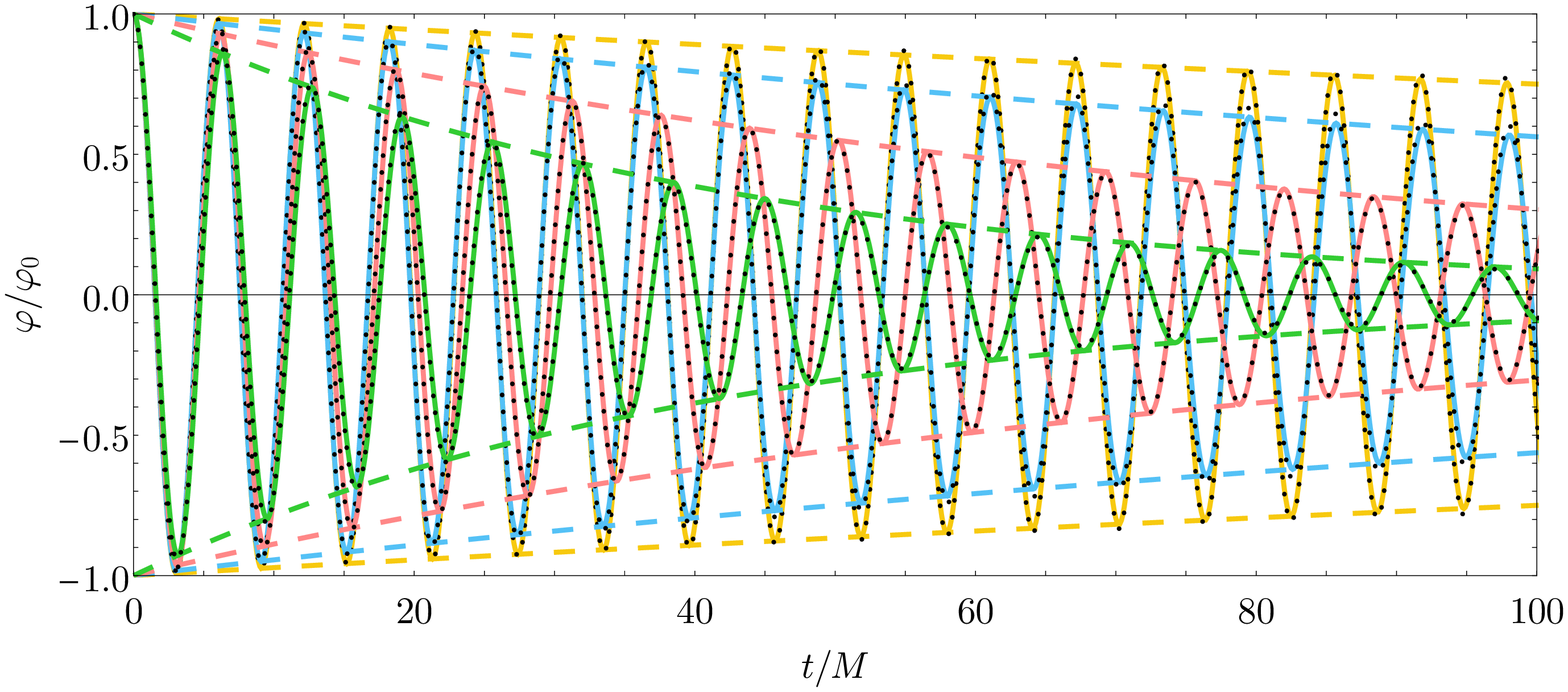}
\\
\vspace{-0.15cm}
\caption{
Solutions of the condensate equation of motion~(\ref{quartic eq}) containing a linear
non-local term and a local cubic term descending from the tree-level quartic potential, where
the self-energy is taken to be (a) the exponential mock kernel from~(\ref{eq:kernelExp}),
and (b) the Lorentzian mock kernel from~(\ref{eq:kernelLor}), for different choices of parameters
where~$\Lambda \varphi_0^2 / M^2 \!=\! 0.5$ in all cases.
Analytic approximations~(\ref{ex2 solution}) are drawn in solid coloured curves with respective amplitudes 
in dashed coloured curves; corresponding numerical solutions are indicated by black dotted curves.
}
\label{quartic_plot}
\end{figure}

\noindent Comparison between the approximate
solution~(\ref{ex2 solution}) and direct numerical solutions of Eq.~(\ref{quartic eq})
is given in Fig.~\ref{quartic_plot}, showing excellent agreement.
%
%
%
The local damping rate in this case is constant,
\begin{equation}
\Upsilon(t) = \frac{\gamma}{2} \, ,
\end{equation}
just as in the case of the linear equation from the preceding section.
This is not surprising since only the linear non-local term contributes to dissipation,
while the local quartic potential does not.
In fact, in the absence of the non-local term, it is well known that the quartic potential 
only shifts the oscillation frequency~(cf. Duffing's equation in~\cite{Bender}), 
which is clearly seen in the early time limit,
\begin{equation}
\varphi(t)
\underset{t \, \ll \, 1/\mu}{\widesim[2.5]{t \, \ll \, 1/\gamma }}
    A_0
	\cos\biggl[ f_0
	    + \Bigl( M 
	        \!+\! \frac{\Lambda A_0^2}{16 M } \Bigr) t \,
		\biggr] \, ,
\end{equation}
These oscillations do not continue at this frequency
forever. As time progresses the amplitude is reduced due to dissipation from the 
non-local term. As the amplitude is reduced, so is the influence of the quartic
potential, which leads to a time-varying effective frequency,
\begin{equation}
\Omega(t) = \frac{d}{dt} \biggl[ 
	(M \!+\! \mu)t + \frac{\Lambda A_0^2}{16 M \gamma} \bigl( 1\!-\!e^{-\gamma t} \bigr) \biggr]
	= M + \mu + \frac{\Lambda A_0^2}{16 M} e^{-\gamma t} \, ,
\end{equation}
which at late times becomes the frequency of the linear oscillator 
from Sec.~\ref{subsec: Case 1: Linear equation}. In fact, at late times
the only information about the quartic potential that survives is the constant phase shift,
\begin{equation}
\varphi(t)
\widesim[2.]{t \, \gg \, 1/\gamma }
    A_0 e^{- \frac{1}{2} \gamma t}
	\cos\biggl[ f_0 + ( M \!+\! \mu ) t
		+ \frac{\Lambda A_0^2}{16 M \gamma}  \biggr] \, .
\end{equation}
%

\subsection{Case 3: Cubic non-locality}
\label{subsec: Case 3: Cubic non-locality}

Consider now the case where the linear non-local correction
in~(\ref{condensate eom}) is either absent or negligible,
and where we keep only the cubic non-local correction,
\begin{equation}
\ddot{\varphi}(t) + M^2 \varphi(t)
	+ \frac{\varphi(t)}{6}
	\int_{t_0}^{t}\! dt' \, v_{\scr R}(t\!-\!t') \varphi^2(t') = 0 \, .
\label{ex3 eom}
\end{equation}
Approximate solution to this equation is constructed by applying the algorithm outlined
in Sec.~\ref{subsec: Case 1: Linear equation}.
\begin{enumerate}[label=(\roman*)]

\item
Consider the non-local term to be a small correction,
\begin{equation}
\ddot{\varphi}(t) + M_\phi^2 \varphi(t)
	+ \varepsilon \frac{\varphi(t)}{6}
	\int_{t_0}^{t}\! dt' \,
	v_{\scr R}(t\!-\!t') \varphi^2(t') = 0 \, .
\end{equation}

\item
Assume the solution depends on the two time
scales~$t$ and~$\tau\!=\!\varepsilon t$,
\begin{equation}
\varphi(t) = F(t,\tau;\varepsilon) \, ,
\end{equation}
so that the equation of motion is,
\begin{align}
\MoveEqLeft[10]
\frac{\partial^2 F(t,\tau;\varepsilon)}{\partial t^2}
	+ 2 \varepsilon \frac{\partial^2 F(t,\tau;\varepsilon) }{\partial t \partial \tau}
	+ \varepsilon^2 \frac{\partial^2 F(t,\tau;\varepsilon)}{\partial\tau^2}
	+ M^2 F(t,\tau;\varepsilon)
\nonumber \\
&
	+ \frac{ \varepsilon }{6} F(t,\tau;\varepsilon)
	    \int_{t_0}^{t} \! dt' \,
		v_{\scr R}(t\!-\!t') \bigl[ F(t',\tau';\varepsilon) \bigr]^2 = 0 \, .
\end{align}

\item
Assume a perturbative power series solution of the form,
\begin{equation}
F(t,\tau;\varepsilon)
	= F_0(t,\tau)
	+ \varepsilon F_1(t,\tau)
	+ \varepsilon^2 F_2(t,\tau) + \dots
\end{equation}

\item
The~$\tau$-dependence of the integrand in the non-local term is expanded
in~$\varepsilon$ as,
\begin{align}
\MoveEqLeft[5]
\varepsilon F(t,\tau;\varepsilon) \int_{t_0}^{t} \! dt' \,
	v_{\scr R}(t\!-\!t') \bigl[ F(t',\tau';\varepsilon) \bigr]^2
\nonumber \\
={}& 
	\varepsilon F(t,\tau;\varepsilon) \int_{t_0}^{t} \! dt' \,
	v_{\scr R}(t\!-\!t') \bigl[ F(t',\tau;\varepsilon) \bigr]^2
	+ \mathcal{O}(\varepsilon^2) \, .
\end{align}

\item
The equation of motion organized in powers of~$\varepsilon$ reads,
\begin{align}
\MoveEqLeft[3]
\biggl[ \frac{\partial^2 F_0(t,\tau) }{ \partial t^2 } + M^2 F_0(t,\tau) \biggr]
	+ \varepsilon \biggl[
	\frac{\partial^2 F_1(t,\tau)}{\partial t^2} + M^2 F_1(t,\tau)
	+ 2 \frac{\partial^2 F_0(t,\tau) }{ \partial t \partial\tau }
\qquad
\nonumber \\
&
	+ \frac{F_0(t,\tau)}{6}
	    \int_{t_0}^{t} \! dt'\, v_{\scr R}(t\!-\!t') \bigl[ F_0(t',\tau) \bigr]^2
	\biggr]
	+ \mathcal{O}(\varepsilon^2)
	= 0 \, .
\end{align}

\item
The solution of the leading equation is,
\begin{equation}
F_0(t,\tau) = {\rm Re} \Bigl[ R(\tau) e^{-iM t} \Bigr] \, .
\end{equation}

\item
The subleading equation is,
\begin{align}
\MoveEqLeft[5]
\frac{\partial^2 F_1(t,\tau)}{ \partial t^2} + M^2 F_1(t,\tau)
\nonumber \\
={}&
	2 \, {\rm Re} \biggl\{ e^{-iM t}  \biggl[ i M \frac{dR(\tau)}{d\tau}
		- \frac{1}{24} \bigl[ R(\tau) \bigr]^2 R^*(\tau)  \int_{t_0}^{t} \! dt'\, v_{\scr R}(t\!-\!t')
\nonumber \\
&	\hspace{3cm}
		- \frac{1}{48} \bigl[ R(\tau) \bigr]^2 R^*(\tau) \int_{t_0}^{t} \! dt'\, v_{\scr R}(t\!-\!t') e^{2iM(t-t')}
		 \biggr] \biggr\}
\nonumber \\
&
	- \frac{1}{24} \, {\rm Re} \biggl\{ \bigl[ R(\tau) \bigr]^3 e^{-3iM t} \int_{t_0}^{t} \! dt'\, v_{\scr R}(t\!-\!t') e^{2iM(t-t')} \biggr\}
\, .
\end{align}
We neglect early time transient effects by taking~$t_0\!\to\!-\infty$,
\begin{equation}
\int_{t_0}^{t} \! dt'\, v_{\scr R}(t\!-\!t')  e^{i\omega(t-t')}
	\approx
	\int_{-\infty}^{t} \! dt'\, v_{\scr R}(t\!-\!t')  e^{i\omega(t-t')}
	=
	\widetilde{v}_{\scr R}(\omega) \, ,
\end{equation}
where we recognized the Fourier transform from~(\ref{PiTildeDef}), so that the subleading equation reads,
\begin{align}
\frac{\partial^2 F_1(t,\tau)}{ \partial t^2} + M^2 F_1(t,\tau)
={}&
	2 {\rm Re} \biggl\{ i M e^{-iM t}  \biggl[ \frac{dR(\tau)}{d\tau}
		+ \Bigl( \frac{\sigma}{2}  \!+\! i \alpha \Bigr) \bigl[ R(\tau) \bigr]^2 R^*(\tau)
		 \biggr]
		 \biggr\}
\nonumber \\
&
	+ 2 {\rm Re} \biggl\{ i M e^{-3iM t}  \Bigl( \frac{\sigma}{2} \!+\! i \alpha_2 \Bigr) \bigl[ R(\tau) \bigr]^3 \biggr\}
\, ,
\end{align}
where we have defined,
\begin{equation}
\alpha_0 = \frac{ {\rm Re}
	    \bigl[ \widetilde{v}_{\scr R}(0) \bigr] }{24M} \, ,
\qquad
\alpha_2 = \frac{ {\rm Re}
	    \bigl[ \widetilde{v}_{\scr R}(2M) \bigr] }{48M} \, ,
\qquad
\sigma =
    - \frac{ {\rm Im}
        \bigl[ \widetilde{v}_{\scr R}(2M) \bigr] }{24 M} \, ,
\end{equation}
and~$\alpha\!=\!\alpha_0\!+\!\alpha_2$.
Note that the imaginary part of the physical proper four-vertex is 
antisymmetric~$ {\rm Im} \bigl[ \widetilde{v}_{\scr R}(-\omega) \bigr] 
	\!=\! - {\rm Im} \bigl[ \widetilde{v}_{\scr R}(\omega) \bigr]$,
so that~${\rm Im}\bigl[ \widetilde{v}_{\scr R}(0) \bigr]  \!=\! 0$.

\item
The terms in the first four lines on the right hand side
cause spurious resonances in the subleading correction
if allowed to appear, and we remove them by requiring,
\begin{equation}
\frac{dR(\tau)}{d\tau}
		+ \Bigl( \frac{\sigma}{2} \!+\! i \alpha_2 \Bigr) \bigl[ R(\tau) \bigr]^2 R^*(\tau)
	= 0 \, ,
\label{ex3 A condition}
\end{equation}
Adopting a polar representation for the solution,~$R(\tau) \!=\! A(\tau) e^{-if(\tau)}$,
breaks up the complex equation above into two real ones,
\begin{equation}
\frac{dA(\tau)}{d\tau}
	= - \frac{\sigma}{2} \bigl[ A(\tau) \bigr]^3 \, ,
\qquad \qquad
\frac{df(\tau)}{d\tau}
	= \alpha \bigl[ A(\tau) \bigr]^2 \, .
\label{ex3 A f eqs}
\end{equation}
Integrating these equations gives,
\begin{equation}
A(\tau) = \frac{A_0}{ \sqrt{ 1 \!+\! \sigma A_0^2 \tau } } \, ,
\qquad \quad
f(\tau) = f_0 + \frac{\alpha}{\sigma}
	\ln\bigl( 1 \!+\! \sigma A_0^2 \tau \bigr) \, ,
\end{equation}
where~$A_0$ and~$f_0$ are constants of integration.

\item
The leading approximation for the condensate evolution in this case is therefore,
\begin{equation}
\varphi(t) =
	\frac{A_0}{ \sqrt{ 1 \!+\! \sigma A_0^2 t } }
		\cos\biggl[ f_0 + M t + \frac{\alpha}{\sigma}
			\ln\bigl( 1 \!+\! \sigma A_0^2 t \bigr) \biggr]
	 \, ,
\label{ex3 solution}
\end{equation}
where the constants of integration are related to the initial conditions via,
\begin{align}
0 ={}&
\bigl( \varphi^2_0 \!-\! A_0^2 \bigr)
	\bigl( M \!+\! \alpha A_0^2 \bigr)^2 
	+ \biggl( \dot{\varphi}_0 \!+\! \frac{\sigma A_0^2}{2} \varphi_0 \biggr)^{\!2}
             \, ,
\\
f_0 ={}& - \arctan\biggl[
    \frac{ \frac{\sigma A_0^2 }{2} \!+\! \frac{\dot{\varphi}_0}{\varphi_0} }{ M\!+\! \alpha A_0^2} \biggr] \, .
\end{align}

\end{enumerate}

\medskip

\begin{figure}[h!]
\footnotesize
\centering
(a) Exponential kernel \hfill
\setlength{\tabcolsep}{6.5pt}
\hfill
\begin{tabular}{c c c c c c}
	&	$a_1/M$		&	$c_1\varphi_0^2/M^2$	&	$\alpha\varphi_0^2/M$	&	$\sigma\varphi_0^2/M$
\\
\cline{1-5}
$\color{yellowish} \blacksquare$		
    &	$1$		&	$1$	&	$-0.039$	&	$0.0067$
\\
$\color{bluish} \blacksquare$
    &	$1$		&	$2$	&	$-0.078$	&	$0.0133$
\\
$\color{reddish} \blacksquare$
    &	$4$		&	$1$	&	$-0.052$	&	$0.0267$
\\
$\color{greenish} \blacksquare$
    &	$4$		&	$2$	&	$-0.103$	&	$0.0533$
\end{tabular}
\includegraphics[width=\linewidth]{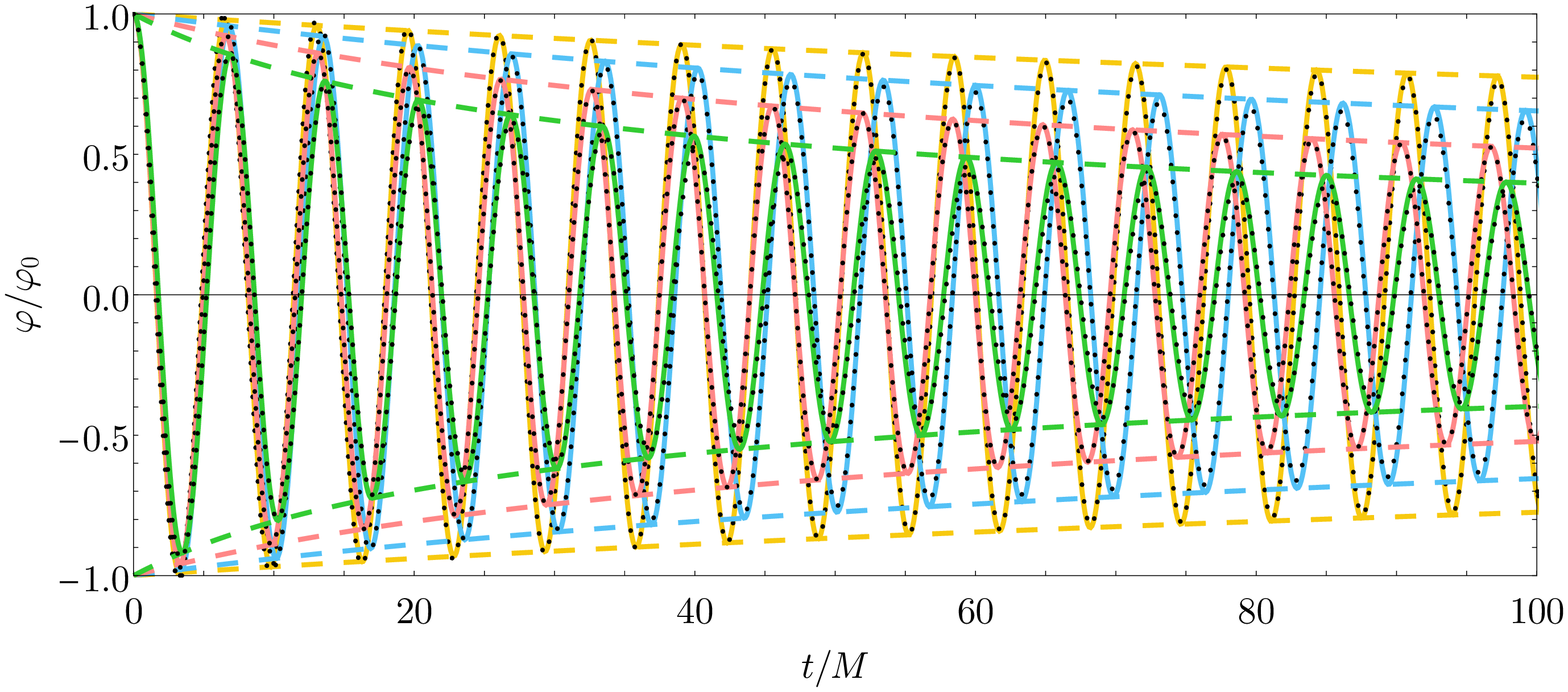}
\\
\vspace{0.35cm}
(b) Lorentzian kernel
\hfill
\begin{tabular}{c c c c c }
    &	$a_2/M$		&	$c_2\varphi_0^2/M^2$	&	$\alpha\varphi_0^2/M$	&	$\sigma\varphi_0^2/M$
\\
\hline
$\color{yellowish} \blacksquare$
    &	$0.5$	&	$1$	    &	$-0.039$	&	$0.0048$
\\
$\color{bluish} \blacksquare$
    &	$0.5$	&	$2$ 	&	$-0.078$	&	$0.0096$
\\
$\color{reddish} \blacksquare$
    &	$2$		&	$1$	    &	$-0.049$	&	$0.0241$
\\
$\color{greenish} \blacksquare$
    &	$2$		&	$2$	    &	$-0.098$	&	$0.0482$
\end{tabular}
\\
\includegraphics[width=\linewidth]{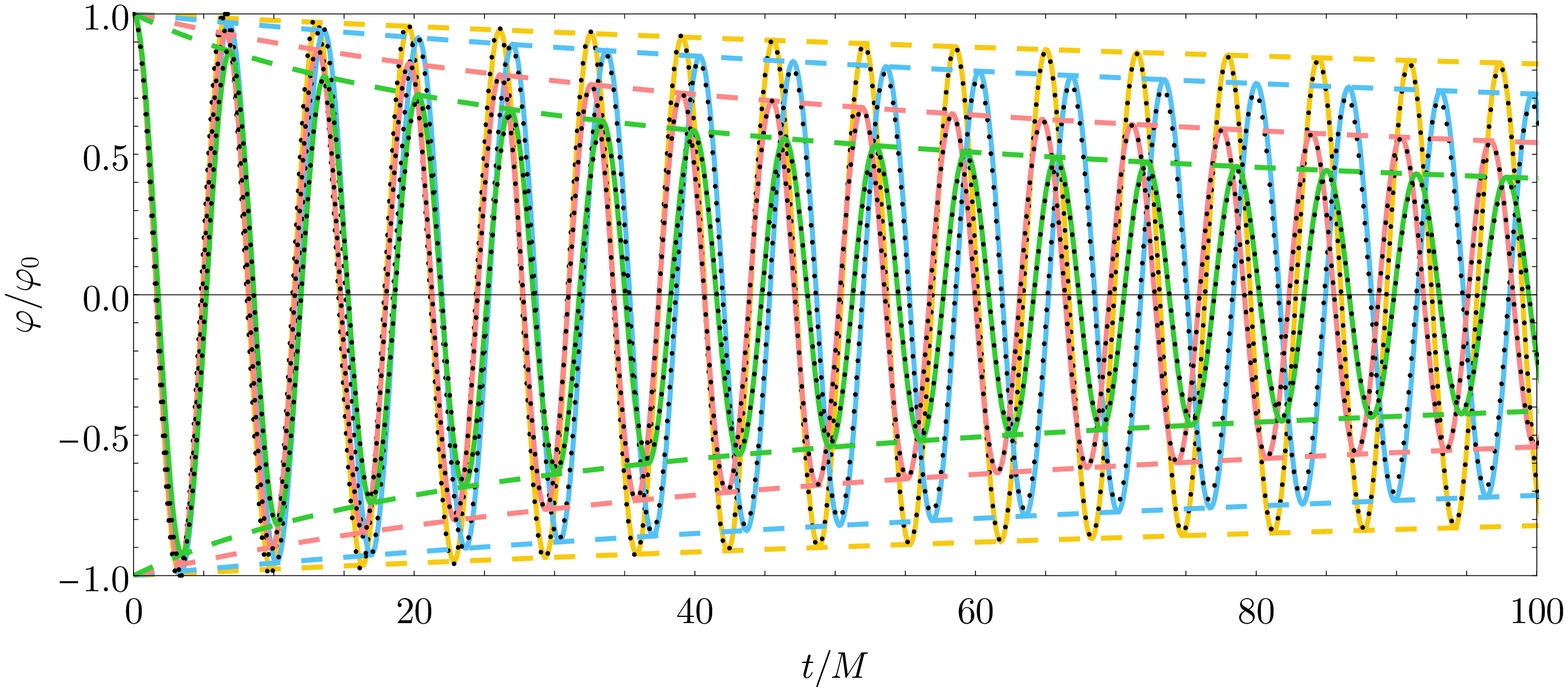}
\caption{
Solutions of the condensate equation of motion~(\ref{ex3 eom}) containing only a cubic non-local term, 
where the proper four-vertex is taken to be (a) the exponential mock kernel from~(\ref{eq:kernelExp}),
and (b) the Lorentzian mock kernel from~(\ref{eq:kernelLor}), for different choices of parameters.
Analytic approximations~(\ref{ex3 solution}) are drawn in solid coloured curves with respective amplitudes 
in dashed coloured curves; corresponding numerical solutions are indicated by black dotted curves.
}
\label{cubic_plot}
\end{figure}

\noindent Comparison between the approximate
solution~(\ref{ex2 solution}) and direct numerical solutions of Eq.~(\ref{quartic eq})
is given in Fig.~\ref{cubic_plot}, showing excellent agreement.
%
%
The solution in~(\ref{ex3 solution}) exhibits 
behaviour rather different than the solution of the
linear equation~(\ref{ex1 solution}). 
Firstly, it oscillates with a continuously changing frequency,
\begin{equation}
\Omega(t) =
    \frac{d}{dt} \biggl[ M t + \frac{\alpha}{\sigma}
			\ln\bigl( 1 \!+\! \sigma A_0^2 t \bigr) \biggr]
	= M + \frac{\alpha A_0^2}
	    {1 \!+\! \sigma A_0^2 t } \, .
\end{equation}
Secondly, and more strikingly, the oscillations are damped, but the
attenuation is a power-law, giving a decreasing local damping rate,
\begin{equation}
\Upsilon(t) 
    = \frac{ \frac{1}{2} \sigma A_0^2}{1 \!+\! \sigma A_0^2 t } \, .
\end{equation}
This signals that the condensate dissipates
less and less as time progresses, which is not strange if one notes that the
non-local correction in the condensate
equation~(\ref{ex3 eom}) gets smaller as the
condensate relaxes to smaller amplitudes.
We emphasize that
the exponential damping (constant local damping rate) is not a general feature
of a dissipating system, but a feature of a linearized
dissipative system, i.e. of linear response theory.
For non-linear systems the form of the damping depends on
particular interactions giving rise to dissipation.

\subsection{Case 4: Full equation}
\label{subsec: Full equation}

Lastly, we consider the full condensate equation of
motion~(\ref{condensate eom}),
\begin{equation}
\ddot{\varphi}(t) + M^2 \varphi(t)
    + \frac{\Lambda}{6} \varphi^3(t)
	+ \int_{t_0}^{t}\! dt' \, \pi_{\scr R}(t\!-\!t') \varphi(t')
	+ \frac{\varphi(t)}{6}
	    \int_{t_0}^{t}\! dt' \, v_{\scr R}(t\!-\!t') \varphi^2(t') = 0 \, ,
\label{ex4 eom}
\end{equation}
and construct the approximate solution by applying the algorithm outlined
in Sec.~\ref{subsec: Case 1: Linear equation}.
\begin{enumerate}[label=(\roman*)]
\item
Treat all non-local and non-linear terms as
small perturbations,
\begin{align}
\MoveEqLeft[3]
\ddot{\varphi}(t) + M^2 \varphi(t)
    + \varepsilon \frac{\Lambda}{6}
    \varphi^3(t)
\nonumber \\
&
	+ \varepsilon\int_{t_0}^{t}\! dt' \, \pi_{\scr R}(t\!-\!t') \varphi(t')
	+ \varepsilon \frac{\varphi(t)}{6}
	\int_{t_0}^{t}\! dt' \, v_{\scr R}(t\!-\!t') \varphi^2(t') = 0 \, .\label{Case4epsilon}
\end{align}

\item
Assume the solution depends on the two time
scales~$t$ and~$\tau\!=\!\varepsilon t$,
\begin{equation}
\varphi(t) = F(t,\tau; \varepsilon) \, ,
\end{equation}
so that the equation of motion is,
\begin{align}
\MoveEqLeft[1]
\frac{\partial^2 F(t,\tau; \varepsilon)}{\partial t^2}
	+ 2 \varepsilon \frac{\partial^2 F(t,\tau; \varepsilon) }{\partial t \partial \tau}
	+ \varepsilon^2 \frac{\partial^2 F(t,\tau; \varepsilon) }{\partial\tau^2}
	+ M^2 F(t,\tau)
	+ \varepsilon \frac{\Lambda}{6} \bigl[ F(t,\tau; \varepsilon) \bigr]^3
\nonumber  \\
&
	+ \varepsilon\int_{t_0}^{t}\! dt' \, \pi_{\scr R}(t\!-\!t') F(t',\tau'; \varepsilon)
	+ \frac{ \varepsilon }{6} F(t,\tau; \varepsilon)
	    \int_{t_0}^{t} \! dt' \, v_{\scr R}(t\!-\!t')
		\bigl[ F(t',\tau'; \varepsilon) \bigr]^2 = 0 \, .
\end{align}

\item
Assume a perturbative series solution of the form,
\begin{equation}
F(t,\tau;\varepsilon)
	= F_0(t,\tau)
	+ \varepsilon F_1(t,\tau)
	+ \varepsilon^2 F_2(t,\tau) + \dots
\end{equation}

\item
The~$\tau$ dependence in the integrands of the non-local
terms is expanded in~$\varepsilon$ as,
\begin{align}
&
\varepsilon  \int_{t_0}^{t} \! dt' \, \pi_{\scr R}(t\!-\!t') F(t',\tau';\varepsilon) 
	= \varepsilon \int_{t_0}^{t} \! dt' \, \pi_{\scr R}(t\!-\!t') F(t',\tau;\varepsilon)
	+ \mathcal{O}(\varepsilon^2)  \, ,
\\
&
\varepsilon F(t,\tau;\varepsilon) \int_{t_0}^{t} \! dt' \, v_{\scr R}(t\!-\!t') \bigl[ F(t',\tau';\varepsilon) \bigr]^2
\\
&	\hspace{3cm}
	= \varepsilon F(t,\tau;\varepsilon) 
		\int_{t_0}^{t} \! dt' \, v_{\scr R}(t\!-\!t') \bigl[ F(t',\tau;\varepsilon) \bigr]^2
	+ \mathcal{O}(\varepsilon^2)  \, .
\nonumber 
\end{align}

\item
The equation of motion organized in powers of~$\varepsilon$,
\begin{align}
\MoveEqLeft[3]
\biggl[ \frac{\partial^2 F_0(t,\tau) }{ \partial t^2 } + M^2 F_0(t,\tau) \biggr]
	+ \varepsilon \biggl[
	\frac{\partial^2 F_1(t,\tau)}{\partial t^2} + M^2 F_1(t,\tau)
\nonumber \\
&
	+ 2 \frac{\partial^2 F_0(t,\tau) }{ \partial t \partial\tau }
	+ \frac{\Lambda}{6} \bigl[ F_0(t,\tau) \bigr]^3
	+ \int_{t_0}^{t} \! dt'\, \pi_{\scr R}(t\!-\!t') F_0(t',\tau)
\nonumber \\
&
	\hspace{1.5cm}
	+ \frac{F_0(t,\tau) }{6}
	    \int_{t_0}^{t} \! dt'\, v_{\scr R}(t\!-\!t') \bigl[ F_0(t',\tau) \bigr]^2
	\biggr]
	+ \mathcal{O}(\varepsilon^2)
	= 0 \, .
\end{align}

\item
The solution of the leading equation is,
\begin{equation}
F_0(t,\tau) = {\rm Re} \Bigl[ R(\tau) e^{-iM t} \Bigr] \, .
\end{equation}

\item
The subleading equation is,
\begin{align}
\MoveEqLeft[1]
\frac{\partial^2 F_1(t,\tau)}{\partial t^2} + M^2 F_1(t,\tau)
	=
\nonumber \\
&
	2 \, {\rm Re} \biggl\{ e^{-i M t} \biggl[ iM \frac{d R(\tau)}{d\tau}
	- \frac{\Lambda }{16} \bigl[ R(\tau) \bigr]^2 R^*(\tau)
	- \frac{R(\tau)}{2} \int_{t_0}^{t} \! dt'\,
		\pi_{\scr R}(t\!-\!t') e^{i M (t-t') }
\nonumber \\
&	\hspace{0.25cm}
	- \frac{ \bigl[ R(\tau) \bigr]^2 R^*(\tau) }{48}
	    \int_{t_0}^{t} \! dt'\, v_{\scr R}(t\!-\!t')  e^{2iM (t-t')}
	- \frac{ \bigl[ R(\tau) \bigr]^2 R^*(\tau) }{24} \int_{t_0}^{t} \! dt'\, v_{\scr R}(t\!-\!t')
	\biggr]
	\biggr\}
\nonumber \\
&
	- 2 \, {\rm Re} \biggl\{ e^{- 3 i M t}
	    \frac{ \bigl[ R(\tau) \bigr]^3}{48}
	    \biggl[
		\Lambda + \int_{t_0}^{t} \! dt'\, v_{\scr R}(t\!-\!t') e^{2iM (t-t')}
		\biggr]
		\biggr\} \, .
\end{align}
We neglect early time transient effects by taking~$t_0\!\to\!-\infty$,
\begin{subequations}
\begin{align}
&
\int_{t_0}^{t} \! dt'\, \pi_{\scr R}(t\!-\!t')  e^{i\omega(t-t')}
	\approx
	\int_{-\infty}^{t} \! dt'\, \pi_{\scr R}(t\!-\!t')  e^{i\omega(t-t')}
	=
	\widetilde{\pi}_{\scr R}(\omega) \, ,
\\
&
\int_{t_0}^{t} \! dt'\, v_{\scr R}(t\!-\!t')  e^{i\omega(t-t')}
	\approx
	\int_{-\infty}^{t} \! dt'\, v_{\scr R}(t\!-\!t')  e^{i\omega(t-t')}
	=
	\widetilde{v}_{\scr R}(\omega) \, ,
\end{align}
\end{subequations}
where we recognized the Fourier transforms from~(\ref{PiTildeDef}), so that the subleading equation reads,
\begin{align}
\MoveEqLeft[1.5]
\frac{\partial^2 F_1(t,\tau)}{\partial t^2} + M^2 F_1(t,\tau)
\nonumber  \\
={}&
	2 \, {\rm Re} \biggl\{ i M e^{-i M t} \biggl[ \frac{d R(\tau)}{d\tau}
		+ \Bigl( \frac{\gamma}{2} \!+\! i \mu \Bigr) R(\tau) 
		+ \Bigl( \frac{\sigma}{2} \!+\! \frac{i \Lambda }{16M} \!+\! i \alpha \Bigr)
			\bigl[ R(\tau) \bigr]^2 R^*(\tau)
		\biggr]
		\biggr\}
\nonumber \\
&
	+ 2 \, {\rm Re} \biggl\{ i M e^{- 3 i M t} \Bigl( \frac{\sigma}{2} \!+\! \frac{i\Lambda}{48M} \!+\! i \alpha_2 \Bigr)
		\bigl[ R(\tau) \bigr]^3 \biggr\} \, ,
\end{align}
where we defined,
\begin{subequations}
\begin{align} 
&
\mu = \frac{ {\rm Re} \bigl[ \widetilde{\pi}_{\scr R}(M) \bigr]}{2M}  \, ,
\qquad
\gamma = - \frac{ {\rm Im} \bigl[ \widetilde{\pi}_{\scr R}(M) \bigr]}{M}  \, ,
\\
&
\alpha_0 = \frac{ {\rm Re}
	    \bigl[ \widetilde{v}_{\scr R}(0) \bigr] }{24M} \, ,
\qquad
\alpha_2 = \frac{ {\rm Re}
	    \bigl[ \widetilde{v}_{\scr R}(2M) \bigr] }{48M} \, ,
\qquad
\sigma =
    - \frac{ {\rm Im}
        \bigl[ \widetilde{v}_{\scr R}(2M) \bigr] }{24 M} \, ,
\end{align}
\label{full Greek}%
\end{subequations}
and~$\alpha\!=\!\alpha_0 \!+\! \alpha_2$.

\item
Spurious resonances are removed from the subleading correction by requiring,
\begin{equation}
\frac{d R(\tau)}{d\tau}
		+ \Bigl( \frac{\gamma}{2} \!+\! i \mu \Bigr) R(\tau) 
		+ \Bigl( \frac{\sigma}{2} \!+\! \frac{i \Lambda }{16M} \!+\! i \alpha \Bigr)
			\bigl[ R(\tau) \bigr]^2 R^*(\tau)
	= 0 \, .
\label{case 4 removal}
\end{equation}
Adopting a polar representation
for the solution,~$R(\tau) \!=\! A(\tau) e^{-if(\tau)}$, breaks up this complex 
equation into two real ones,
\begin{align}
\frac{dA(\tau)}{d\tau}
	= - \frac{\gamma}{2 } A(\tau)
	- \frac{ \sigma }{2 } \bigl[ A(\tau) \bigr]^3 \, ,
\qquad
\frac{df(\tau)}{d\tau}
	=
	\mu
	+ \Bigl( \frac{\Lambda}{16M }\!+\! \alpha \Bigr) \bigl[ A(\tau) \bigr]^2 \, .
\end{align}
These are readily integrated to give,
\begin{align}
A(\tau) ={}& \frac{A_0 e^{- \frac{1}{2} \gamma \tau } }
	{ \sqrt{ 1 + \frac{\sigma A_0^2}{ \gamma } \bigl( 1 \!-\! e^{-\gamma \tau} \bigr) } } \, ,
\\
f(\tau) ={}& f_0 + \mu \tau
	+ \frac{1}{\sigma} \Bigl(
	    \frac{\Lambda}{16M}
	        \!+\! \alpha \Bigr)
		\ln\biggl[ 1 + \frac{ \sigma A_0^2 }{ \gamma }
		    \bigl( 1 \!-\! e^{-\gamma\tau} \bigr) \biggr] \, ,
\end{align}
where~$A_0$ and~$f_0$ are constants of integration.

\item
The leading approximation for the solution
of Eq.~(\ref{ex4 eom}) is,
\begin{align}
\varphi(t) \approx{}&
	\frac{A_0 e^{- \frac{1}{2} \gamma t } }
	{ \sqrt{ 1 + \frac{\sigma A_0^2}{ \gamma }
	    \bigl( 1 \!-\! e^{-\gamma t} \bigr) } } \,
\label{ex4 solution}
\\
&	\hspace{0.5cm}
	\times
	\cos\biggl\{ f_0 + ( M \!+\! \mu ) t
	+ \frac{1}{\sigma} \Bigl(
	    \frac{\Lambda}{16M }
	    \!+\! \alpha \Bigr)
		\ln\biggl[ 1 + \frac{\sigma A_0^2}{ \gamma } \bigl( 1 \!-\! e^{-\gamma t} \bigr) \biggr] \biggr\} \, ,
\nonumber
\end{align}
where~$A_0$ and~$f_0$ are related to initial conditions
%
%
\begin{figure}[h!]
\footnotesize
\centering
(a) Exponential kernel \hfill
\setlength{\tabcolsep}{3.5pt}
\begin{tabular}{c c c c c | c c c c}
&	\multicolumn{4}{c}{\tt self-energy}	&	\multicolumn{4}{c}{\tt proper 4-vertex}
\\
	&	$a_1/M$		&	$c_1$	&	$\mu/M$	&	$\gamma/M$
	&	$a_1/M$		&	$c_1 \varphi_0^2/M^2$	&	$\alpha\varphi_0^2/M$	&	$\sigma\varphi_0^2/M$
\\
\hline
$\color{yellowish} \blacksquare$	
&	$0.1$	&	$0.05$	&	$0.0002$	&	$0.0001$ 	
&	$1$		&	$2$		&	$-0.078$		&	$0.013$
\\
$\color{bluish} \blacksquare$	
&	$0.1$	&	$0.05$	&	$0.0002$	&	$0.0001$	
&	$3$		&	$2$		&	$-0.094$		&	$0.053$
\\
$\color{reddish} \blacksquare$		
&	$2$		&	$0.05$	&	$-0.0120$	&	$0.0320$	
&	$1$		&	$2$		&	$-0.078$	&	$0.013$
\\
$\color{greenish} \blacksquare$	
&	$2$		&	$0.05$	&	$-0.0120$	&	$0.0320$	
&	$3$		&	$2$		&	$-0.094$		&	$0.053$
\end{tabular}
\includegraphics[width=\linewidth]{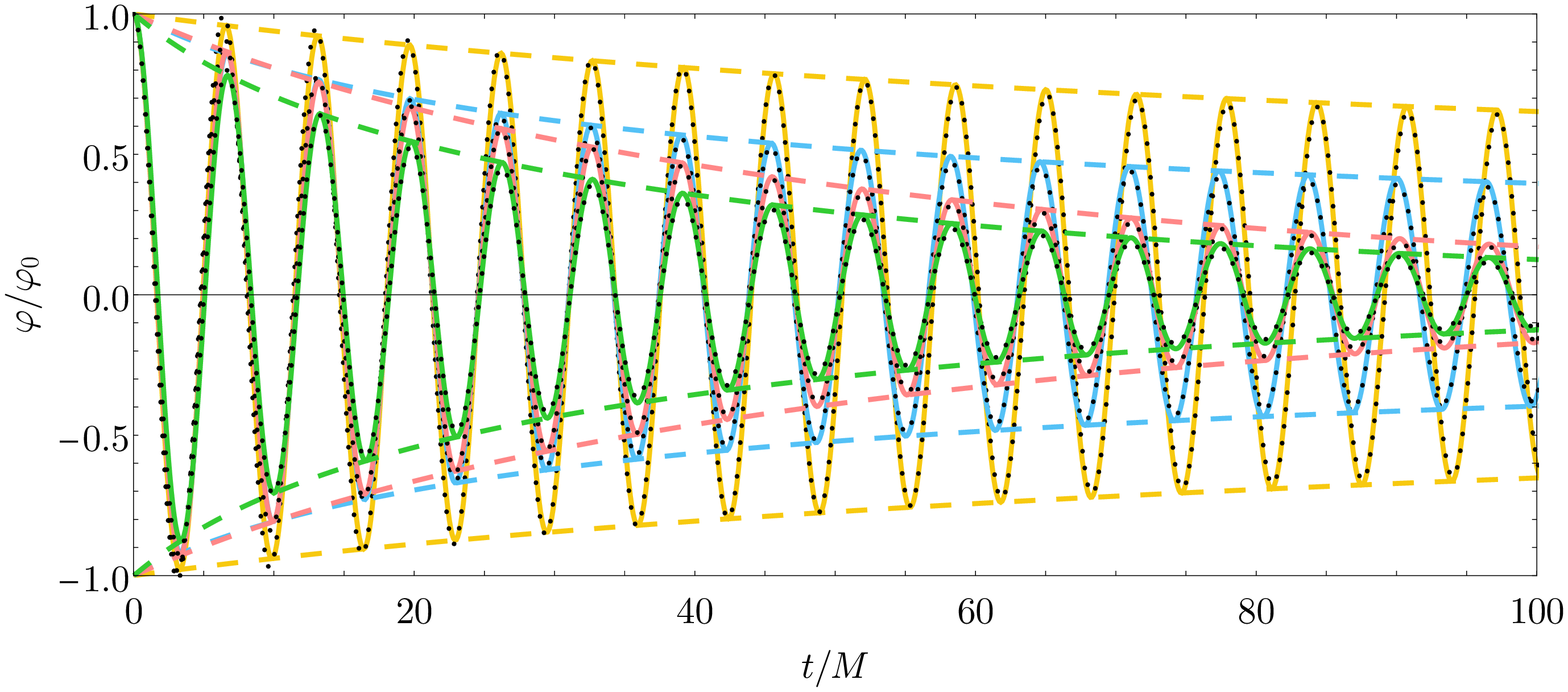}
\\
\vspace{0.1cm}
(b) Lorentzian kernel
\hfill
\begin{tabular}{c c c c c | c c c c}
&	\multicolumn{4}{c}{\tt self-energy}	&	\multicolumn{4}{c}{\tt proper 4-vertex}
\\
	&	$a_1/M$		&	$c_1$	&	$\mu/M$	&	$\gamma/M$
	&	$a_1/M$		&	$c_1 \varphi_0^2/M^2$	&	$\alpha\varphi_0^2/M$	&	$\sigma\varphi_0^2/M$
\\
\hline
$\color{yellowish} \blacksquare$		
&	$0.1$	&	$0.05$	&	$0.0006$	&	$0.00004$ 
&	$0.5$	&	$2$		&	$-0.078$	&	$0.0096$
\\
$\color{bluish} \blacksquare$			
&	$0.1$	&	$0.05$	&	$0.0006$	&	$0.00004$
&	$3$		&	$2$		&	$-0.107$	&	$0.0448$
\\
$\color{reddish} \blacksquare$			
&	$2$		&	$0.05$	&	$-0.0175$	&	$0.02382$
&	$0.5$	&	$2$		&	$-0.078$	&	$0.0096$
\\
$\color{greenish} \blacksquare$		
&	$2$		&	$0.05$	&	$-0.0175$	&	$0.02382$
&	$3$		&	$2$		&	$-0.107$	&	$0.0448$
\end{tabular}
\\
\includegraphics[width=\linewidth]{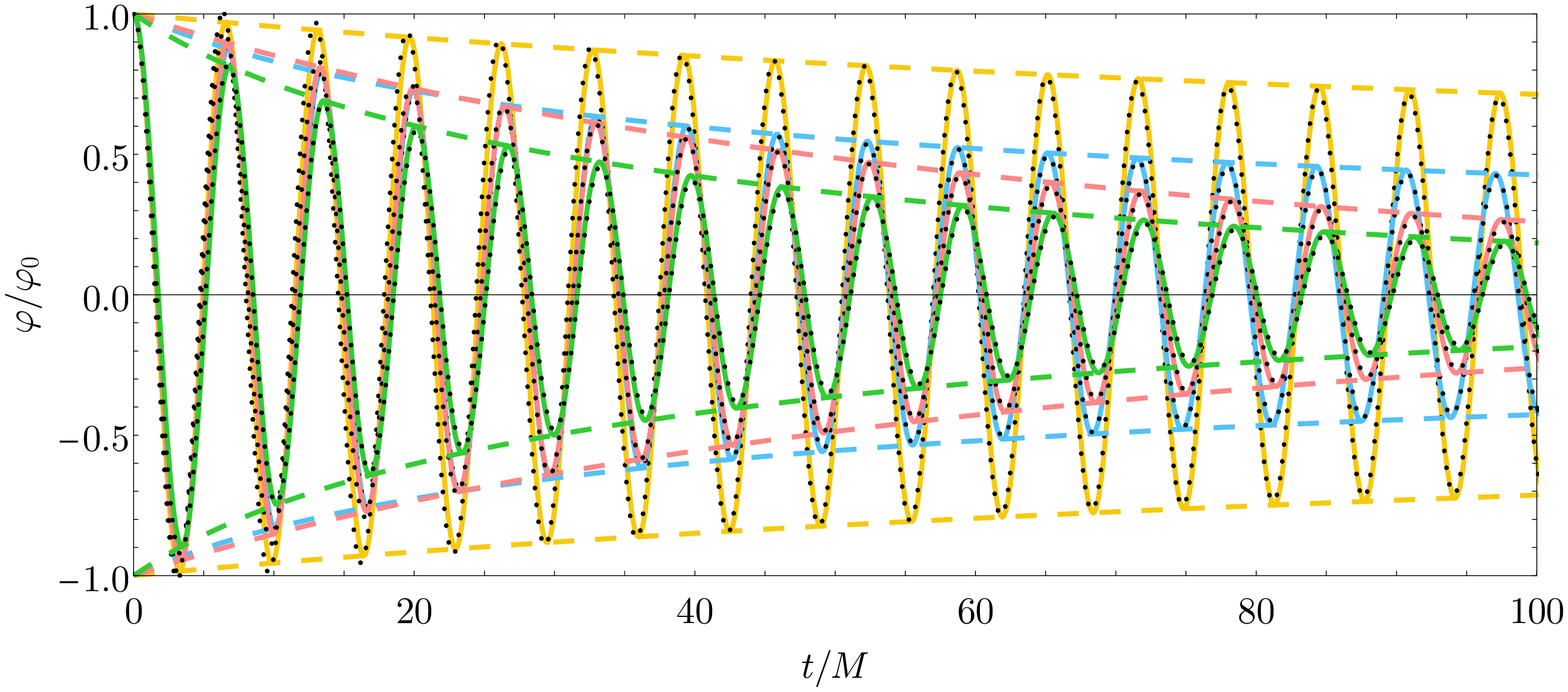}
\\
\vspace{-0.55cm}
\caption{
Solutions of the condensate equation~(\ref{ex4 eom}) containing a local cubic term
and both linear and cubic non-local terms. The self-energy and the proper 4-vertex are taken to be 
(a) the exponential mock kernels~(\ref{eq:kernelExp}),
and (b) the Lorentzian mock kernels~(\ref{eq:kernelLor}), for different choices of 
parameters;~$\Lambda \varphi_0^2 / M^2 \!=\! 0.5$ in all cases.
Analytic approximations~(\ref{ex4 solution}) are drawn in solid coloured curves with respective amplitudes 
in dashed coloured curves; corresponding numerical solutions are indicated by black dotted curves.
}
\label{full_plot}
\end{figure}
%
for the condensate,
\begin{align}
&
\hspace{-0.2cm}
0 =
\bigl( \varphi_0^2 \!-\! A_0^2 \bigr)
    \biggl( M \!+\! \mu
    \!+\! \Bigl( \frac{\Lambda}{16 M} \!+\! \alpha \Bigr) A_0^2 \biggr)^{\!2}
    \! +
    \biggl(
    \dot{\varphi}_0
    \!+\! \frac{ \gamma \!+\! \sigma A_0^2 }{2} \varphi_0
    \biggr)^{\!2}
    \, ,
\\
&
\hspace{-0.25cm}
f_0 = - \arctan\biggl[ \frac{ \frac{\gamma + \sigma A_0^2}{2}
    	\!+\! \frac{\dot{\varphi}_0}{\varphi_0} }
    { M \!+\! \mu \!+\! \bigl( \frac{\Lambda}{16M}
        \!+\! \alpha \bigr) A_0^2 } \biggr] \, .
\end{align}

\end{enumerate}

\medskip

\begin{figure}[h!]
\centering

\includegraphics[width=\linewidth]{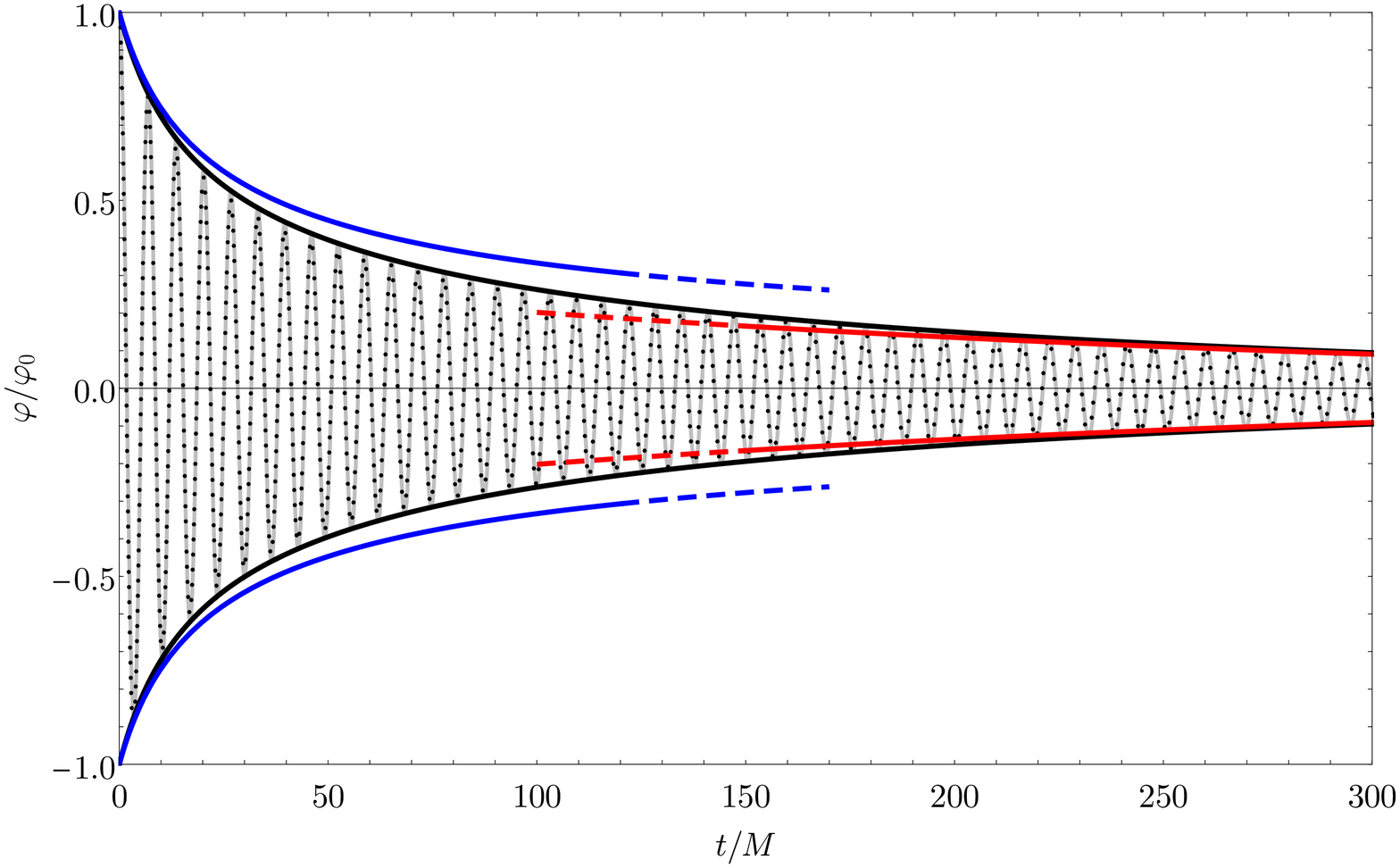}
\\
\vspace{0.4cm}
\includegraphics[width=\linewidth]{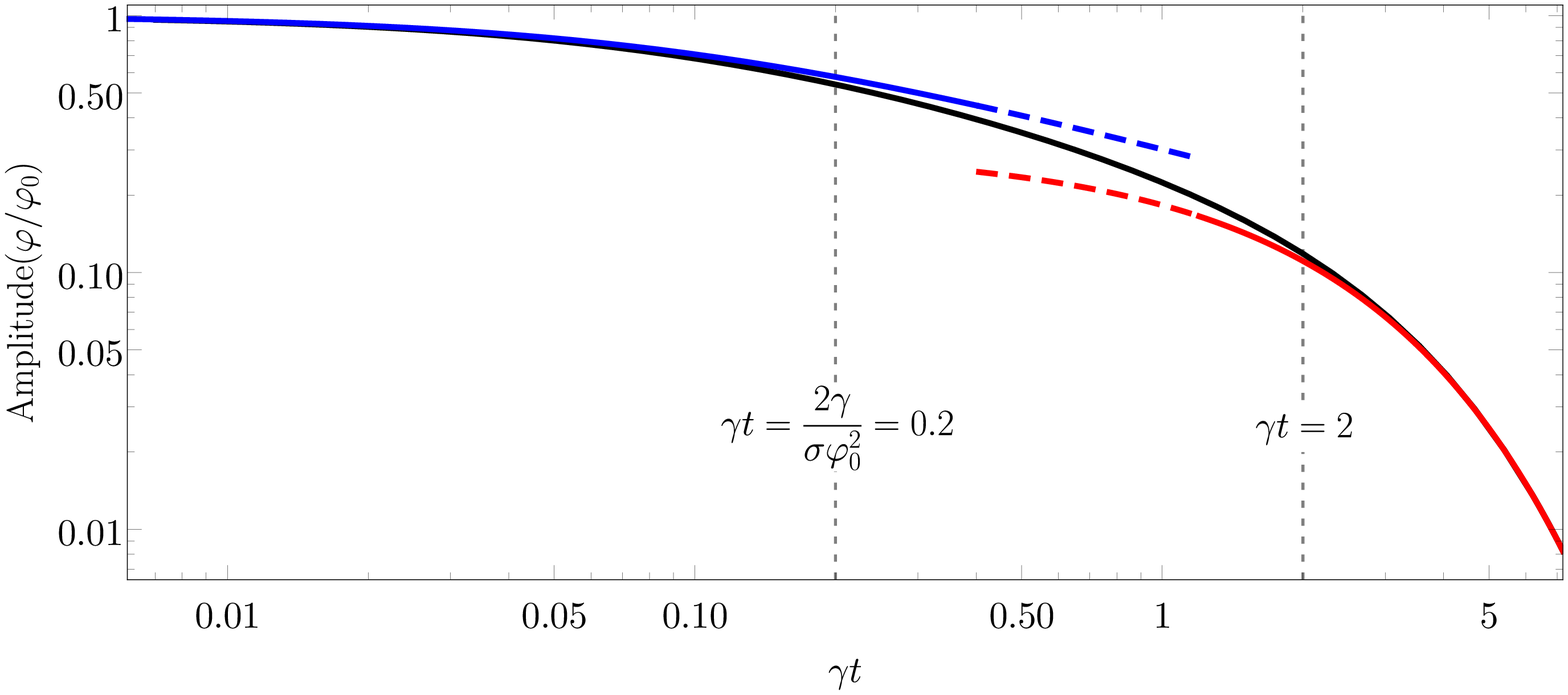}
\caption{
{\it Top:} Plot of the solution for damped oscillations for particular choice of parameters in Eq.~(\ref{ex4 eom});
{\it Bottom:} Log-log plot of the amplitude of oscillations.
Numerical solution for oscillations in dotted black; analytic approximation 
is in solid grey; full amplitude is in solid black; early time limiting amplitude from~(\ref{ex4 early solution}) is in blue; 
late time limiting amplitude from~(\ref{ex4 late solution}) is in red.
Both mock kernels are exponential ones from~(\ref{eq:kernelExp}), with 
parameters~$a_1/M\!=\!0.5$ and~$c_1\!=\!0.05$ for self-energy, 
and~$a_1/M\!=\!4$ and~$c_1\varphi_0^2/M^2\!=\!3$ for proper 4-vertex;
quartic coupling is~$\Lambda \varphi_0^2 / M^2 \!=\! 0.5$.
These choices of parameters imply~$ \sigma \varphi_0^2 /\gamma \!=\! 10$, allowing
for power-law  damping for~$t\!<\! 2/( \sigma \varphi_0^2)$, and
exponential damping for~$t\!>\!2/\gamma$, with a transient period in between,
as depicted on the bottom plot.
}
\label{envelopes}
\end{figure}

\noindent
This approximate solution captures very well the behaviour
of the system, as it shows excellent agreement with numerical solutions given
in Fig.~\ref{full_plot}.
The solution~(\ref{ex4 solution}) incorporates properties
from all three previous cases. Most notably,
the condensate exhibits both the exponential and the
power-law damping, and the interplay of the two depends
on the ratio between the two inverse time
scales~$\gamma/(\sigma A_0^2)$, as is seen from the local damping rate,
\begin{equation}
\Upsilon(t) 
    = \frac{\gamma}{2} + \frac{ \frac{1}{2} e^{-\gamma t} \sigma A_0^2  }
        {1 + \frac{\sigma A_0^2 }{\gamma}
            \bigl( 1 \!-\! e^{-\gamma t} \bigr) } \, .
\label{fullUpsilon}
\end{equation}
At early times the damping is dominated by the
cubic non-locality,
\begin{equation}
\varphi(t)
\widesim[2.]{t \, \ll \, 1/\gamma }
\frac{ A_0 }{ \sqrt{ 1 \!+\! \sigma A_0^2 t } } \,
    \cos\biggl[
    f_0 + (M\!+\!\mu)t
    + \frac{1}{\sigma}
        \Bigl( \frac{\Lambda}{16M} \!+\! \alpha \Bigr)
        \ln\bigl( 1 \!+\! \sigma A_0^2 t \bigr)
    \biggr] \, ,
\label{ex4 early solution}
\end{equation}
which matches the result~(\ref{ex3 solution})
from the preceding section, with the effect of the
local quartic potential included.
Eventually the power-law
damping is shut off as the amplitude of oscillations
becomes smaller, and only the exponential damping
is left.
This is exhibited by the late-time
limit,
\begin{equation}
\varphi(t)
\widesim[2.]{t \, \gg \, 1/\gamma }
	\frac{A_0 e^{- \frac{1}{2} \gamma t } }
	{ \sqrt{ 1 + \frac{\sigma A_0^2}{ \gamma } } } \,
	\cos\biggl[ f_0 + ( M \!+\! \mu ) t
	+ \frac{1}{\sigma} \Bigl(
	    \frac{\Lambda}{16M }
	    \!+\! \alpha \Bigr)
		\ln\Bigl( 1 \!+\! \frac{A_0^2 \sigma}{ \gamma }
		\Bigr) \biggr] \, ,
\label{ex4 late solution}
\end{equation}
taking the form of the solution to the linear
equation~(\ref{ex1 solution}). It is clear that the
non-linear terms in the condensate equation~(\ref{ex4 eom})
affect the transient behaviour, and that the memory of
this period is retained at late times in two aspects:
a constant phase shift, and, more importantly, 
an additional 
factor~$\bigl( 1 \!+\! \frac{\sigma A_0^2}{\gamma} \bigr)^{\!-1/2}$,
which can be relevant depending on the condensate
amplitude at the beginning of the oscillatory period.
The local oscillation frequency,
\begin{equation}
\Omega(t) = M + \mu 
	+ 
	\frac{ \Bigl( \frac{\Lambda A_0^2}{ 16 M} \!+\! \alpha A_0^2 \Bigr)
		e^{-\gamma t} }{1 + \frac{ \sigma A_0^2}{ \gamma} \bigl( 1 \!-\! e^{-\gamma t} \bigr) } \, .
\label{ex4 frequency}
\end{equation}
exhibits similar behaviour where the effects of the cubic non-locality and the quartic
potential are eventually shut off as the condensate relaxes.\
Fig.~\ref{envelopes} illustrates distinct different regimes in 
case of a hierarchy between the dissipation coefficients,~$\sigma \varphi_0^2/\gamma \sim 10$.

%
%
%

An important check of the approximate solutions we found in the four cases are the two assumptions
mentioned at the very beginning of the section,
\begin{itemize}
\item
For all the non-local and non-linear effects to be small enough to consider them as perturbations
we must have that the amplitude and the frequency of oscillations do not change much during one 
oscillation, which implies,~$\gamma, \mu, \sigma \varphi_0^2, \alpha \varphi_0^2 \!\ll\! M$.

\item
The window of the kernels has to be small enough so that slow processes can be neglected within
it (steps (v) in the algorithm), meaning that the amplitude of the solution has to vary only slightly 
within the~$1/a_{1,2}$ extend of mock kernels. This is true 
if~$\gamma, \mu, \sigma \varphi_0^2, \alpha \varphi_0^2 \!\ll\! a_{1,2}$.
\end{itemize}
Both conditions are satisfied for all the plots shown.

\section{Comparison to Markovianised equations}
\label{sec: Comparison to Markovian equations}

In this section we consider a different approach to the non-local equation~(\ref{FullEoM}),
which first calls for approximating it by a Markovian equation. In the first part of this section
we examine in detail
the Markovianisation method appropriate for an oscillating scalar system, and find it 
comes with certain intrinsic limitations stemming from the inability to formulate it
as a systematic expansion in some parameter, and with an ambiguity related to perturbative
manipulations of the local equation of motion. The latter issues can be removed by other 
considerations.

The second part of the section is devoted to solving the Markovianised equation using the
method of multiple-scale perturbation theory developed in the preceding section. We find that
the leading approximate solution matches exactly the one found by solving the non-local
equation directly in Sec.~\ref{subsec: Full equation}. This suggests Markovian equations can
be used if we limit ourselves to precision where subleading corrections are neglected.

\subsection{Markovainising non-local terms}
\label{subsec: Markovainizing non-local terms}

Markovainising (localising) the non-local terms in the condensate equation~(\ref{FullEoM})
is performed utilizing the approach from~\cite{Greiner:1996dx}, which relies on assuming that 
the kernels of non-local terms,~$\pi_{\scr R}(t\!-\!t')$ and~$v_{\scr R}(t\!-\!t')$, 
provide effective windows of some typical width~$\overline{t}$,
beyond which they are suppressed, the precise details depending on the particular model and 
interactions~(cf.~e.g.~\cite{Gautier:2012vh,Drewes:2012qw,Buldgen:2019dus} for a discussion).
This approach was applied in~\cite{Mukaida:2013xxa,Mukaida:2015nos} to the 
equation~(\ref{FullEoM}) we study here, 
but with some differences in the resulting Markovian equation which we comment on below.
Adopting the assumptions given at the beginning of Sec.~\ref{sec: Solving condensate equations}
--- that field elongations are restricted to the mildly non-linear regime,
that coupling constants are perturbatively small, and that kernel windows span no more than 
several oscillations
---
then, within a given window being integrated over, the condensate can be approximated 
by the tree-level solution,
\begin{equation}
\varphi(t') \approx
	\overline{\varphi}_0 \cos(Mt') + \frac{\dot{\overline{\varphi}}_0}{M} \sin(Mt') \, ,
\label{linear expansion}
\end{equation}
where~$\overline{\varphi}_0$ and~$\dot{\overline{\varphi}}_0$ are initial conditions 
at the beginning of  the window. Starting from this assumption, it is a simple 
matter of making use of addition formulas for trigonometric functions to derive the relation,
\begin{equation}
\varphi(t') = \varphi\bigl( t \!-\! (t\!-\!t') \bigr)
	\approx
	\varphi(t) \cos\bigl[ M(t\!-\!t') \bigr]
	- \frac{\dot{\varphi}(t)}{M} \sin\bigl[ M(t\!-\!t') \bigr] \, ,
\label{varphi expansion}
\end{equation}
which effectively Markovianises the non-local terms.
For the linear term this is rather
straightforward.
\begin{align}
\int_{t_0}^{t} \! dt' \, \pi_{\scr R}(t\!-\!t') \, \varphi(t')
	\approx
	2 M \mu\varphi(t)
	+ \gamma \dot{\varphi}(t) \, ,
\end{align}
where the two parameters are,
\begin{align}
&
\mu = \frac{ {\rm Re} \bigl[ \widetilde{\pi}_{\scr R}(M) \bigr] }{2M} \, ,
\qquad \quad
\gamma = - \frac{ {\rm Im}
    \bigl[ \widetilde{\pi}_{\scr R}(M) \bigr] }{M} \, ,
\end{align}
with the Fourier transform defined in~(\ref{PiTildeDef}), and 
where we have neglected early time transient effects by taking~$t_0\!\to\!-\infty$.
In the same manner as in step (vii) of the algorithm outlined in Sec.~\ref{sec: Solving condensate equations}. 

\bigskip

Markovianising the non-linear term superficially seems just as straightforward,
as it involves only a square of the Markovianising expression~(\ref{varphi expansion}),
\begin{align}
\varphi^2(t')
	\approx{}&
	\frac{1}{2} \varphi^2(t)  \biggl( 1 + \cos\bigl[ 2M(t\!-\!t') \bigr] \biggr)
\nonumber \\
&
	- \frac{\varphi(t) \dot{\varphi}(t)}{M} \sin\bigl[ 2M(t\!-\!t') \bigr]
	+ \frac{\dot{\varphi}^2(t) }{2M^2}
		\biggl( 1 - \cos\bigl[ 2M(t\!-\!t') \bigr] \biggr) \, ,
\label{Markovianizing square}
\end{align}
which Markovianises the non-local term,
\begin{equation}
\frac{\varphi(t)}{6} \! \int_{t_0}^{t} \! dt' \, v_{\scr R}(t\!-\!t') \, \varphi^2(t')
	\approx
	2 M \bigl( \alpha_0 \!+\! 2 \alpha_2 \bigr) \varphi^3(t)
	+ 4 \sigma \varphi^2(t) \dot{\varphi}(t)
	+ \frac{ 2 \bigl( \alpha_0 \!-\! 2 \alpha_2 \bigr) }{M}
		\varphi(t) \dot{\varphi}^2(t)
 \, ,
\end{equation}
where the three parameters are
\begin{equation}
\alpha_0 =
	\frac{ {\rm Re} \bigl[\widetilde{v}_{\scr R}(0) \bigr]}{24M}
	\, ,
\qquad \quad
\alpha_2 =
	\frac{ {\rm Re} \bigl[ \widetilde{v}_{\scr R}(2M) \bigr] }{48M} \, ,
\qquad \quad
\sigma = - \frac{ {\rm Im} \bigl[
        \widetilde{v}_{\scr R}(2M) \bigr] }{24M} \, ,
\label{SigmaFromSelfEnergy}
\end{equation}
with the Fourier transform defined in~(\ref{PiTildeDef}), and 
where we have neglected early time transient effects by taking~$t_0\!\to\!-\infty$.
We shall see though there are subtleties involved when considering non-linear
non-local terms.

Having localised both non-local terms we have derived a Markovian condensate equation,
\begin{align}
\MoveEqLeft[5]
\ddot{\varphi}(t)
	+ \frac{ 2 \bigl( \alpha_0 \!-\! 2 \alpha_2 \bigr) }{M}
		\varphi(t) \dot{\varphi}^2(t)
	+ \Bigl[ \gamma 	+ 4 \sigma \varphi^2(t) \Bigr] \dot{\varphi}(t)
\nonumber \\
&
	+ \bigl( M^2 \!+\! 2M\mu \bigr) \varphi(t)
	+ \biggl[ \frac{\Lambda}{6}
		+ 2 M \bigl( \alpha_0 \!+\! 2 \alpha_2 \bigr) \biggr] \varphi^3(t)
	= 0 \, .
\label{non-conserving equation}
\end{align}
This equation exhibits some peculiar properties. To understand what the issue is,
consider the special case of vanishing parameters~$\gamma\!=\!\sigma\!=\!0$ in the dissipative terms,
corresponding
to the physical situation where all on-shell decays are  kinematically forbidden,
and where energy of the oscillating condensate is conserved. However, there seems to be no
conserved energy associated to this equation. One way to see this is to note that 
conservative equations follow from an action principle, and there is no action that would produce the
term~$\varphi\dot{\varphi}^2$, but not simultaneously also~$\varphi^2 \ddot{\varphi}$.
This issue can be addressed by considering more closely the square of the Markovianizing
condensate expansion~(\ref{Markovianizing square}), and realising that at the level of this
approximation we are allowed to freely interchange~$\varphi \!\to\! - \ddot{\varphi} / M^2 $
on the account of the linear tree-level equation, which allows us to write~(\ref{Markovianizing square})
with one free parameter~$\xi$,~\footnote{Considering higher derivatives would 
introduce unphysical behaviour due to the Ostrogradsky instability~\cite{Woodard:2015zca}.
Even though such equations
can still be made sense of in the effective sense~\cite{Glavan:2017srd}, this would defeat any
utility of Markovian equations here.}
\begin{align}
\varphi^2(t')
	\approx{}&
	\biggl[ \frac{(1\!-\!\xi)}{2} \varphi^2(t)
		- \frac{\xi}{2} \frac{\varphi(t) \ddot{\varphi} (t) }{M^2}
		\biggr] \Bigl( 1 + \cos\bigl[ 2M(t\!-\!t') \bigr] \Bigr)
\nonumber \\
&
	- \frac{\varphi(t) \dot{\varphi}(t)}{M} \sin\bigl[ 2M(t\!-\!t') \bigr]
	+ \frac{\dot{\varphi}^2(t) }{2M^2}
		\Bigl( 1 - \cos\bigl[ 2M(t\!-\!t') \bigr] \Bigr) \, ,
\label{quadratic expansion}
\end{align}
leading then to the Markovian condensate equation with the same free parameter,
\begin{align}
\MoveEqLeft[3]
\biggl[ 1 - \frac{2\xi \bigl( \alpha_0 \!+\! 2 \alpha_2 \bigr) }{M}
		\varphi^2(t)\biggr] \ddot{\varphi} (t)
	+ \frac{ 2 \bigl( \alpha_0 \!-\! 2 \alpha_2 \bigr) }{M}
		\varphi(t) \dot{\varphi}^2(t)
	+ \Bigl[ \gamma 	+ 4 \sigma \varphi^2(t) \Bigr] \dot{\varphi}(t)
\nonumber \\
&
	+ \bigl( M^2 \!+\! 2 M \mu \bigr) \varphi(t)
	+ \biggl[
		\frac{\Lambda}{6}
		+
		2(1\!-\!\xi) M \bigl( \alpha_0 \!+\! 2 \alpha_2 \bigr) \biggr] \varphi^3(t)
		= 0 \, .
\label{local eom}
\end{align}
This equation allows to address the energy non-conserving issue with
equation~(\ref{non-conserving equation}) by
choosing~$\xi \!=\! - (\alpha_0 \!-\! 2\alpha_2) /(\alpha_0 \!+\! 2 \alpha_2)$,
which indeed has a conserved energy associated to it
when~$\gamma\!=\!\sigma\!=\!0$.~\footnote{The conserved energy functional in this case is
\begin{equation*}
E =
  \frac{1}{2} \biggl[ 1
    + 2 \frac{\bigl( \alpha_0 \!-\!2 \alpha_2 \bigr)}{M} \varphi^2(t) \biggr] \dot{\varphi}^2(t)
  + \frac{1}{2} \bigl( M^2 \!+\! 2 M \mu \bigr) \varphi^2(t)
  + \biggl[
    \frac{\Lambda}{24}
    \!+\!
    \alpha_0 M  \biggr] \varphi^4(t) \, .
\end{equation*}
}
Nevertheless, we keep the arbitrary parameter in the Markovian Eq.~(\ref{local eom})
when solving it in the following section.
In~\cite{Mukaida:2013xxa,Mukaida:2015nos} the equation~(\ref{local eom}) above
is derived, but with keeping only~$\gamma$ and~$\sigma$ coefficients and neglecting 
coefficients~$\mu$, $\alpha_0$, and~$\alpha_2$. The former are responsible
for dissipative behaviour as they are not time-reversal invariant, while the latter are
responsible for time-dependent frequency corrections to the condensate oscillations,
as can be seen in Sec.~\ref{sec: Solving condensate equations} and as 
we demonstrate below. Even though the physical behaviour
induced by these coefficients is different, they do arise as corrections of the same order 
to the evolution equation for the condensate.

\subsection{Solving Markovian equation}
\label{subsec: Solving Markovian equation}

We found in Sec~\ref{subsec: Full equation} that the approximate solution obtained using
multiple-scale perturbation theory captures the evolution of the system very well up to very late times.
Here we apply the same algorithm to solve the Markovian equation~(\ref{local eom}).

\begin{enumerate}[label=(\roman*)]

\item
Identify small terms by introducing~$\varepsilon$ in equation of motion~(\ref{local eom}),
\begin{align}
\MoveEqLeft[3]
\biggl[ 1 - 2 \varepsilon \xi \frac{ \bigl( \alpha_0 \!+\! 2 \alpha_2 \bigr) }{M}
		\varphi^2(t)\biggr] \ddot{\varphi} (t)
	+ 2 \varepsilon \frac{ \bigl( \alpha_0 \!-\! 2 \alpha_2 \bigr) }{M}
		\varphi(t) \dot{\varphi}^2
	+ \varepsilon \Bigl[ \gamma + 4 \sigma \varphi^2(t) \Bigr] \dot{\varphi}(t)
\nonumber \\
&
	+ \bigl( M^2 \!+\! 2 \varepsilon M \mu \bigr) \varphi(t)
	+ \varepsilon \biggl[
		\frac{\Lambda}{6}
		+
		2(1\!-\!\xi) M \bigl( \alpha_0 \!+\! 2 \alpha_2 \bigr) \biggr] \varphi^3(t)
		= 0 \, .\label{Markovianepsilon}
\end{align}

\item
Assume two time scales~$t$ and~$\tau\!=\!\varepsilon t$ \, ,
\begin{equation}
\varphi(t) = F(t,\tau; \varepsilon) \, ,
\end{equation}
and expand derivatives using the chain rule,
\begin{align}
\dot{\varphi}(t) ={}&
	\frac{\partial F(t,\tau; \varepsilon)}{\partial t} 
		+ \varepsilon \frac{\partial F(t,\tau; \varepsilon) }{ \partial \tau} \, ,
\\
\ddot{\varphi}(t) ={}&
	\frac{\partial^2 F(t,\tau; \varepsilon)}{\partial t^2}
	+ 2 \varepsilon \frac{\partial^2 F(t,\tau; \varepsilon) }{\partial t \partial \tau}
	+ \varepsilon^2 \frac{\partial^2 F(t,\tau; \varepsilon)}{\partial\tau^2} \, .
\end{align}

\item
Assume a perturbative series solution of the form,
\begin{equation}
F(t,\tau; \varepsilon) = F_0(t,\tau) + \varepsilon F_1(t,\tau) + \varepsilon^2 F_2(t,\tau) + \dots
\end{equation}

\item
The step of approximating the non-locality is already accomplished by the 
Markovianisation relation~(\ref{quadratic expansion}).

\item
Organize the equation in powers of~$\varepsilon$.
\begin{align}
\MoveEqLeft[2]
\biggl[ \frac{ \partial^2 F_0(t,\tau)}{ \partial t^2} + M^2 F_0(t,\tau) \biggr]
	+ \varepsilon \Biggl\{
	\frac{ \partial^2 F_1(t,\tau)}{ \partial t^2} + M^2 F_1(t,\tau)
	+ 2 \frac{\partial^2 F_0(t,\tau)}{\partial t \partial \tau}
\nonumber \\
&
	+ 2 M \mu F_0(t,\tau)
	+ \biggl[ \frac{\Lambda}{6} + 2 M \bigl( \alpha_0 \!+\! 2 \alpha_2 \bigr) \biggr] \bigl[ F_0(t,\tau) \bigr]^3
\nonumber \\
&
	+ \Bigl( \gamma + 4 \sigma \bigl[ F_0(t,\tau) \bigr]^2 \Bigr) \frac{\partial F_0(t,\tau)}{\partial t}
	+ 2 \frac{ \bigl( \alpha_0 \!-\! 2 \alpha_2 \bigr) }{M}
		F_0(t,\tau) \Bigl[ \frac{\partial F_0(t,\tau)}{\partial t} \Bigr]^{\!2}
\nonumber \\
&
	- 2 \xi \frac{\bigl( \alpha_0 \!+\! 2\alpha_2 \bigr) }{ M } \bigl[ F_0(t,\tau) \bigr]^2 
		\biggl[ \frac{\partial^2 F_0(t,\tau) }{\partial t^2} + M^2 F_0(t,\tau) \biggr]
	\Biggr\}
	+
	\mathcal{O}(\varepsilon^2)
	= 0 \, .
\end{align}

\item
The solution of the leading equation is,
\begin{equation}
F_0(t,\tau) = {\rm Re} \Bigl[ A(\tau) e^{-iM t} \Bigr] \, .
\end{equation}

\item
The step of neglecting early time transient effects is already 
incorporated into the Markovianisation of non-local terms.

\item
The subleading equation is,
\begin{align}
\MoveEqLeft[1.5]
\frac{\partial^2 F_1}{\partial t^2} + M^2 F_1
\nonumber \\
={}&
	2\, {\rm Re} \biggl\{ i M e^{-iMt} \biggl[
	\frac{d A(\tau)}{d \tau}
	+ \Bigl( \frac{\gamma}{2} \!+\! i \mu \Bigr) A(\tau)
	+ \Bigl( \frac{\sigma}{2} \!+\! \frac{ i \Lambda}{16M} \!+\! \alpha \Bigr)
		\bigl[ A(\tau) \bigr]^2 A^*(\tau)
	\biggr] \biggr\}
\nonumber \\
&
	+ 2 \, {\rm Re} \biggl\{ i M e^{-3iMt} \Bigl(
		\frac{\sigma}{2} + \frac{ i \Lambda}{48M} \!+\! i \alpha_2 \Bigr)
		\bigl[ A(\tau)\bigr]^3 
		\biggr\}  \, .
\end{align}
Note that the dependence on the arbitrary parameter~$\xi$ completely disappears
at this level after solving the leading equation.
Removing spurious resonances from the subleading equation  implies,
\begin{equation}
\frac{d A(\tau)}{d \tau}
	=
	- \Bigl( \frac{\gamma}{2} \!+\! i \mu \Bigr) A(\tau)
	- \Bigl( \frac{\sigma}{2} \!+\! \frac{ i \Lambda}{16M} \!+\! i \alpha \Bigr)
		\bigl[ A(\tau) \bigr]^2 A^*(\tau) \, .
\end{equation}
This is the exact equation for~$A(\tau)$ we derived in Sec.~\ref{subsec: Full equation} when solving the 
non-local equation directly. In fact, the entire subleading equation is exactly the same. Therefore, 
the leading order approximate solution is the same as in~(\ref{ex4 solution}),
\begin{align}
\varphi(t) ={}&
	\frac{R_0 e^{- \frac{1}{2} \gamma t } }
	{ \sqrt{ 1 + \frac{\sigma R_0^2}{ \gamma } \bigl( 1 \!-\! e^{-\gamma t} \bigr) } } \,
\\
&	\hspace{0.cm}
	\times
	\cos\biggl\{ f_0 + ( M \!+\! \mu ) t
	+ \frac{1}{\sigma} \Bigl(
	    \frac{\Lambda}{16M }
	    \!+\! \alpha \Bigr)
		\ln\biggl[ 1 + \frac{R_0^2 \sigma}{ \gamma } \bigl( 1 \!-\! e^{-\gamma t} \bigr) \biggr] \biggr\} \, .
\nonumber
\end{align}

\end{enumerate}
We see that, even though the Markovianized condensate equation~(\ref{local eom}) contains ambiguities,
and even though it is unclear how to write its subleading corrections, it does capture well
the behaviour of the system when solved using the methods from
Sec.~\ref{sec: Solving condensate equations}. The ambiguity is fixed by requiring the
equation to follow from the action principle, in which case it takes the form,
\begin{equation}
\mathcal{K}(\varphi) \ddot{\varphi} 
	- \frac{1}{2} \mathcal{K}'(\varphi) \dot{\varphi}^2
	+ \Gamma(\varphi) \dot{\varphi}
	+ \mathcal{V}'(\varphi) \, ,
\label{Markov equation}
\end{equation}
which is precisely equation~(\ref{proper local eq}) advertised in the introductory section, 
with the three field-dependent functions,
\begin{subequations}
\begin{align}
\mathcal{K}(\varphi) ={}& 1 + \frac{2 \bigl( \alpha_0 \!-\! 2\alpha_2 \bigr) }{M} \varphi^2 \, ,
\\
\Gamma(\varphi) ={}& \gamma + 4 \sigma \varphi^2 \, ,
\label{damping coefficient sec4}
\\
\mathcal{V}(\varphi) ={}&
	\frac{1}{2} \bigl( M^2 \!+\! 2M\mu \bigr) \varphi^2
	+ \Bigl( \frac{\Lambda}{4!} \!+\! 2 M \alpha_2 \Bigr) \varphi^4 \, .
\end{align}
\label{Markov functions}%
\end{subequations}
The issues with not accounting for subleading corrections is tied to the Markovianising relation~(\ref{quadratic expansion}),
which we do not know how to generalize past leading order. Accounting for subleading corrections would require finding a 
more sophisticated Markovianising relation, which likely would require an analysis closely related to directly solving the
non-local equation using multiple-scale perturbation theory, where at each level of perturbation the equations are local.
Nevertheless, with this restriction in mind, Markovianised equations can be used to capture the behaviour of the full non-local
system.

\section{Comparison to time-dependent perturbation theory}
\label{sec: Limitations of time-dependent perturbation theory}


Particle production from an oscillating scalar is often described by the
standard time-dependent perturbation theory. This approach has e.g.~been used to study reheating in~\cite{Ichikawa:2008ne}.
While this method does have its range of applicability, it misses several important effects if applied beyond this range.
Namely, this method applies only to early times, as it neither captures the thermal effects nor the backreactions on the condensate oscillations.
Therefore, in general it is not possible to infer how the condensate dissipates.

In this section we illustrate explicitly the limitations of the time-dependent perturbation theory.
To this end we consider a simplified system in~(\ref{action}),
\begin{equation}
S[\Phi, \chi]  = \int\! d^4 x \, \biggl[
	\frac{1}{2} (\partial_\mu \Phi) (\partial^\mu\Phi)
	+ \frac{1}{2} (\partial_\mu \chi) (\partial^\mu \chi)
	- \frac{m_\phi^2}{2} \Phi^2
	- \frac{m_\chi^2}{2} \chi^2
	- \frac{g}{4} \Phi^2 \chi^2
	\biggr] \, ,
\label{simpler action}
\end{equation}
where the self-interaction coupling constants are set to zero,~$\lambda_\phi\!=\!\lambda_\chi\!=\!0$,
and only the coupling constant~$g$ of the bi-quadratic interaction is kept.
We briefly go through the derivation of how one infers the dissipation of the condensate
by considering the perturbative particle production as a power series in the coupling constant~$g$.
The computation is done at zero temperature
for the sake of simplicity.
Considering thermal effects within this approach is cumbersome at best, and they are often 
neglected within it.
However, reformulating the approach in terms of particle distributions offers a way 
to incorporate thermal effects (cf. e.g.~\cite{Emond:2018ybc,Moroi:2020bkq}),
even though one has to pay attention to the early time restriction. 
Going beyond these requires additional assumptions,
which start resembling more and more the approach we set up in 
sections~\ref{sec: 2PI-resummed effective action}
and~\ref{sec: Solving condensate equations}.

\label{LeadinOrderTimeDependentPertTheo}

The~$\Phi$ field in the action~(\ref{simpler action}) is assumed to have a homogeneous condensate,
which is very close to the classical condensate~$\varphi_\text{cl}$ satisfying the classical equation of motion,
\begin{equation}\label{varphiclansatz}
\ddot{\varphi}_\text{cl}(t) + m_\phi^2 \varphi_\text{cl}(t) = 0 
\qquad \Rightarrow \qquad
\varphi_\text{cl}(t) = \varphi_0 \cos(m_\phi t) \, .
\end{equation}
So it is expanded as~$\Phi(x)\!=\! \varphi_\text{cl}(t) \!+\! \phi(x)$, where~$\phi$ captures its deviation
from the classical behaviour. The~$\chi$ field is assumed to be a spectator at the classical level,
meaning it has no classical condensate (nor can it acquire
it due to quantum corrections because of the~$\mathbb{Z}_2$ symmetry).
The resulting action reads,
\begin{align}
S[\phi, \chi]  ={}& \int\! d^4 x \, \biggl[
	\frac{1}{2} (\partial_\mu \phi) (\partial^\mu\phi)
	+ \frac{1}{2} (\partial_\mu \chi) (\partial^\mu \chi)
	- \frac{m_\phi^2}{2} \phi^2
	- \frac{m_\chi^2}{2} \chi^2
\nonumber \\
&	\hspace{2cm}
	- \frac{g}{4} \varphi_\text{cl}^2 \chi^2
	- \frac{g}{2} \varphi_\text{cl} \phi \chi^2
	- \frac{g}{4} \phi^2 \chi^2
	\biggr] \, .
\label{fluctuation action}
\end{align}
The assumption of the time-dependent perturbation theory is that~$\phi$ and~$\chi$ can be treated
as small perturbations. This is already where the restriction of describing the system at early times
comes in. Quantum effects will introduce dissipation into the evolution of the condensate, which
inevitably leads to large deviations from its classical behaviour. What we will be able to describe
is the early-time damping of the condensate, where its amplitude is attenuated linearly in time.
This is a generic feature of early time dissipation, as is evident from Taylor-expanding all the solutions
from Sec.~\ref{sec: Solving condensate equations} for early times. Unfortunately, it is not possible to
extract how the condensate dissipates at later times from this limited information, as it is not possible to
resum the first two terms of the Taylor series into the full function. Often it is assumed that the condensate
amplitude dissipates exponentially, which is only valid for linear systems, but not in general.
This assumption is based on our intuition on the damped harmonic oscillator, whose evolution is linear.
We should not rely on this intuition to guide our conclusions in non-linear cases.

The evolution of the system in~(\ref{fluctuation action}) is solved for as a power series in~$g$.
The leading order then corresponds to treating~$\phi$ and~$\chi$ as free quantum fields,
that are described in the standard canonical quantization, and their space of states given in the
Fock basis of particle excitations. The free~$\chi$ field operator takes the form,
\begin{equation}
\hat{\chi}_0(t,\bb{x}) =
		\int\! \frac{d^3k}{(2\pi)^{3/2}} \,
	\biggl[ \frac{e^{ - i \omega_{\bb{k} } t + i\bb{k}\cdot\bb{x}} }{\sqrt{ 2 \omega_{\bb{k} } } }
		\hat{a}_{\bb{k}}
		+
		\frac{e^{ i \omega_{\bb{k} } t - i\bb{k}\cdot\bb{x}} }{\sqrt{ 2 \omega_{\bb{k} } } }
			\hat{a}_{\bb{k}}^\dag \biggr] \, ,
\end{equation}
where~$\omega_{\bb{k} } \!=\! \sqrt{ \bb{k}^2 \!+\! m_\chi^2 }$, and where~$\hat{a}_{\bb{k}}$
and~$\hat{a}^\dag_{\bb{k}}$ are annihilation and creation operators satifying canonical
commutation relations,~$ \bigl[ \hat{a}_{\bb{k}} , \hat{a}_{\bb{q}}^\dag \bigr] \!=\! \delta^3(\bb{k} \!-\! \bb{q})$,
and which define the Fock space of particle states in the standard way.
Analogous decomposition applies to the~$\phi$ field operator. 

We proceed by assuming that the initial state is the vacuum of the free theory, and ask the question
of what is the probability that the evolution of the system excites particles from the
vacuum, i.e.\ for the vacuum state to decay. This will be caused by the interaction terms
in the second line of~(\ref{fluctuation action}). In fact,
at leading order in~$g$ only the first
of these interaction terms will contribute,
as that is the only one associated with a kinematically
allowed process.
As this term depends on the classical oscillating
condensate, we will interpret it as the classical oscillations induced decay of the vacuum
into particle states. The basic quantity needed to describe this decay is the transition amplitude
for the initial vacuum state at~$t_0\!=\!0$ to get excited to some other state at
a later time~$t$,
%
\begin{equation}\label{InOut}
\mathcal{A} \Bigl( \big| 0 ;t_0 \big\rangle \to \big| \text{out} ; t \big\rangle \Bigr)
	= \big\langle \text{out}; t \big| \hat{U}_I(t;t_0)  \big| 0 ; t_0 \big\rangle \, ,
\end{equation}
where Dyson's time-evolution operator is,
\begin{align}
\hat{U}_I(t;t_0)
	={}& \mathcal{T} \exp\biggl[ -i \int_{t_0}^{t} \! dt' \, \! : \! \hat{H}_I(t') \! : \biggr]
\nonumber \\
	={}& \bb{1}
		 -i \int_{t_0}^{t} \! dt' \, \! : \!\hat{H}_I(t') \! :
		+ \frac{(-i)^2}{2} \int_{t_0}^{t} \! dt' \, dt'' \,
			{\mathcal{T}} \Bigl\{ : \!\hat{H}_I(t') \! : \, : \! \hat{H}_I(t'') \! : \Bigr\}
			+ \dots \, ,
\label{Dyson series}
\end{align}
and where~$: \ :$ denotes normal ordering of operators that the interaction Hamiltonian~$\hat{H}_I(t)$
is composed of,
\begin{equation}
\hat{H}_I(t) = \int\! d^3x \, \biggl[
	\frac{g}{4} \varphi_\text{cl}^2(t) \hat{\chi}_0^2(t,\vec{x})
	+ \frac{g}{2} \varphi_\text{cl}(t) \hat{\phi}_0(t,\vec{x}) \hat{\chi}_0^2(t,\vec{x})
	+ \frac{g}{4} \hat{\phi}_0^2(t,\vec{x}) \hat{\chi}_0^2(t,\vec{x})
	\biggr] \, .
\end{equation}
At leading order in the coupling~$g$ only the first term above contributes, and limits the
possible~\textit{out} states to the two-$\chi$-particle subspace. The only non-vanishing
transition amplitude at leading order is computed straightforwardly making use of Wick contractions,
\begin{align}
\MoveEqLeft[4]
\mathcal{A} \Bigl( \bigl| 0; t_0 \bigr\rangle \to \bigl| 1^\chi_{\bb{k}} \,1^\chi_{\bb{q}} ; t \bigr\rangle \Bigr)
	= \big\langle 1^\chi_{\bb{k}} \,1^\chi_{\bb{q}} \big|
	(-i) \int_{t_0}^{t} \! dt' \! :\!\hat{H}_I(t') \!: \! \big| 0 \big\rangle
	+ \mathcal{O}(g^2)
\nonumber \\
={}&
	- \frac{ig}{4} \int_{t_0}^{t} \! dt'
		\int\! d^3x
		\, \varphi_\text{cl}^2(t') \,
		\big\langle 0 \big| \hat{a}_{\bb{k}}
			\hat{a}_{\bb{q}} : \! \hat{\chi}_0^2(t',\bb{x}) \!: \! \big| 0 \big\rangle
\nonumber \\
={}&
	\frac{ g \varphi_0^2 }{ 32 \omega_{\bb{k}}  } \delta^3(\bb{k} \!+\! \bb{q})
	\Biggl[
		\frac{ 1 \!-\! e^{ 2 i ( \omega_{\bb{k}} - m_\phi ) t  } }{  \omega_{\bb{k}} - m_\phi  }
		+ \frac{ 1 \!-\! e^{ 2 i ( \omega_{\bb{k}} + m_\phi ) t } }{  \omega_{\bb{k}} + m_\phi  }
		+ \frac{ 2 ( 1 \!-\! e^{ 2 i \omega_{\bb{k}} t } ) }{ \omega_{\bb{k}} } \Biggr] \, ,\label{VacuumToTwo}
\end{align}
where~$\delta^3(\bb{k} \!+ \bb{q})$ signifies spatial momentum conservation and we have substituted $t_0=0$.

One of the usual quantities of interest is the probability density rate for the vacuum
to decay to any other particle state,
\begin{equation}
\frac{d}{dt} \frac{P}{V} = \frac{1}{V} \frac{d}{dt}
	\sum_{|\text{out} \rangle \neq |0\rangle}
	\Bigl| \mathcal{A}\Bigl( \bigl| 0; t_0 \bigr\rangle \to \bigl| \text{out}; t \bigr\rangle \Bigr) \Bigr|^2 \, ,
\label{P general}
\end{equation}
where the 3-dimensional volume is~$V\!=\!(2\pi)^3 \delta^3(\bb{0})$. As already mentioned, at leading order
the sum over the~$out$ states is restricted to the 2-$\chi$-particles subspace, on which the
decomposition of the identity operator reads,
\begin{equation}
\bb{1} \, \Big|_{\text {2-$\chi$-pt.}} = \frac{1}{2} \int \! d^3k \, d^3q \,
	\Bigl( \hat{a}_{\bb{k}}^\dag \hat{a}_{\bb{q}}^\dag \big| 0 \big\rangle \Bigr)
	\Bigl( \big\langle 0 \big| \hat{a}_{\bb{k}} \hat{a}_{\bb{q}} \Bigr) \, ,
\end{equation}
In fact, having cosmological applications in mind, it is better to consider
the rate of change of the expectation value of energy density of produced~$\chi$
particles, which is then given at leading order by,
\begin{equation}
\dot{\rho}_\chi = \frac{1}{2V} \frac{d}{dt}
	\int\! d^3k \, d^3 q \, (\omega_{\bb{k}} \!+\! \omega_{\bb{q}}) \,
	\Bigl| \mathcal{A}\Bigl( \bigl| 0; t_0 \bigr\rangle \to
		\bigl| 1^\chi_{\bb{k}} \, 1^\chi_{\bb{q}}; t \bigr\rangle \Bigr) \Bigr|^2 \, .
\label{rho_chi def}
\end{equation}
These quantities are proportional to each other in the limit of Fermi's golden rule.

Computing the quantity in~(\ref{rho_chi def}) is now done by assuming the time~$t$ to be
much larger than one classical oscillation period,~$t\!\gg\!2\pi/m_\phi$, which is the limit of
Fermi's golden rule, and implies that we are considering quantities averaged over several
oscillation periods. In that limit most of the terms in~(\ref{rho_chi def}) average to zero, and the
only non-vanishing contribution comes from the following
term in the integrand,
\begin{equation}
\frac{\sin[2(\omega_{\bb{k}} \!-\! m_\phi)t]}{2( \omega_{\bb{k}} \!-\! m_\phi) }
\widesim[2.5]{t \, \gg \, 2\pi/m_\phi } \,
	\frac{\pi\delta( \omega_{\bb{k}} \!-\! m_\phi )}{2}\, ,\label{OriginOfFermiRule}
\end{equation}
where the limiting definition for the~$\delta$-function has been
used,~$\sin(zx)/(\pi z) \!\xrightarrow{x\to\infty}\!\delta(z)$.
Making use of this, the rate of change of energy density is,
\begin{align}
\MoveEqLeft[3]
\bigl\langle \dot{\rho}_\chi(t) \bigr\rangle_\text{osc}
	\, \widesim[2.5]{t \, \gg \, 2\pi/m_\phi } \,
	\frac{ g^2 \varphi_0^4 }{256 (2\pi)^3} \int \!
		d^3k \, \biggl\{
		\frac{ \sin[ 2( \omega_{\bb{k}} \!-\! m_\phi ) t ] }
			{ \omega_{\bb{k}} ( \omega_{\bb{k}} \!-\! m_\phi ) }
		+
		\Bigl( \begin{matrix}\text{\ttfamily terms negligible} \\ {\text{\ttfamily after averaging}} \end{matrix} \Bigr)
		\biggr\}
\nonumber \\
\approx{}& \frac{ g^2\varphi_0^4 }{ 512 \pi m_\phi }
    \int_{0}^{\infty}\! dk \, k^2  \,
		\delta( \omega_{\bb{k}} \!-\! m_\phi )
	=
	\theta(m_\phi \!-\! m_\chi) \times
	    \frac{ g^2 \varphi_0^4 }{ 512 \pi }
		\sqrt{ m_\phi^2 - m_\chi^2} \, ,
\label{rho chi result}
\end{align}
where the averaging over classical oscillations is defined as
\begin{equation}
\bigl\langle f(t) \bigr\rangle_\text{osc}
    \equiv \frac{m_\phi}{2\pi}
    \int_{t-\frac{2\pi}{m_\phi}}^{t} \! dt' \,
        f(t') \, .
\end{equation}
This result is, in essence, Fermi's golden rule, in a slightly different form. What this result tells us
is that a condensate oscillating with frequency~$m_\phi$ can produce two~$\chi$ particles
each of energy~$m_\phi\!=\! \sqrt{ m_\chi^2 \!+\! \bb{k}^2 }$, and with spatial momenta~$\pm \bb{k}$.
This can only happen if the kinematic conditions allow, which implies~$m_\phi \!>\! m_\chi$, as indicated
by the step function in the result above.

Note that the energy density of the produced~$\chi$ particles grows linearly in time,
as suggested by the result in~(\ref{rho chi result}). Clearly this cannot proceed forever, and the
result is limited to the regime where the energy density of produced particles is only a fraction
of the energy density~$\rho_\varphi$ stored in the oscillating
condensate,~$\rho_\chi \!\ll\! \rho_\varphi$. Moreover,
this condition has to be satisfied over many oscillation periods for the approximations going into
the result~(\ref{rho chi result}) to be valid. Our next task is to infer what is the correction to the
evolution of the scalar condensate in this regime, so we can compare it to the results of
Sec.~\ref{subsec: Case 3: Cubic non-locality}. This can be done by appealing to energy conservation,
which tells us that,
\begin{equation}
\dot{\rho}_\varphi = - \dot{\rho}_\chi \, .
\end{equation}
Assuming that the energy density of the condensate~$\varphi$ is given by the classical functional,
\begin{equation}
\rho_\varphi = \frac{1}{2} \dot{\varphi}^2 + \frac{m_\phi^2}{2} \varphi^2 \, ,
\end{equation}
and that condensate evolution at leading order in~$g$ takes the form,
\begin{equation}\label{varphiansatz}
\varphi(t) = \varphi_\text{cl}(t) + \delta\varphi(t) \, ,
\qquad \quad
\varphi_\text{cl}(t) = \varphi_0 \cos(m_\phi t) \, ,
\end{equation}
where~$\delta\varphi(t)$ is of order~$g^2$, we get by linearizing the energy conservation equation
in~$\delta\varphi$,
\begin{equation}
\Bigl\langle \dot{\varphi}_\text{cl} \bigl( \delta \ddot{\varphi} + m_\phi^2 \delta\varphi \bigr)
	\Bigr\rangle_\text{osc}
	= - \bigl\langle \dot{\rho}_\chi \bigr\rangle_\text{osc}
	+ \mathcal{O}(g^4) \, ,
\label{averaged eq}
\end{equation}
where the left hand side is averaged over several oscillations in the same sense as the result
in~(\ref{rho chi result}), and where we have used that the purely classical contribution to the
energy density is conserved by itself, and that the classical contribution to the condensate satisfies
the classical equation of motion.
One can see that the averaged equation~(\ref{averaged eq})
is solved by the following ansatz,
\begin{equation}
\delta\varphi (t)
    =
    a t \times \varphi_\text{cl}( t)
    +
    \frac{b t}{m_\phi} \times \dot{\varphi}_\text{cl}(t)\,,
    \qquad
    a,b = \text{const.}
    \, ,
\label{ansatz}
\end{equation}
that vanishes at initial time.
This collapses the equation into an algebraic one,
\begin{equation}
- \bigl\langle \dot{\rho}_\chi \bigr\rangle_\text{osc} =
    2 a \bigl\langle \dot{\varphi}_\text{cl}^2(t)
        \bigr\rangle_\text{osc}
    + \frac{ 2b }{m_\phi} \bigl\langle \dot{\varphi}_\text{cl}(t)
        \ddot{\varphi}_\text{cl}(t)
        \bigr\rangle_\text{osc}
    = m_\phi^2 \varphi_0^2 a \, ,\label{AlgebraitParticleProduction}
\end{equation}
providing a solution for one of the constants,
\begin{equation}
a =
    - \frac{ \bigl\langle \dot{\rho}_\chi \bigr\rangle_\text{osc}}
        { m_\phi^2 \varphi_0^2 }
    = - \frac{ g^2 \varphi_0^2 }{ 512 \pi m_\phi }
		\sqrt{ 1 - \frac{m_\chi^2}{m_\phi^2} } \, .
\end{equation}
Note that due to averaging over oscillations it is not possible to determine
constant~$b$ from ansatz~(\ref{ansatz}).
Therefore, the evolution of the condensate we are able to infer
using the method of time-dependent perturbation theory is
\begin{equation}
\varphi(t) =
    \Biggl[ 1
        - t \!\times\! \frac{ g^2 \varphi_0^2 }{ 512 \pi m_\phi }
		\sqrt{ 1 - \frac{m_\chi^2}{m_\phi^2} } \ \Biggr] \varphi_0 \cos(m_\phi t)
	+
	\varphi_0 \sin(m_\phi t) \! \times \! bt
	\, .\label{PhitInTimeDEpPertTheo}
\end{equation}
Yet again it is clear that this result applies to early times only, in particular it applies in the
regime where
\begin{equation}\label{timerandepertthery}
t \ll \frac{512 \pi m_\phi^2}{ g^2 \varphi_0^2 \sqrt{ m_\phi^2 \!-\! m_\chi^2 } } \, .
\end{equation}
%
It is worth noting that the second term~$b$ is often not
considered at all, in which case the first term is readily
identified as damped classical oscillations.
Comparing this result to the zero temperature limit of the result~(\ref{ex3 solution}) from
Sec.~\ref{subsec: Case 3: Cubic non-locality}, expanded at early times,
\begin{equation}
\varphi(t)
\, \widesim[2.5]{t \, \ll \, 1/(\sigma \varphi_0^2) } \,
\varphi_0 \biggl[ 1 - t \times \frac{\sigma \varphi_0^2}{2} \biggr] \cos(m_\phi t)
	- \varphi_0 \sin(m_\phi t) \times \alpha \varphi_0^2 t \, ,
\end{equation}
we see that~(i) the early time correction that we
were not able to determine is non-vanishing and
corresponds to frequency correction of the condensate
oscillations,
and (ii) the early time correction we did determine indeed
corresponds to damping provided that~$\sigma$
from Sec.~\ref{subsec: Case 3: Cubic non-locality} matches the appropriate coefficient computed here,
\begin{equation}\label{TimeDependentPerturbatnionTheoryDamping}
\sigma = \frac{2 \dot{\rho}_\chi }{ \varphi_0^4 m_\phi^2 }
	= \frac{ g^2 }{ 256 \pi m_\phi }
		\sqrt{ 1 - \frac{m_\chi^2}{m_\phi^2} }  \, ,
\end{equation}
which indeed is the case when the zero temperature limit of the imaginary part
of the proper 4-vertex diagram~(\ref{Im vR omega}) is plugged into the definition of~$\sigma$.
This illustrates that the method of time-dependent perturbation theory
    can capture the behaviour at early times.
    However, it neglects any kind of feedback effects,
    including the decrease in the condensate amplitude due to dissipation,
    the time-dependent frequency shifts due to the non-linearities in the system,
    and the impact of the produced particles on the damping rate.


The restriction to early times comes from assuming that the condensate evolution is approximately the
classical one at leading order.
Inferring the evolution at late times, when the
condensate has dissipated appreciably,  would require resumming an infinite subclass of subleading
corrections in the Dyson series~(\ref{Dyson series}).
This could possibly be circumvented by making more sophisticated assumptions.
Namely, that the condensate evolution has to take the form~$\varphi(t) \!=\! A(t) \varphi_\text{cl}(t)$,
and that the amplitude correction can be effectively treated as approximately constant locally, i.e.,
inside of the integral in~\eqref{VacuumToTwo}.
This assumption is very close to the assumption of step (iv) in the
algorithm of Sec.~\ref{sec: Solving condensate equations}, but it still neglects the frequency correction.
On top of this assumption, one would still need to construct an equation of motion for the condensate
amplitude~$A(t)$, which would yield the late-time evolution. 
Whatever this procedure might be,
if correctly done, this equation
has to take the form of equation for~$A(\tau)$ in~(\ref{ex3 A f eqs}) which follows from multiple-scale
perturbation theory. It becomes clear that accomplishing this would be far more cumbersome
than simply setting up the problem in the non-equilibrium 2PI formalism of
Sec.~\ref{sec: 2PI-resummed effective action}, and solving it using methods of
Sec.~\ref{sec: Solving condensate equations} or Sec.~\ref{sec: Comparison to Markovian equations}.
Moreover, the time-dependent perturbation theory as presented here assumes an initial state without particles in~\eqref{InOut}. This cannot account for important effects that modify the time evolution of $\varphi$, including quantum statistical effects, screening, and scatterings of particles off the field $\varphi$.
Implementing these here would be cumbersome, but the 2PI approach systematically incorporates
them from the beginning through the quantum statistical averages introduced in~\eqref{AverageDef}. 

\bigskip 

By construction
the result~(\ref{PhitInTimeDEpPertTheo}) has a simple physical interpretation
as a decay of the ``vacuum" $\bigl| 0; t_0 \bigr\rangle$ to a two-particle
state~$\bigl| 1^\chi_{\bb{k}} \,1^\chi_{\bb{q}} ; t \bigr\rangle$, induced by the external
field~$\varphi_\text{cl}(t)$. However, the
split into the classical condensate and fluctuations,~$\Phi(x)\!=\! \varphi_{\rm cl}(t) \!+\! \phi(x)$,
is not a physical one since~$\bigl\langle \phi\bigr\rangle\!\neq\!0$. This suggests that we 
view~$\varphi_{\rm cl}$ as a leading contribution to the full 
condensate~$\varphi \!=\! \bigl\langle \Phi \bigr\rangle$. The classical condensate is then seen as 
quantum coherent state built out of a large number of zero mode quanta~$\upphi$,
with an average number of quanta per unit volume being,
\begin{equation}
\overline{n} = \frac{m_\phi \varphi_0^2}{2} \, .
\end{equation}
Now, the production of~$\chi$ particles induced by the external field can better be seen as
the zero mode quanta~$\upphi$ annihilating into~$\chi$ particles,~$\upphi \upphi \!\to\! \chi\chi$.
This process depletes the coherent state making up the condensate, whose evolution then
 necessarily must deviate from~$\varphi_{\rm cl}$ after some time.
The zero mode quanta can also be seen as particles at rest with their energy equalling their mass~$m_\phi$
(but they also have to be seen as completely delocalised). Thus the energy of the process where
two~$\upphi$ quanta annihilate is~$2m_\phi$, which explains the 
kinematic threshold in~\eqref{VacuumToTwo}. The coefficient~$\sigma$ from~(\ref{TimeDependentPerturbatnionTheoryDamping})
carries the information about the damping of the oscillating condensate amplitude at early times,
and we have just argued it is constructed from an amplitude for~$\upphi\upphi \!\to\! \chi\chi$ decay.
This is a special case of a general result of cutting rules that follow from the optical theorem.
The coefficient~$\sigma$ is proportional to the imaginary part of the second diagram
in~(\ref{quartic non-local}), which precisely corresponds to this on-shell decay.

\section{Discussion and conclusion}
\label{sec: Discussion and conclusions}

\noindent {\bf Summary.}
In this work we studied the damping of scalar condensate oscillations in the mildly non-linear regime
in the formalism that naturally accommodates quantum statistical effects of the thermalised 
medium where dissipation takes place.
Thermal corrections to the damping rate are known to be important for applications in cosmology, in particular for the thermal history of the reheating epoch~\cite{Boyanovsky:1995ema,Kolb:2003ke,Yokoyama:2004pf,Mukaida:2012bz,Drewes:2013iaa,Drewes:2014pfa,Drewes:2015coa,Drewes:2017fmn,Co:2020xaf,Lei:2021rbl},
or the production of axions~\cite{Salvio:2013iaa,Carenza:2019vzg} and sterile neutrinos~\cite{Drewes:2015eoa}.
Our starting point was the non-local equation of motion~(\ref{FullEoM})
following from the 2PI-resummed effective action~(\ref{truncated eff action}), 
derived as a loop expansion and a small
field expansion in the 2PI effective action formalism of non-equilibrium quantum field theory.
Even though in Sec.~\ref{sec: 2PI-resummed effective action} we 
derived Eq.~(\ref{FullEoM}) for a particular two-scalar model given by action~(\ref{action}),
the form of the equation is rather generic, and captures a broad class of interactions 
with both bosons and fermions.
The information about particular interactions is encoded in the integral 
kernels~$\pi_{\scr R}(t\!-\!t')$ and~$v_{\scr R}(t\!-\!t')$, which are closely related to 
the retarded self-energies and proper four-vertex functions in a given particle physics model.
In Sec.~\ref{sec: Solving condensate equations} 
we have developed a systematic approximation scheme to directly solve the non-local 
equation~(\ref{FullEoM}) describing a damped oscillating scalar condensate.
We accomplished this by applying multiple-scale perturbation theory,
and were able to find a leading-order analytic approximation
in~(\ref{ex4 solution}) that captures the full solution very well, as demonstrated 
by comparing to exact numerical solutions in Fig.~\ref{full_plot}. 
However, solving non-local equations directly is notoriously difficult, and it is often 
advantageous to approximate 
them by local equations. For that reason in Sec.~\ref{sec: Comparison to Markovian equations}
we considered the Markovianisation procedure~\cite{Greiner:1996dx} applied to our
non-linear equation~(\ref{FullEoM}).
We found an ambiguity in the resulting equation that we were able to fix by requiring 
the equation to follow from an action principle. Solving the resulting Markovian equation using
the same methods of multiple-scale perturbation theory from Sec.~\ref{sec: Solving condensate equations} produces the same 
leading order solution~(\ref{ex4 solution}), justifying the use of Markovian equations.
Examining the more common time-dependent perturbation theory approach
in Sec.~\ref{sec: Limitations of time-dependent perturbation theory},
we found it is neither capable of describing the evolution
beyond early times when feedback effects are very small,
nor does it naturally incorporate thermal statistical
effects as the non-equilibrium 2PI formalism of Sec.~\ref{sec: 2PI-resummed effective action} does.

\bigskip

\noindent {\bf Analytic approximate solution.}
The main feature of the approximate solution~(\ref{ex4 solution}) of the non-linear non-local 
equation~(\ref{FullEoM}) exhibits
are the two competing regimes of damping --- a well known exponential one dominated by 
linear evolution at late times, and a power-law one dominated by non-linear evolution at early and 
intermediate times --- conveniently expressed in terms of a local damping rate,
defined in~(\ref{upsilon definition}) and computed in~(\ref{fullUpsilon}), 
\begin{equation}
\Upsilon(t) 
    \approx \frac{\gamma}{2} + \frac{ \frac{1}{2} e^{-\gamma t} \sigma \varphi_0^2 }
        {1 + \frac{\sigma \varphi_0^2}{\gamma}
            \bigl( 1 \!-\! e^{-\gamma t} \bigr) } \, .
\label{damping rate discussion}
\end{equation}
Another noteworthy feature is the time-dependent frequency
of oscillations, given in Eq.~(\ref{ex4 frequency}).
Depending on the parameters of the
model the power-law damping regime can persist for a significant amount of time, before
the condensate amplitude attenuates enough, and oscillations inevitably transition 
into the exponential damping regime.
Thus, the power-law damping regime can leave imprints on the late time linear
evolution in the form of amplitude correction and frequency shift.
An example of such transition between damping regimes in given in Fig.~\ref{envelopes}.
We emphasise that damping of the oscillations beyond the linear regime in general differs from the exponential damping found for linear oscillators, and depends on the particular dissipative interactions.

\bigskip

\noindent {\bf Thermal effects and microscopic interpretation.}
The damping of the condensate~(\ref{damping rate discussion}) is a macroscopic effect,
but it derives from microscopic dissipative processes. 
At the end of Sec.~\ref{sec: Limitations of time-dependent perturbation theory}
a picture of the condensate made up of a large number of zero mode quanta~$\upphi$
decaying into~$\chi$-particles was presented.
A similar, but more intricate picture applies to the condensate in a thermal medium
of non-vanishing particle numbers that is characterised by a statistical density operator $\varrho$. 
In such a medium the physical excitations are quasi-particles 
that have modified properties compared to particles in vacuum;
in Sec.~\ref{sec: 2PI-resummed effective action} we have recounted how the
first such important modifications arise ---
thermally generated quasi-particle masses, that can significantly modify
kinematics of various processes.
Moreover, new channels of dissipation open up in a thermal medium due to the possibility that zero-mode
quanta~$\upphi$ making up the condensate interact with quasi-particles in the medium.
The connection between the macroscopic behaviour of the condensate and the microphysical
dissipative processes is given by the 
{\it dissipation coefficients}~$\gamma$ and~$\sigma$ appearing in~(\ref{damping rate discussion}),
which are related to imaginary parts
of the self-energy and the proper four-vertex as in~(\ref{full Greek}). The physical interpretation of 
these quantities is systematised by the
cutting rules at finite temperature~\cite{Weldon:1983jn,Kobes:1985kc,Kobes:1986za,Landshoff:1996ta,Gelis:1997zv,Bedaque:1996af}.~\footnote{To some degree these considerations can be 
generalised to genuine non-equilibrium situations, cf.~e.g.~\cite{Arnold:2002zm} and references therein.}
In short, these cutting rules applied to diagrams in~(\ref{quadratic non-local}) related
to the self energy~$\pi_{\scr R}$,
and the ones in~(\ref{quartic non-local}) 
related to the proper four-vertex~$v_{\scr R}$ 
suggest that processes contributing to the~$\gamma$ dissipation coefficient
consist of~$\upphi \!\rightleftarrows\! \chi\chi\phi$
and~$\upphi \rightleftarrows \phi\phi\phi$ and rearrangements thereof 
(e.g.~scattering~$\chi\upphi \!\to\! \chi \phi$),
and that processes contributing to the~$\sigma$ dissipation coefficient consist
of~$\upphi\upphi \!\rightleftarrows\! \chi \chi$,~$\upphi\upphi \!\rightleftarrows\! \phi \phi$
and permutations thereof.
The use of the 2PI formalism for non-equilibrium quantum field theory ensures all the processes contribute
with appropriate weights determined by the Bose-Einstein distribution.
The number of zero mode quanta participating in the microscopic processes can also be
seen from the arguments of~$\widetilde{\pi}_{\scr R}$ and~$\widetilde{v}_{\scr R}$ in~(\ref{full Greek}), 
where it is
the multiple of the tree-level frequency~$M$ that determines the number of quanta.
Even though this discussion is based on the simple two-scalar model in~(\ref{action}),
it is easily generalised to many other models and interactions since all the microphysical processes 
are encoded by the two integral kernels~$\pi_{\scr R}$ and~$v_{\scr R}$ appearing in the
condensate equation of motion~(\ref{FullEoM}). The microphysical interpretation of these coefficients in terms of cutting rules holds irrespectively of the particle content.

\bigskip

Naturally, the result for damping~(\ref{damping rate discussion}) we found is an approximate one,
and that is the reason why only processes involving one and two condensate quanta contribute 
to it. The evolution of the system is non-linear, and will necessarily generate contributions where many more 
condensate quanta decay into two or three~$\phi$ and~$\chi$ particles,
(e.g.~processes such as~$\upphi\upphi\upphi\upphi \!\to\! \chi\chi$ or~$\upphi\upphi\upphi \!\to\! \chi\chi\phi$), that would lead to 
the appearance of higher dissipation coefficients associated with higher number of quanta 
participating in the microphysical process. 
One way to see how higher dissipation coefficients arise is to continue
with the algorithm in Sec.~\ref{sec: Solving condensate equations} 
beyond leading order and work out subleading corrections.
 It should be clear after some inspection that higher harmonics are generated in
 the condensate's evolution and that this leads to 
 non-local terms generating higher dissipation coefficients
at subleading orders.
In addition, were we to include higher loops in 
the effective action~(\ref{truncated eff action}) we would also account for higher order processes
where one or two zero mode quanta decay into four or more~$\phi$ and~$\chi$ quasi-particles,
producing further corrections to dissipation coefficients.
 However, the contributions of higher processes, either as corrections to dissipation coefficients 
 or by generating higher dissipation coefficients,
 are expected to be subleading compared to the leading
order approximation~(\ref{damping rate discussion}), and likely negligible in most applications. 
It is nonetheless important to note that they are present in principle, because it is due
to them that a self-interacting scalar oscillating in the mildly non-linear regime will relax to the 
minimum via dissipating into fluctuations.

The Markovianised equation~(\ref{Markov equation}), on the other hand, will not capture 
the higher dissipation coefficients. By construction, it contains
only~$\gamma$ and~$\sigma$ coefficients inside of the damping coefficient 
in~(\ref{damping coefficient sec4}),
\begin{equation}
\Gamma(\varphi) = \gamma + 4 \sigma \varphi^2 \, ,
\label{damping coefficient discussion}
\end{equation}
and solving it to subleading orders will never generate any
other dissipation coefficients apart from the products of these two.
Going past this would require generalising the main Markovianisation
relation~(\ref{varphi expansion}) to include subleading corrections, 
which is not clear how it should be accomplished.
Nevertheless, Markovianised equation captures the behaviour of the system very well, and as long
as extreme precision is not required they are more than adequate to describe the system, as long as
we fix the ambiguities arising by requiring that it follows from an action principle.

\bigskip

\noindent{\bf Damping coefficient vs.~dissipation rate.}
It is very important to note that the local damping rate~$\Upsilon(t)$ given 
in~(\ref{damping rate discussion})
is {\it not} the same thing as the damping coefficient~$\Gamma(\varphi)$
in~(\ref{damping coefficient discussion}) above,
that appears in the Markovianised equation, and is of key interest in deriving effective 
quantum kinetic equations that generalise the standard Boltzmann equations.
The two are sometimes conflated since in the
case of linear condensate evolution, that is most often studied, 
the two are just constants of the same dimension,
and thus proportional to each 
other,~$\Gamma(\varphi) \big|_{\sigma = 0} \!=\! 2 \Upsilon(t) \big|_{\sigma = 0}\!=\! \gamma$.
However, in general the local damping rate is a property of the solution and
therefore is a function of time,
while the damping coefficient is a property of the Markovian equation of motion and 
therefore is a function of the
condensate. Whenever a diagram that contributes to dissipation descends from an interaction term 
containing more than one condensate field it is crucial to draw this distinction.~\footnote{It also
makes no sense to try to establish a connection between and~$\Upsilon(t)$ and~$\Gamma(\varphi)$
by plugging the solution~$\varphi(t)$ into the latter and averaging over oscillations. For once, we would
already need to have a solution to start with, and would not need any other information, and moreover
the result of that procedure would not match~$\Upsilon(t)$. }

While solving the non-local equation directly and solving the Markovianised equation both
produce the same local damping rate~(\ref{damping rate discussion}) and the same
local frequency~(\ref{ex4 frequency}), 
the time-dependent perturbation theory we examined in 
Sec.~\ref{sec: Limitations of time-dependent perturbation theory} cannot 
describe the evolution of the condensate beyond very early times where damping is almost
negligible and only linear in time. This is evident from 
the fact that the method produces a constant local damping rate,
\begin{equation}
\Upsilon = \frac{\sigma \varphi_0^2}{2} \, ,
\end{equation}
which is a zero temperature and early time limit of~(\ref{damping rate discussion}). In addition to
not incorporating thermal statistical effects (except if formulated in terms of occupation 
numbers), this method also cannot account for feedback effects and damping. 
Both shortcomings make the application of time-dependent perturbation theory for the computation of transport coefficients in quantum kinetic equations questionable.
Nevertheless,
this method is sometimes used to work out the damping coefficients in Markovianised equations.
While this works in cases where condensate evolution is linear (most often cubic 
interactions~$\Phi\chi^2$ are considered), in non-linear cases such as one considered in 
Sec.~\ref{sec: Limitations of time-dependent perturbation theory}
the heuristic 
prescription~$\Gamma(\varphi) \!=\! 2\Upsilon \big|_{\varphi_0 \to \varphi} \!=\! \sigma \varphi^2$
definitively fails, as it is off by a factor of 4 compared to the correct 
result~(\ref{damping coefficient discussion}).

\bigskip

\noindent {\bf Limits of applicability.}
Having discussed the physical interpretation of our results, and different methods and
approaches of obtaining them, it is important to discuss the limitations of the results ---
the restriction to
the mildly non-linear regime, and perturbative processes, 
implemented via the small field expansion introduced
in Sec.~\ref{sec: 2PI-resummed effective action}. By construction,
this expansion neglects the non-perturbative effect of resonant particle 
production~\cite{Shtanov:1994ce,Kofman:1994rk,Kofman:1997yn}, 
which puts a restriction on the physical systems we can describe.
Even though when present the resonant particle production 
 typically is the dominant mechanism of dissipation in oscillating systems, 
there are potentially interesting regimes where this effect is subdominant,
and where dissipation is dominated by perturbative processes. Such regimes
are well described by the formalism outlined in Sec.~\ref{sec: 2PI-resummed effective action}
--- the small field expansion within the 2PI formulation of the Schwinger-Keldysh
formalism for non-equilibrium quantum field theory. The conditions restricting a given system
to such a regime are illustrated on the example of the simple two-scalar model~(\ref{action})
we considered here, but are generalised to more complex models in a straightforward manner.

There are two similar, but in principle distinct conditions which determine the applicability 
of our description. To understand both it is useful to note that the full effective 
masses,~$\mathcal{M}_{\phi,\chi}^2\!=\! M_{\phi, \chi}^2\!+\!\Delta m_{\phi,\chi}^2$,
of~$\phi$ and~$\chi$ quasi-particles do not consist solely of constant
tree-level and thermal contributions~$M_{\phi,\chi}^2$ from~(\ref{thermal masses}), but also receive time-dependent 
contributions~$\Delta m_{\phi,\chi}^2$ on the account of coupling to the oscillating
condensate~$\varphi(t)$ at the linear level. These can be read off from the inverse tree-level 
propagators in~(\ref{Gs}),
\begin{equation}
\Delta m_\phi^2 \! =\! \frac{\lambda_\phi \varphi^2}{2} \, ,
\qquad \quad
\Delta m_\chi^2 \! =\! \frac{g \varphi^2}{2} \, .
\end{equation}
The small field expansion used throughout this paper
assumes these time-dependent contributions are much smaller than the
respective tree-level and thermal masses,
\begin{equation}
\frac{ \Delta m_\phi^2 }{ M_\phi^2 } \ll 1 \, ,
\qquad \quad
\frac{ \Delta m_\chi^2 }{ M_\chi^2 } \ll 1 \, .
\label{small delta m}
\end{equation}
These restrictions justify the solutions  for two-point functions
in~(\ref{solid propagator solution}) and~(\ref{dashed propagator solution})
being given as a small field expansion, which is controlled by the 
small parameters in~(\ref{small delta m}) or~(\ref{par limit small}).
Since for simplicity we had presented the formalism 
in Sec.~\ref{sec: 2PI-resummed effective action} in the high temperature
limit, where thermal masses dominate over the tree-level ones, 
we quote the restrictions in the same limit of the two-scalar model, 
where the small field conditions above
translate into
\begin{equation}
\frac{\varphi_0^2}{T^2} 
	\ll \frac{1}{12} \Bigl( 1 \!+\! \frac{g}{\lambda_\phi} \Bigr)
	\, ,
\qquad \quad
\frac{\varphi_0^2}{T^2} 
	\ll \frac{1}{12} \Bigl( 1 \!+\! \frac{\lambda_\chi}{g} \Bigr)
	\, .
\label{par limit small}
\end{equation}
for the particular two-scalar model, and are expressed as restrictions on the amplitude of oscillations
in units of the temperature. The range of values that the initial condensate amplitude~$\varphi_0$
is allowed to take crucially depends on the coupling constants and possible hierarchy between them.

In addition to assuming the small field expansion, we are neglecting the non-perturbative effect
of resonant particle production, which is absent provided the effective quasi-particle
masses change only adiabatically,
\begin{equation}
\frac{\dot{\mathcal{M}}_\phi }{ \mathcal{M}_\phi^2 } \ll 1 \, ,
\qquad \quad
\frac{\dot{\mathcal{M}}_\chi }{ \mathcal{M}_\chi^2 } \ll 1 \, .
\end{equation}
For the simple two-scalar model this implies
\begin{equation}
\frac{\varphi_0^2}{T^2} \ll \frac{1}{12} \Bigl( 1 \!+\! \frac{g}{\lambda_\phi} \Bigr) \, ,
\qquad \quad
\frac{\varphi_0^2}{T^2} \ll \frac{1}{12} \Bigl( 1 \!+\! \frac{\lambda_\chi}{g} \Bigr) 
	\sqrt{ \frac{ \lambda_\chi \!+\! g }{ \lambda_\phi \!+\! g } }
	\, ,
\label{par limit adiabatic}
\end{equation}
where we also made use of the already assumed small field restriction~(\ref{small delta m}).
Note that the left condition above coming from~$\phi$ quasi-particles is the same as the 
respective condition from the small field expansion~(\ref{par limit small}), 
while the one coming from~$\chi$ quasi-particles differs. The reason behind this is
that the frequency at which the effective masses oscillate are set by the thermal mass of the condensate,
which equals the thermal mass of~$\phi$ quasi-particles, but differs in general from the one 
of~$\chi$ quasi-particles.

It is interesting to ask whether the transition between damping regimes seen in 
Fig.~\ref{envelopes}, which requires~$\sigma \varphi_0^2/\gamma \!\gg\! 1$, 
could be supported by the simple two-scalar model~(\ref{action}). 
In answering this question we are limited by the knowledge of two-loop self-energy
from~(\ref{PiDef}). We only have an estimate for the part coming from
self-interactions governed by the coupling constant~$\lambda_\phi$,
in the upper case of~(\ref{ThermalScatterings}), and that only in the high temperature limit.
That is the dominant contribution to self-energy if we assume a
hierarchy~$\lambda_\phi \!\gg\! g$. In that case conditions~(\ref{par limit small})
and~(\ref{par limit adiabatic}) imply~$\varphi_0^2/T^2 \!\ll\! \sqrt{g/\lambda_\phi}$, and the relevant ratio
 is~$\sigma \varphi_0^2/\gamma \!\sim\! [\varphi_0^2/T^2] \!\times\! [g^2/\lambda_\phi^2] \!\times\! \lambda_\phi^{-1/2}$, which could possibly be larger than one if the coupling~$\lambda_\phi$ is small enough
to overcome the other two small ratios, confirming the relevance of our
result~(\ref{ex4 solution}) even in this simple model. On the other hand, assuming a different
hierarchy~$\lambda_\chi \!\gg\! g \!\gg\! \lambda_\phi$ would allow the 
for~$\varphi_0^2/T^2 \!\gtrsim\! 1 $, and, assuming that the self-energy contribution governed by the
coupling constant~$g$ would take the same form as the one in~(\ref{ThermalScatterings}), we 
get for the relevant ratio to be of the order~$\sigma \varphi_0^2/\gamma \!\sim\! [\varphi_0^2/T^2] \!\times\! \lambda_\phi^{-1/2}$,
which could indeed be very large. However, the kinematic condition
from~(\ref{Im vR omega T}) in this case is not met, and because of it~$\sigma$ 
dissipation coefficient actually vanishes.
Nevertheless, assuming this hierarchy for a system in vacuum (at vanishing temperature) might
lead to different conclusions, as the restrictions~(\ref{par limit small}) 
and~(\ref{par limit adiabatic}) would be modified and
the kinematic condition in~(\ref{Im vR omega}) could be met by tree-level masses, 
though some of the approximations we make to derive
the results might not apply in that case~\cite{Boyanovsky:1994me}. More complex models 
should allow for interesting regimes more readily. Our work provides a description of 
the evolution of such systems with an oscillating scalar condensate, once the appropriate 
self-energies and proper four-vertices are computed in a given particle physics model.
Detailed studies of phenomenological applications of the results
presented here are left for future work.

\bigskip

\noindent {\bf Conclusion.}
Non-linear effects in the evolution of an scalar condensate oscillating in a medium can be relevant 
even in  the regime where non-perturbative resonant particle production is negligible.
This regime, which we refer to as the mildly non-linear regime, has not been investigated 
nearly as much as the linear regime or the non-perturbative regime, and this work helps to 
fill this gap~(see also~\cite{Mukaida:2013xxa} for a recent study). 
The non-equilibrium and the non-linear nature of the problem necessitates a judicious choice
of formalism and approximation methods for describing the evolution. The 2PI formalism 
in the Schwinger-Keldysh formulation we utilize is appropriate for describing quantum and statistical
effects, and yields an effective  non-local equation of motion for the scalar condensate~(\ref{FullEoM}). 
This macroscopic equation is connected to elementary microscopic processes via the self-energy
and the proper four-vertex calculable in a given particle physics model. We examined solutions of
Eq.~(\ref{FullEoM}) and methods of obtaining them. The most reliable, though most difficult, approach
is solving the non-local equation directly. This we accomplished using multiple-scale perturbation theory,
appropriate for time-dependent problems,
to find the effects of non-linear evolution in solution~(\ref{ex4 solution}) --- power-law 
damping~(\ref{damping rate discussion})
and time-dependent frequency shifts --- in addition to the well known effects of exponential damping 
and constant frequency shifts known from linear evolution. A considerably simpler and far more
used approach to solving equations such as~(\ref{FullEoM}) is approximating them by Markovian
counterparts. We found this approach to describe the system at leading order equally well as solving
the non-local equation directly, though some care is necessary when attempting more precise
descriptions. Standard time-dependent perturbation theory fails to describe systems for long
times, and is able to describe the oscillating condensate only during early times
while its damping is negligible.
Our results provide a guideline for the appropriate choice of methods when describing dissipative effects and particle production in the early Universe, e.g.~during reheating, Dark Matter production or moduli decay. Since they were obtained with minimal asumptions, they may be applied to a wide range of other systems 
that can effectively be described by an oscillating scalar field.

\acknowledgments

We are grateful to Gilles Buldgen for his contributions to this article,
and we thank Oleg Lebedev for inspiring discussions that motivated this project. 
This work was partially supported
by the FSR Postdoc incoming fellowship of UCLouvain;
by the Czech Science Foundation (GA\v{C}R) grant 20-28525S;
by the Swiss National Science Foundation (project number 200020/175502).


\bibliographystyle{JHEP}
\bibliography{bibliography.bib}

\end{document}